\documentclass[epj,nopacs]{svjour}

\usepackage{graphicx}
\usepackage{atlasphysics}
\usepackage{subfigure}
\usepackage{hyperref} 
\begin{document}

%%%%%%%%%%%%%%%%%%%%%%%%%%%%%%%%%%%%
%          Definitions             %
%%%%%%%%%%%%%%%%%%%%%%%%%%%%%%%%%%%%
\def\invpb{\ensuremath{\mathrm{pb^{-1}}}}
\def\mee{\ensuremath{m_{ee}}}
\def\MT{\ensuremath{m_\mathrm{T}}}

\def\etconecorr20{\ensuremath{\ET^\mathrm{cone0.2,\,corr}}}
\def\etcone30{\ensuremath{\ET^\mathrm{cone0.3}}}
\def\ptcone30{\ensuremath{\pT^\mathrm{cone0.3}}}
\def\Isol{\ensuremath{\textit{I}_{\Delta R=0.4}}}

\def\chargeMisID{\ensuremath{\epsilon_\mathrm{QmisID}}}
\def\ETtag{\ensuremath{\ET^\mathrm{tag}}}

\def\loose{\ensuremath{\textit{loose}}}
\def\medium{\ensuremath{\textit{medium}}}
\def\tight{\ensuremath{\textit{tight}}}
\def\forwardloose{\ensuremath{\textit{forward loose}}}
\def\forwardtight{\ensuremath{\textit{forward tight}}}
\def\tag{\ensuremath{\mathrm{tag}}}
\def\tags{\ensuremath{\mathrm{tags}}}
\def\probe{\ensuremath{\mathrm{probe}}}
\def\probes{\ensuremath{\mathrm{probes}}}
\def\TandP{\ensuremath{\mathrm{T\&P}}}
\def\StoB{\ensuremath{\mathrm{S/B}}}
\def\SS{\ensuremath{\mathrm{SS}}}
\def\OS{\ensuremath{\mathrm{OS}}}
\def\endcap{\ensuremath{\mathrm{endcap}}}
\def\Endcap{\ensuremath{\mathrm{Endcap}}}
\def\endplate{\ensuremath{\mathrm{endplate}}}
\def\etaphi{\ensuremath{\eta\times\phi}}

%%%%%%%%%%%%%%%%%%%%%%%%%%%%%%%%%%%%
%           Title page             % 
%%%%%%%%%%%%%%%%%%%%%%%%%%%%%%%%%%%%
\titlerunning{Electron performance measurements
with the ATLAS detector}
\title{\vspace{-3.0cm}\flushleft{\normalsize\normalfont{CERN-PH-EP-2011-117}} 
\flushright{\vspace{-0.86cm}\normalsize\normalfont{Submitted to Eur. Phys. J. C}} \\
[3.0cm] \flushleft{Electron performance measurements
with the ATLAS detector 
using the 2010 LHC proton-proton collision data}}
\author{The ATLAS Collaboration}
\institute{}

%\date{\today}
\date{Received: October 14, 2011 / Revised version: February 17, 2012}

\abstract{
  Detailed measurements of the electron performance of the ATLAS detector 
  at the LHC are reported,
  using decays of the \Zboson, \Wboson\ and \Jpsi\ particles.
  Data collected in 2010 at \rts=7 TeV are used, 
  corresponding to an integrated luminosity of almost 40~\invpb. 
  The inter-alignment of the inner detector and the electromagnetic calorimeter,
  the determination of the electron energy scale and resolution, 
  and the performance in terms of response uniformity and linearity
  are discussed.  
  The electron identification, reconstruction and trigger efficiencies,
  as well as the charge misidentification probability, are also presented.  
\PACS{{}{}}
} 

%\PACS{
%      {PACS-key}{discribing text of that key}   \and
%      {PACS-key}{discribing text of that key}
%     } % end of PACS codes

\maketitle
\sloppy

%%%%%%%%%%%%%%%%%%%%%%%%%%%%%%%%%%%%
%            Content               % 
%%%%%%%%%%%%%%%%%%%%%%%%%%%%%%%%%%%%

\section{Introduction}
\label{sec:Intro}
The precise determination of the electron performance of the ATLAS detector at
the LHC is essential  both for Standard Model measurements and for searches for
Higgs bosons and other new phenomena. 
Physics processes of prime interest at the LHC are expected to produce electrons from a few GeV
to several TeV. Many of them, such as Higgs-boson production, have small cross-sec\-tions
and suffer from large background, typically from jets of hadrons. 
Therefore an excellent electron identification capability, with high efficiency and
high jet rejection rate, is required over a broad energy range 
to overcome the low signal-to-background ratio.
For example, 
in the moderate transverse energy region $\ET=20-50$~GeV a jet-rejection factor of about $10^5$
is desirable to extract a pure signal of electrons above the residual 
background from jets faking electrons. In the central region up to $|\eta|<2.5$,
this challenge is faced by using a powerful combination of detector technologies: 
silicon detectors, a transition radiation tracker and a longitudinally layered 
electromagnetic calorimeter system with fine lateral segmentation.

A further strength of the ATLAS detector is its ability to reconstruct and
identify electrons outside the tracking coverage up to $|\eta|<4.9$. This brings
several advantages. For example, it improves the sensitivity of the measurement of
forward-backward asymmetry,  and therefore the weak mixing angle, in \Zee\ 
events, 
and it enlarges the geometrical acceptance of searches for Higgs bosons
and other new particles.

To realize the full physics potential of the LHC, 
the electron energy and momentum must be precisely measured.
Stringent requirements on the alignment 
and on the calibration of the calorimeter come, 
for example, from the goal of a high-precision \Wboson\ mass measurement.

This paper describes the measure\-ments 
of the electron energy scale and resolution and of the efficiency
to trigger, reconstruct and identify electrons using \Zee, \Wen\ and \Jee\
events observed in the data collected in 2010 at a centre-of-mass energy of
$\rts = 7$~TeV, corresponding to an integrated luminosity of almost 40~\invpb. 
As the available statistics are significantly lower for isolated electrons from
\Jee\ decays and these electrons are also more difficult to extract, only a
subset of the measurements were performed in this channel. 

The structure of the paper is 
the following. 
In Section~\ref{sec:Detector},
a brief reminder of the inner detector and calorimeter
system is presented. 
The data and Monte Carlo (MC) samples used in this work are summarized
in Section~\ref{sec:Samples}. 
Section~\ref{sec:Algorithms} starts with
the introduction of the trigger, reconstruction and
identification algorithms 
and then proceeds by presenting the
inclusive single and dielectron spectra in Subsection~\ref{sec:Spectra}. 
The inter-alignment of the inner
detector and the electromagnetic (EM) calorimeter is discussed in
Subsection~\ref{sec:Alignment}. 
The in-situ calibration of the electron energy
scale is described in Section~\ref{sec:CalibrationTop} followed by its
performance in terms of resolution, linearity in energy, 
and uniformity in $\phi$.
The measurement of the electron selection efficiencies with the tag-and-probe
technique is presented in Section~\ref{sec:Efficiencies}. 
The identification efficiency determination is discussed in 
detail in Subsection~\ref{sec:IDPerf}, 
and the differences observed between data and MC
predictions are attributed to imperfections of the MC description of  
the main discriminating variables. 
The reconstruction efficiency is reported in Subsection~\ref{sec:RecoPerf}, 
followed by the charge misidentification
probability in Subsection~\ref{sec:ChargeMisID}, and the trigger efficiency  
in Subsection~\ref{sec:TriggerPerf}.
Conclusions and an outlook are given in Section~\ref{sec:Conclusion}.

\section{The ATLAS detector}
\label{sec:Detector}
A complete description of the ATLAS detector is provided in Ref.~\cite{detpaper}. 

ATLAS uses a right-handed coordinate system
with its origin at 
the nominal $pp$ interaction point at the centre of the detector.
The positive $x$-axis is defined by the direction
from the interaction point to the centre of the LHC ring, with the
positive $y$-axis pointing upwards, while the
beam direction defines the $z$-axis. The azimuthal angle $\phi$ is measured
around the beam axis and the polar angle $\theta$ is the angle from
the $z$-axis. The pseudorapidity is defined as $\eta = -\ln
\tan(\theta/2)$.

The inner detector (ID) provides a precise 
reconstruction of tracks within $|\eta| < 2.5$. It consists of three layers of pixel
detectors close to the beam-pipe, four layers of 
silicon microstrip detector modules with pairs of single-sided sensors glued back-to-back
(SCT) providing eight hits per track at intermediate radii, and a transition
radiation tracker (TRT) at the outer radii, providing about 35 hits per track
(in the range $|\eta| < 2.0$). The TRT offers substantial discriminating
power between electrons and charged hadrons over a wide energy range (between 0.5 and 100
GeV) via the detection of X-rays produced by transition radiation.
The inner-most pixel vertexing
layer (also called the {\it b-layer}) is located just outside the beam-pipe at a
radius of 50 mm. It provides precision vertexing and significant rejection of
photon conversions through the requirement that a track has a hit in this 
layer.

A thin superconducting solenoid, contributing 0.66 radiation length at
normal incidence to the amount of passive material before the EM calorimeter,
surrounds the inner detector and provides a 2 T magnetic field. 

The electromagnetic calorimeter system is separated into two parts:
a presampler detector and an EM calorimeter, a lead--liquid-argon (LAr) 
detector with accordion-shaped kapton electrodes and lead absorber plates. 
The EM calorimeter 
has three longitudinal layers (called {\it strip, middle} and {\it back} layers)
and a fine segmentation in the lateral
direction of the showers within the inner detector coverage. At high energy,
most of the EM shower energy is collected in the middle layer which has a
lateral granularity of $0.025 \times 0.025$ in \etaphi\ space.  The
first (strip) layer consists of finer-grained strips in the $\eta$-direction with a
coarser granularity in $\phi$. It offers discrimination against multiple photon showers
(including excellent $\gamma-\pi^0$
separation), a precise estimation of the pseudorapidity of the impact 
point and, in combination with the middle layer, 
an estimation of the photon pointing direction~\cite{cscbook}.
These two layers are complemented by a presampler detector placed
in front with a granularity of $0.025 \times 0.1$ 
covering only the range $|\eta| < 1.8$ to
correct for energy lost in the material before the calorimeter, and by the back
layer behind, which collects the energy deposited in the tail of very high
energy EM showers. The transition region between the barrel (EMB) and \endcap\ 
(EMEC) calorimeters, $1.37 < |\eta| < 1.52$, 
has a large amount of material in front of the first active calorimeter
layer. 
The \endcap\ EM calo\-rim\-e\-ters are divided into two wheels,
the outer (EMEC-OW) and the inner (EMEC-IW) wheels 
covering the ranges $1.375 < |\eta| < 2.5$ and $2.5 <
|\eta| < 3.2$, respectively. 

Hadronic calorimeters 
with at least three longitudinal segments surround the EM
calorimeter and are used in this context to reject hadronic jets.
The forward calorimeters (FCal) cover the range
$3.1 < |\eta| < 4.9$ and also 
have EM shower identification capabilities given
their fine lateral granularity and longitudinal segmentation into three layers.

\section{Data and Monte Carlo samples}
\label{sec:Samples}
The results are based on the proton-proton collision data collected
with the ATLAS detector in 2010 at \rts=7~TeV. 
After requiring good data-quality criteria, in particular those concerning the
inner detector and the EM and hadronic calorimeters,
the total integrated luminosity used for the measurements 
is between 35 and 40~\invpb\ depending on the trigger requirements. 

The measurements are compared to expectations 
from MC simulation.
The \Zee, \Jee\ and \Wen\ MC samples were generated by {\tt
PYTHIA}~\cite{Pythia} and processed through the full ATLAS 
detector simulation~\cite{ATLASGeant4} based on {\tt GEANT4}~\cite{Geant4}. 
To study the effect of multiple proton-proton interactions 
different pile-up configurations with on average about two interactions
per beam crossing were also simulated.

In addition, MC samples were produced with
additional passive material in front of the EM calorimeter representing 
a conservative estimate of the possible increases 
in the material budget based on various studies using 
collision data, 
including studies of track reconstruction 
efficiency~\cite{MinbiasPaper1,MinbiasPaper2,MaterialK0s,MaterialHadronicInt}, 
the measurement of the photon conversion rate~\cite{MaterialConversions}, 
studies of the energy flow in the EM calorimeter~\cite{MaterialEnergyFlow}, 
EM shower-shape variables and the energy to momentum ratio. 
In these samples, the amounts of additional material 
with respect to the nominal geometry, 
expressed in units of radiation length ($X_0$) and given at normal incidence, are
0.05$X_0$ in the inner detector, 0.2$X_0$ in its services,
0.15$X_0$ at the end of the SCT and TRT {\endcap}s and at the ID \endplate,
0.05$X_0$ between barrel presampler detector and the strip layer 
of the EM calorimeter,
and 0.1$X_0$ in front of the LAr EM barrel calorimeter in the cryostat.

\begin{figure*} \begin{center}
\includegraphics[width=0.49\textwidth]{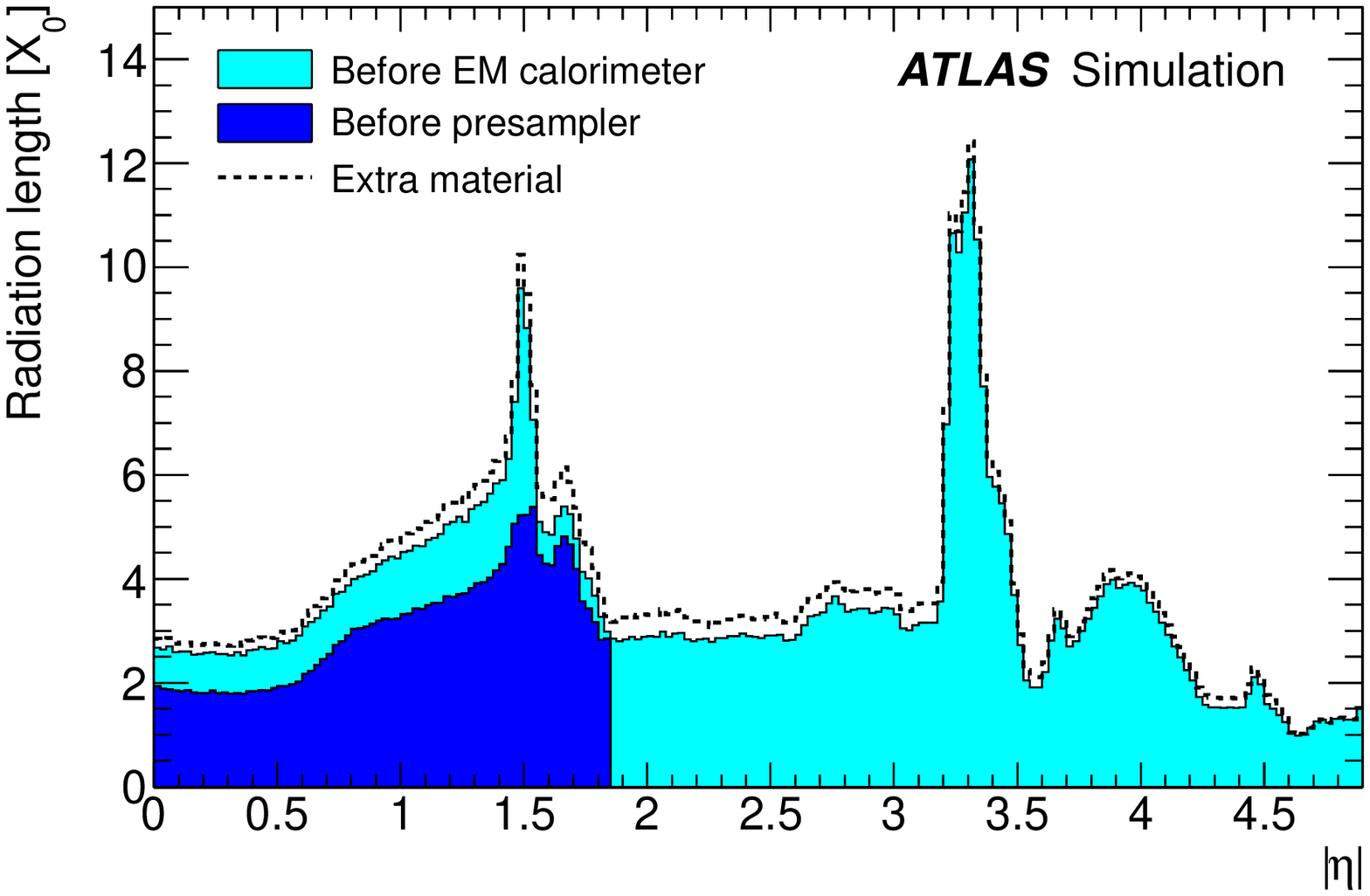}
\includegraphics[width=0.49\textwidth]{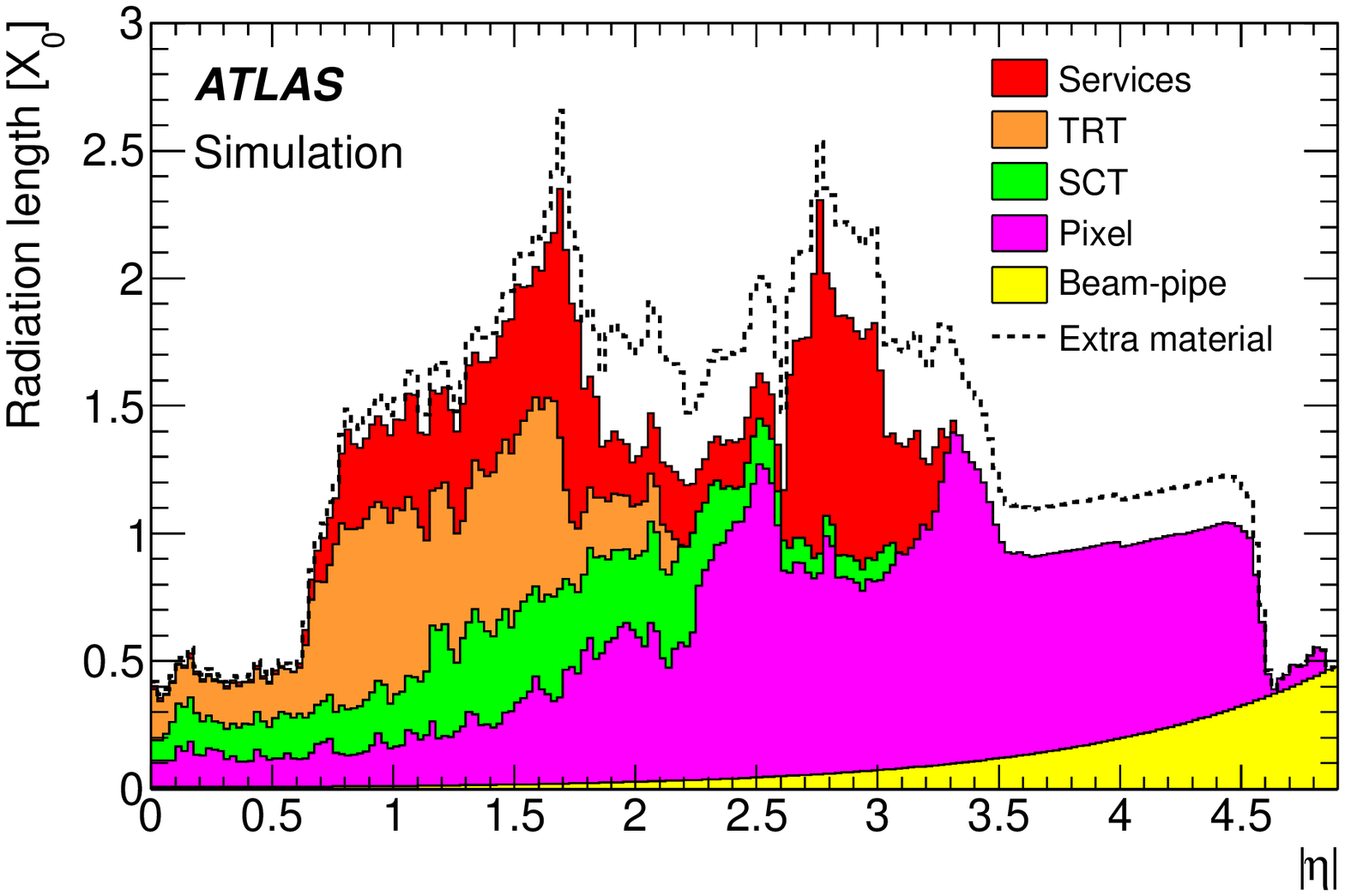}
\end{center} \caption{ Amount of material, in units of radiation length
$X_{0}$,  traversed by a particle as a function of $\eta$: (left) material in
front of the presampler detector and the EM calorimeter,  and (right) material
up to the ID boundaries. The contributions of the different detector elements, 
including the services and thermal enclosures are shown separately by filled
color areas.  The extra material used for systematic studies  is indicated by
dashed lines. The primary vertex position has been smeared along the beamline. }
\label{fig:XMat_Gprime_alpha} \end{figure*}

The distribution of material as a function of $\eta$ 
in front of the presampler detector and the EM calorimeter 
is shown on the left of Figure~\ref{fig:XMat_Gprime_alpha}
for the nominal and extra-material geometries. 
The contributions of the different detector elements up to the 
ID boundaries,
including the services and thermal enclosures,
are detailed on the right.

The peak in the amount of material before the electromagnetic calorimeter at
$|\eta| \approx 1.5$, corresponding to the transition region between the barrel
and endcap EM calorimeters, is due to the cryostats, the corner of the barrel
electromagnetic calorimeter, the inner detector services and the tile
scintillator. 
The sudden increase of material at $|\eta| \approx 3.2$, corresponding to the
separation between the endcap calorimeters and the FCal, is mostly due to the
cryostat that acts also as a support structure. It runs almost projective
at the low radius part of EMEC IW.

\section{Electron trigger, reconstruction and identification}
\label{sec:Algorithms}

\subsection{Trigger}
\label{sec:Trigger}
The ATLAS trigger system~\cite{TriggerPaper2010} 
is divided into three levels. The hardware-based first-level trigger (L1) 
performs a fast event selection by searching for high-\pT\ objects and large
missing or total energy using reduced granularity data from the calorimeters
and the muon system and reduces the event rate to a maximum of 75 kHz.
It is followed by the software-based second-level trigger (L2) and
event filter (EF), collectively referred to as the high-level trigger (HLT).
The reconstruction at L2 is seeded by the L1 result. It uses, with full 
granularity and precision, all the available detector data (including
the information from the inner detector) 
but only in the regions identified by the L1 as Regions of Interest (RoI).
After L2 selection, the event rate is about 3 kHz. 
In the EF, more complex algorithms seeded by the L2 results and profiting from 
offline-like calibration and alignment are used to reduce the event rate to 
about 200 Hz.

At L1, electromagnetic objects are selected if the total transverse
energy deposited in the EM
calorimeter in two adjacent towers of 
$\Delta\eta \times \Delta\phi = 0.1 \times 0.1$ size is above a certain 
threshold. Fast calorimeter and tracking reconstruction algorithms are
deployed at L2. The L2 calorimeter reconstruction is very similar to the offline
algorithm, with the notable difference that clusters are seeded by the highest \ET\
cell in the middle calorimeter layer instead of applying the full offline
sliding-window algorithm described in Subsection~\ref{sec:Reco}. The L2 track
reconstruction algorithm was developed independently 
to fulfill the more stringent timing requirements.  The EF uses the offline
reconstruction and identification algorithms described in 
Subsections~\ref{sec:Reco} and \ref{sec:ID}. It applies similar (typically  
somewhat looser) cuts in order to remain fully efficient for objects 
identified offline. 

During the 2010 proton-proton collision data taking period, 
the trigger menu continuously
evolved in order to fully benefit from the increasing LHC luminosity. 
Initially, the trigger relied on the L1 decision only while the HLT
decisions were recorded but not used to reject events. As the luminosity increased,
the HLT began actively rejecting events with higher and higher \ET\ thresholds and
more stringent selections. 
A detailed description of the trigger configuration and selection criteria applied 
in 2010 can be found in Refs.~\cite{TriggerPaper2010,TrigEGConf2010}.

\subsection{Reconstruction} 
\label{sec:Reco}
Electron reconstruction~\cite{ElectronNote} in the {\it central region} of
$|\eta|<2.47$ starts from energy deposits
(clusters) in the EM calorimeter which are then associated to reconstructed
tracks of charged particles in the inner detector. 

To reconstruct the EM clusters, seed clusters of longitudinal towers
with total transverse energy above 2.5 GeV are
searched for by a {\it sliding-window} algorithm. 
The window size is $3 \times 5$ in units of 0.025$\times$0.025 
in \etaphi\ space,
corresponding to the granularity of the calorimeter middle layer.
The cluster reconstruction is expected to be very efficient
for true electrons. In MC simulations, the efficiency is about 95\% at 
$\ET=5$~GeV and 
100\% for electrons with $\ET>15$~GeV from \Wboson\ and \Zboson\ decays.

In the tracking volume of $|\eta| < 2.5$, reconstructed
tracks  extrapolated from their last measurement point to the middle layer of 
the calorimeter are very loosely matched to the seed clusters. 
The distance between the track impact point and the
cluster position is required to satisfy $\Delta\eta<0.05$.
To account for bremsstrahlung losses, the size of the sign corrected 
$\Delta\phi$ window is 0.1 on the side where the extrapolated track bends
as it traverses the solenoidal magnetic field and is 0.05 on the other side.
An electron is reconstructed if at least one
track is matched to the seed cluster. 
In the case where several tracks are matched to the same cluster,
tracks with silicon hits are preferred, and
the one with the smallest $\Delta R = \sqrt{\Delta
\eta^2+\Delta \phi^2}$ distance to the seed cluster is chosen. 

The electron cluster is then rebuilt using $3 \times 7$ ($5 \times 5$) 
longitudinal towers of cells in the barrel (endcaps). 
These lateral cluster sizes were optimized to take into account the different
overall energy distributions in the barrel and endcap calorimeters.
The cluster energy is then determined~\cite{cscbook} by 
summing four different contributions: 
(1) the estimated energy deposit in the material in front
of the EM calorimeter, 
(2) the measured energy deposit in the cluster, 
(3) the estimated external energy deposit outside the cluster (lateral leakage), 
and 
(4) the estimated energy deposit beyond the EM calorimeter (longitudinal leakage). 
The four terms are
parametrised as a function of the measured cluster energies in 
the presampler detector
(where it is present) and in the three EM calorimeter longitudinal layers based
on detailed simulation of energy deposition in both active and inactive material
in the relevant detector systems. The good description of the detector in the
MC simulation is therefore essential in order to correctly reconstruct the 
electron energy.

The four-momentum of {\it central electrons}
is computed using information from both the final
cluster  and the best track matched to the original seed cluster. The energy is
given by the cluster energy. The $\phi$ and $\eta$ directions are taken from the
corresponding track parameters at the vertex.

In the {\it forward region}, $2.5<|\eta|<4.9$, 
where there are no tracking detectors, the electron 
candidates are reconstructed only from energy deposits in the calorimeters
by grouping neighbouring cells in three dimensions, based 
on the significance of their energy content with respect to the expected noise.
These {\it topological clusters}~\cite{TopoClusters}  
have a variable number of cells in contrast to the 
fixed-size sliding-window clusters used in the central region. 
The direction of {\it forward electrons} is defined by the barycentre of the
cells belonging to the cluster. 
The energy of the electron is determined simply by summing the
energies in the cluster cells and is then corrected for energy loss 
in the passive material before the calorimeter.
An electron candidate in the forward region is reconstructed
only if it has a small hadronic energy component and 
a transverse energy of $\ET>5$~GeV.

\subsection{Requirements on calorimeter operating conditions}
\label{sec:OQ}
The quality of the reconstructed energy of an electron object relies on the
conditions of the EM calorimeter. Three types of problems arose during data
taking that
needed to be accounted for at the analysis level:

\begin{itemize}

\item 
Failures of electronic front-end boards (FEBs).
A few percent of the cells are not read out because they are
connected to non-functioning FEBs, on which the active
part (VCSEL) of the optical transmitter to the readout boards has failed~\cite{LArPaper}.
As this
can have an important impact on the energy reconstruction in the EM calorimeter,
the electron is rejected if part of the cluster falls into a dead FEB
region in the EM calorimeter strip or middle layer. If the dead region is in the
back layer or in the presampler detector, which in general contain only a small fraction
of the energy of the shower, the object is considered good and an energy
correction is provided at the reconstruction level.  

\item 
High voltage (HV) problems. 
A few percent of the HV sectors are operated under non-nominal high voltage,
or have a zero voltage on one side of the readout electrode (for redundancy,
each side of an EM electrode, which is in the middle of the LAr gap, 
is powered separately)~\cite{LArPaper}.
In the very rare case when a part of the cluster falls
into a dead high-voltage region, the cluster is rejected. Non-nominal voltage
conditions increase the equivalent noise in energy but 
do not require special treatment for the energy reconstruction.

\item Isolated cells producing a high noise signal or no signal at all. These
cells are masked at the reconstruction level, so that their energy is set to
the average of the neighbouring cells. 
Nonetheless an electron is rejected, if any of the cells in its core, defined as
the 3x3 cells in the middle layer, is masked.

\end{itemize}
The loss of acceptance due to these object quality requirements was about 6\% per
electron on
average dominated by losses due to non-functioning FEBs (replaced during the
2010/2011 LHC winter shutdown).

%These requirements are also applied to the MC samples when performing
%comparisons with data. 
%Nonetheless, the treatment of clusters around dead
%FEBs could induce differences between data and MC when the dead area has not been
%simulated since the barycentre of these clusters tends to be shifted. The
%uncertainty  on the identification efficiency due to these edge-effects has been estimated 
%to be about 0.4\% per electron.

These requirements are also applied to the MC samples when performing
comparisons with data. 
Nonetheless, differences arise between data and MC, induced for example by  
the treatment of clusters around dead FEBs. 
While the barycentre of such clusters tends to be shifted in the data,
this behaviour is not fully reproduced by MC when the dead area has not
been simulated. 
The total uncertainty on the loss of acceptance is estimated 
to be about 0.4\% per electron.

\subsection{Identification} 
\label{sec:ID}
\begin{table*}
\caption{Definition of variables used for \loose, \medium\ and 
\tight\ electron identification cuts for the central region of the detector 
with $|\eta|<2.47$.}
\label{tab:IDcuts}
\footnotesize
\begin{center}
\begin{tabular}{|l|l|l|}
\hline
Type & Description & Name \\ \hline
\multicolumn{3}{|l|}{\bf{Loose selection}}\\
\hline
Acceptance & $|\eta|<2.47$ & \\ 
\hline
 Hadronic leakage & Ratio of $E_T$ in the first layer of the hadronic  calorimeter to $E_T$ of & $R_\mathrm{had1}$ \\ 
& the EM cluster  (used over the range $|\eta| < 0.8$ and $|\eta| > 1.37$) &  \\
\cline{2-3}
  & Ratio of $E_T$ in the hadronic calorimeter to $E_T$ of the EM cluster& $R_\mathrm{had}$ \\
 & (used over the range $|\eta| > 0.8$ and $|\eta| < 1.37$) &  \\
\hline
Middle layer of & Ratio of the energy in 3$\times$7 cells over the energy in 7$\times$7 cells & $R_{\eta}$ \\
EM calorimeter & centred at the electron cluster position &  \\
\cline{2-3}
 & Lateral shower width, $\sqrt{(\Sigma E_i \eta_i^2)/(\Sigma E_i) -((\Sigma E_i\eta_i)/(\Sigma E_i))^2}$,& $w_{\eta2}$ \\
 & where $E_i$ is the energy and $\eta_i$ is the pseudorapidity of cell $i$  &  \\
 &  and the sum is calculated within a window of $3 \times 5$ cells &  \\
\hline
 \multicolumn{3}{|l|}{\bf{Medium selection} (includes loose)}\\
\hline
Strip layer of       & Shower width, $\sqrt{(\Sigma E_i (i-i_\mathrm{max})^2)(\Sigma E_i)}$, where $i$ runs over all strips  &  $w_\mathrm{stot}$   \\  
EM calorimeter       & in a window of $\Delta\eta \times \Delta\phi \approx 0.0625 \times 0.2$, corresponding  typically   &                   \\
		     & to 20 strips in $\eta$, and $i_\mathrm{max}$ is the index of the highest-energy strip        &                   \\
\cline{2-3}
        & Ratio of the energy difference between  the largest   and second largest          &  $E_\mathrm{ratio}$   \\
                     &  energy deposits in the cluster      over the sum of these energies         &                   \\
\hline
Track quality        & Number of hits in the pixel detector ($\geq 1$)       &    $n_\mathrm{pixel}$    \\
\cline{2-3}
                     & Number of total hits in the pixel and SCT   detectors ($\geq 7$) &   $n_\mathrm{Si}$                \\
\cline{2-3}
                     & Transverse impact parameter ($|d_0|<$5 mm)                      &       $d_0$    \\
\hline
Track--cluster      & $\Delta\eta$ between the cluster position in the strip layer  and the&   $\Delta\eta$   \\
matching              &  extrapolated track ($|\Delta\eta| < 0.01$) &      \\ 
\hline
\multicolumn{3}{|l|}{\bf{Tight selection} (includes medium)}\\
\hline
Track--cluster     & $\Delta\phi$ between the cluster position in the middle layer and the &  $\Delta\phi$ \\
matching             &     extrapolated track ($|\Delta\phi| <0.02$)    &        \\
\cline{2-3}
                    & Ratio of the cluster energy to the track momentum            &       $E/p$       \\  
\cline{2-3}
                    & Tighter $\Delta\eta$ requirement ($|\Delta\eta| < 0.005$)                         &      $\Delta\eta$   \\
\hline
Track quality       & Tighter transverse impact parameter requirement ($|d_0|<$1 mm)           &       $d_0$      \\
\hline
TRT                 & Total number of hits in the TRT      & $n_\mathrm{TRT} $                \\
\cline{2-3}
                    & Ratio of the number of high-threshold hits to the    total number of                    &    $f_\mathrm{HT}$     \\
                    & hits in the TRT                 &                   \\
\hline
Conversions         & Number of hits in the b-layer ($\geq 1$)               &   $n_\mathrm{BL}$  \\
\cline{2-3}
                    & Veto electron candidates matched to reconstructed photon             &                   \\
                    & conversions             &                   \\
\hline

\end{tabular}
\end{center}
\end{table*}

The baseline electron identification in the central $|\eta|<2.47$ region 
relies on a cut-based selection using
calorimeter, tracking and combined variables that provide good separation between
isolated or non-isolated signal electrons,  background electrons (primarily from
photon conversions and Dalitz decays) and jets faking electrons. The cuts can be
applied independently. Three reference sets of cuts have been defined with
increasing background rejection power: \loose, \medium\ and 
\tight~\cite{ElectronNote} with an expected 
jet rejection of about 500, 5000 and 50000, respectively,
based on MC simulation.
Shower shape
variables of the EM calorimeter middle layer and hadronic leakage variables are 
used in the \loose\ selection. Variables from the EM calorimeter strip layer, 
track quality
requirements and track-cluster matching are added to the \medium\ selection. 
The \tight\ selection adds 
$E/p$, particle identification using the TRT, 
and discrimination against photon conversions 
via a b-layer hit requirement and information about reconstructed conversion vertices~\cite{PhotonNote}.
Table~\ref{tab:IDcuts} lists all variables used in the \loose, \medium\
and \tight\ selections. The cuts are optimised in 10 bins of cluster $\eta$ 
(defined by calorimeter geometry, detector acceptances and regions of increasing 
material in the inner detector) and 11 bins of cluster $E_{T}$ from 5 GeV to  above
80 GeV.

Electron identification in the forward $2.5 < |\eta| < 4.9$ region, where no tracking detectors
are installed, is based solely on cluster moments\footnote{
The cluster moment of degree $n$ for a variable $x$ is defined as:
\begin{equation}
\langle x^n\rangle=\frac{\sum_i{E_i\,x_i^n}}{\sum_i{E_i}},
\end{equation}
where $i$ runs over all cells of the cluster.
}
 and shower shapes~\cite{ElectronNote}. 
These provide efficient discrimination
against hadrons due to the good transverse and longitudinal segmentation of the calorimeters, though
it is not possible to distinguish between electrons and photons.
Two reference sets of cuts are defined, \forwardloose\ and
\forwardtight\ 
selections.
Table~\ref{tab:ForwardIDcuts} lists the identification variables.

\begin{table*}
\caption{Definition of variables used for \forwardloose\ and \forwardtight\ 
electron identification cuts for the $2.5<|\eta|<4.9$ region of the detector.}
\label{tab:ForwardIDcuts}
\footnotesize
\begin{center}
\begin{tabular}{|l|l|l|}
\hline
Type & Description & Name \\ \hline
\multicolumn{3}{|l|}{\bf{Forward loose selection}}\\
\hline
Acceptance & $2.5<|\eta|<4.9$ & \\ 
\hline
Shower depth & Distance of the shower barycentre from the calorimeter 
& $\lambda_\mathrm{centre}$ \\
& front face measured along the shower axis & \\ 
\hline
Longitudinal& Second moment of the distance of each cell 
& $\langle \lambda^2 \rangle$ \\
 second moment & to the shower centre in the longitudinal direction ($\lambda_i$)& \\ 
\hline
Transverse& Second moment of the distance of each cell 
& $\langle r^2 \rangle$ \\
 second moment & to the shower centre in the transverse direction ($r_i$)& \\ 
\hline
\multicolumn{3}{|l|}{\bf{Forward tight selection (includes forward loose)}}\\
\hline
Maximum cell energy & 
Fraction of cluster energy in the most energetic cell & $f_\mathrm{max}$ \\
\hline
Normalized & $w_2$ is the second moment of $r_i$ setting $r_i=0$ 
& $\frac{w_2}{w_2+w_\mathrm{max}}$ \\
lateral moment & for the two most energetic cells, while & \\
& $w_\mathrm{max}$ is the second moment of $r_i$ setting $r_i=4$~cm & \\
& for the two most energetic cells and $r_i=0$ for the others & \\
\hline
Normalized & $l_2$ is the second moment of $\lambda_i$ setting $\lambda_i=0$ 
& $\frac{l_2}{l_2+l_\mathrm{max}}$ \\
longitudinal moment & for the two most energetic cells, while & \\
& $l_\mathrm{max}$ is the second moment of $\lambda_i$ setting $\lambda_i=10$~cm & \\
& for the two most energetic cells and $\lambda_i=0$ for the others & \\
\hline
\end{tabular}
\end{center}
\end{table*}

\subsection{Inclusive single and dielectron spectra} 
\label{sec:Spectra}
\begin{figure*}
  \begin{center}
    \includegraphics[width=0.49\textwidth]{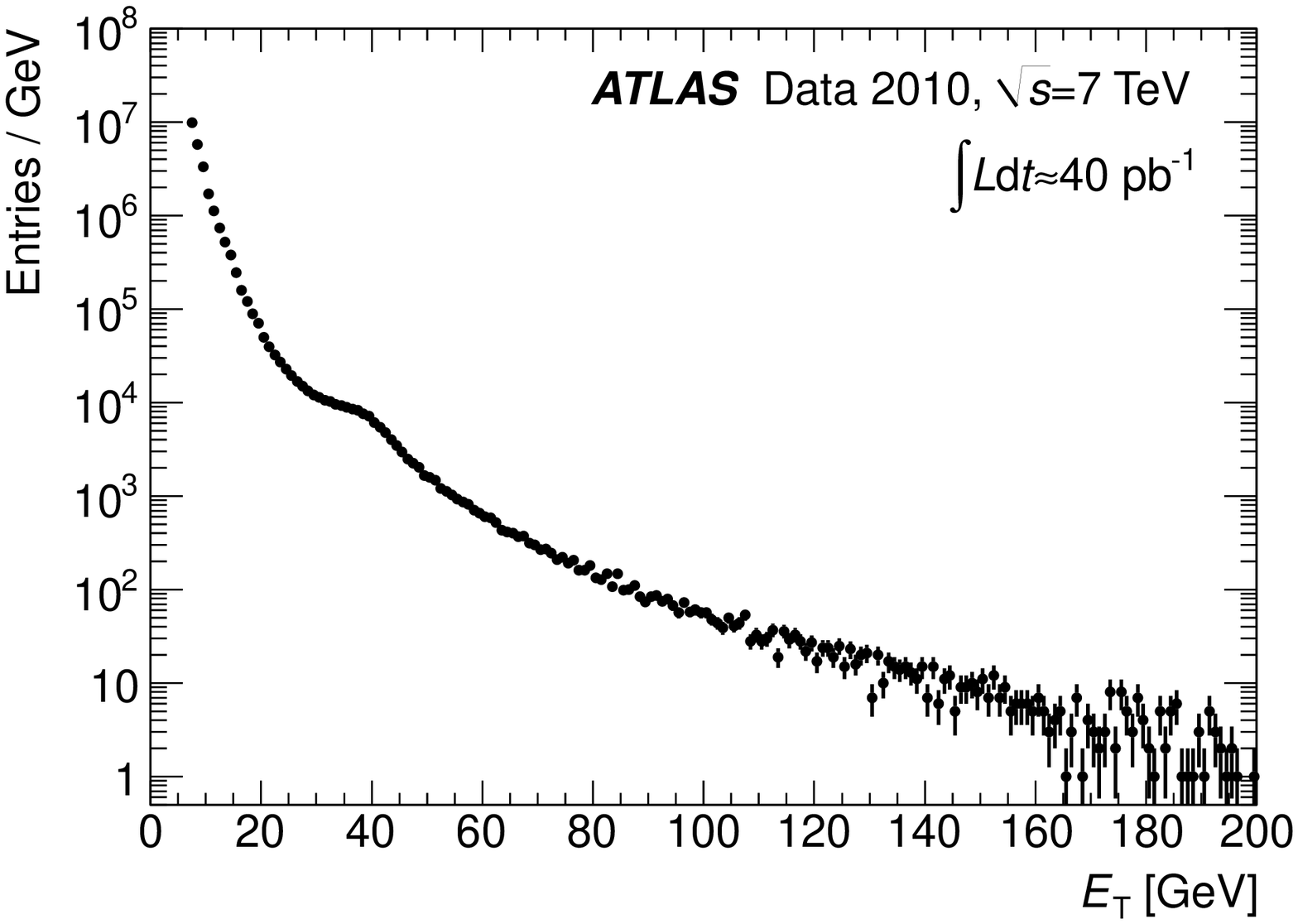}
    \includegraphics[width=0.49\textwidth]{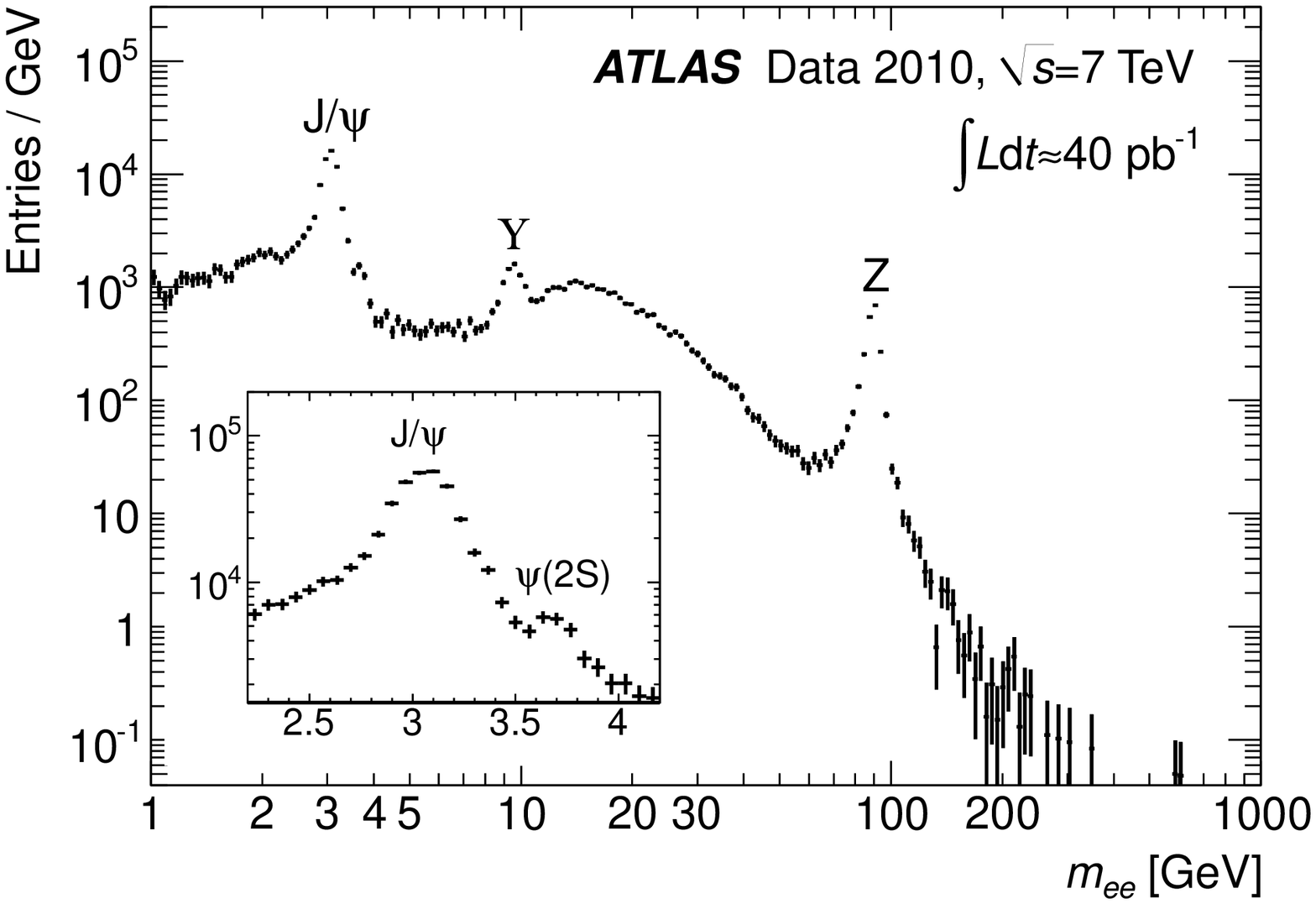}
  \end{center}
  \caption{(left) \ET\ distribution of electron candidates passing the \tight\ identification cuts
  for events selected by single electron triggers with varying \ET\ thresholds. 
  Data with $\ET<20$~GeV correspond to lower integrated luminosity values and
  were rescaled to the full luminosity.
  (right) Reconstructed dielectron mass distribution of electron candidate pairs passing the \tight\ 
  identification cuts
  for events selected by low \ET\ threshold dielectron triggers.
  The number of events is normalised by the bin width.
  Errors are statistical only.}
  \label{fig:Spectra}
\end{figure*}

To illustrate the electron identification performance,  
the left of Figure~\ref{fig:Spectra} 
shows the \ET\ distribution of all electron candidates passing the
 \tight\ identification cuts and having $|\eta|<2.47$ 
excluding the transition region, $1.37<|\eta|<1.52$.
The data sample was collected by 
single electron triggers with varying thresholds.
The Jacobian peak at $\ET\approx 40$~GeV
from \Wboson\ and \Zboson\ decays is clearly visible above the sum of contributions from semi-leptonic
decays of beauty and charm hadrons, electrons from photon conversions and
hadrons faking electrons.

The measurement of known particles decaying into dielectron final states 
is an important ingredient in order to calibrate and
measure the performance of the detector. The dielectron mass spectrum is plotted
on the right of Figure~\ref{fig:Spectra} using a selection of unprescaled, low \ET\ threshold dielectron triggers.
Both electrons are required to pass the \tight\
selection, to be of opposite sign, and to have $\ET>5$~GeV and $|\eta|<2.47$. 
The \Jpsi, \Ups\ and \Zboson\ peaks are clearly visible, and evidence for 
the $\psi$(2S) meson is also apparent. 
The shoulder in the region of $\mee \approx 15$ GeV is caused by
the kinematic selection.

\subsection{Inter-alignment of the inner detector and the electromagnetic calorimeter}
\label{sec:Alignment}
A global survey of the positions of the LAr cryostats
and of the calorimeters inside them was performed with an accuracy of about $1-2$~mm 
during their integration
and installation in the ATLAS cavern\footnote{Measurements were performed when
warm and  predictions are used to estimate the calorimeter positions inside the cryostats when
cold.}.  
Since the intrinsic accuracy of the EM calorimeter shower position measurement is 
expected to be about 200~$\mu$m for high energy electrons\cite{detpaper},
accurate measurements of the in-situ positions of the EM calorimeters are 
prerequisites to precise matching of the extrapolated tracks and the shower barycentres.

For most ATLAS analyses
using track--cluster matching cuts (as described in Table~\ref{tab:IDcuts}), 
or photon pointing, a precision of the order of 
1~mm is sufficient. A precision  as good as 
100~$\mu$m is very valuable to improve bremsstrahlung recovery for 
precision measurements, such as the \Wboson\ mass measurement.

The relative positions of the four independent parts of the EM calorimeter (two
half-barrels and two endcaps) were measured
with  respect to the
inner detector position, assuming that the ID itself is already well-aligned.
About 300000 electron
candidates with $\pT>10$~GeV, passing the \medium\ identification cuts, 
were used. 

The comparison of the cluster position and the extrapolated 
impact point of the electron
track on the calorimeter provides a determination of the calorimeter translations and
tilts with respect to their nominal positions. A correction for the sagging  of
the calorimeter absorbers (affecting the azimuthal measurement of the cluster) has
been included for the barrel calorimeter with an amplitude of 1~mm. The derived
alignment constants are then used to correct the electron cluster positions.

\begin{figure*}
\includegraphics[width=\textwidth]{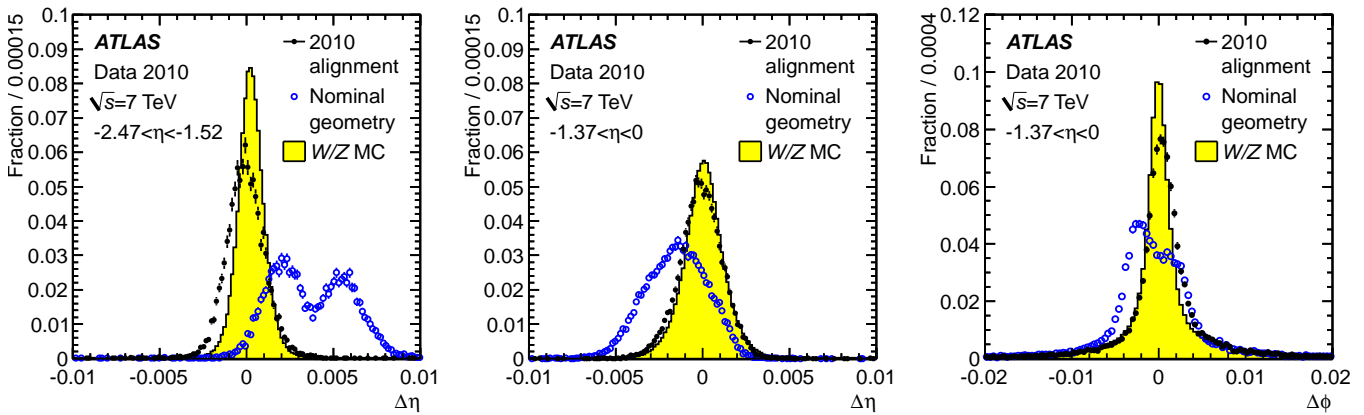}
\caption{Track--cluster matching variables of electron candidates
from \Wboson\ and \Zboson\ decays for reconstruction with nominal geometry
and after the 2010 alignment corrections have been applied:
(left) $\Delta\eta$ distributions for $-2.47<\eta<-1.52$  and 
(middle) $-1.37<\eta<0$;
(right) $\Delta\phi$ distributions for $-1.37<\eta<0$. 
The MC prediction with perfect alignment is also shown.}
\label{fig:DEta}
\end{figure*}

To illustrate the improvements brought by this first alignment procedure, the $\Delta\eta$
track--cluster matching variable used in electron reconstruction and
identification is shown in Figure~\ref{fig:DEta}. Here, a sample of electron
candidates collected at the end of the 2010 data taking period with $\pT>20$~GeV, 
passing the \medium\ identification cuts and 
requirements similar to the ones described in Subsection~\ref{sec:CalibEventSelection}
to select \Wboson\ and \Zboson\ candidates, is used.
The two-peak structure for $-2.47<\eta<-1.52$ visible on the left 
is due to the 
transverse displacement of the \endcap\ by about 5~mm which is then
corrected by the alignment procedure.  
On the right of Figure~\ref{fig:DEta}, $\Delta\phi$ for the barrel
$-1.37<\eta<0$ is also shown.
After including corrections for sagging,
a similar precision is reached in $\phi$ in the endcaps, as well.
After the inter-alignment, 
the \tight\ track--cluster matching cuts 
($|\Delta\eta|<0.005$ and $|\Delta\phi|<0.02$) can be applied with high efficiency.

These inter-alignment corrections are applied for all datasets used in the following
sections.

\section{Electron energy scale and resolution}
\label{sec:CalibrationTop}

\subsection{Electron energy-scale determination}
\label{sec:Calibration}
The electromagnetic calorimeter energy scale was derived
from test-beam measurements. The total uncertainty is 3\% in the central region
covering $|\eta|<2.47$,
and it is 5\% in the forward region covering $2.5<|\eta|<4.9$.
The dominant uncertainty, introduced by the transfer of the test-beam 
results to the ATLAS environment, comes from the LAr absolute temperature
normalization in the test beam cryostat.

Even with the limited statistics of \Zee\ and \Jee\ decays available in the 2010 dataset,
the well known masses of the \Zboson\ and \Jpsi\
particles can be used to improve considerably the knowledge of the electron energy scale and to 
establish the linearity of the response of the EM calorimeter. An alternative strategy to determine
the electron energy scale is to study the ratio of the energy $E$ measured by the EM
calorimeter and the momentum $p$ measured by the inner detector, $E/p$. This technique gives
access to the larger statistics of \Wen\ events but depends on the
knowledge of the momentum scale and therefore the alignment of the inner detector. 

The strategy to calibrate the EM calorimeter is described 
in Refs.~\cite{cscbook,atlastdr}. It was validated using test-beam
data~\cite{Aharrouche:2006nf,Aharrouche:2007nk,Aharrouche:2010zz}. The
energy calibration is divided into three steps:
\begin{enumerate}
\item The raw signal extracted from each cell in ADC counts is converted 
into a deposited energy using the electronic calibration of the EM 
calorimeter~\cite{atlastdr,Aubert:2002mw,Aubert:2002dm}.
\item MC-based calibration~\cite{cscbook} corrections are applied at the
cluster level for energy loss due to absorption in the passive material and 
leakage outside the cluster as discussed in Subsection~\ref{sec:Reco}. For the 
central region, $|\eta|<2.47$, additional fine corrections depending on the 
$\eta$ and $\phi$ coordinates of the electron are made to compensate
for the energy modulation as a function of the impact point.
\item The in-situ calibration using \Zee\ decays 
determines the energy scale and intercalibrates, as described in
Subsection~\ref{sec:CalibEventSelection}, the
different regions of the calorimeters covering $|\eta|<4.9$.
\end{enumerate}

For calibrated electrons with 
transverse energy larger than 20 GeV, 
the ratio between the reconstructed and the
true electron energy is expected to be within 1\% of unity for almost all
pseudorapidity regions.
The energy resolution is better than 2\% for $\ET>25$~GeV
in the most central region, $|\eta|<0.6$,
and only exceeds 3\% close to the transition region of the barrel and \endcap\ 
calorimeters where the amount of passive material in front of the calorimeter
is the largest. 

This section describes the in-situ 
measurement of the electron energy
scale and the determination of the energy resolution.
The in-situ calibration is performed using \Zee\ decays 
both for central and forward electrons.
The linearity of response versus energy is cross-checked in the central region
using \Jee\ and \Wen\ decays, but only with limited accuracy.
Due to the modest \Zee\ statistics in the 2010 data sample, 
the intercalibration is performed only among the calorimeter sectors in $\eta$. 
The non-uniformities versus $\phi$ are much smaller, as expected.
They are shown in Subsection~\ref{sec:CalibUniformity}. 

\subsubsection{Event selection}
\label{sec:CalibEventSelection}

High-\ET\ electrons from \Zboson\ and \Wboson\ decays are collected using
EM triggers requiring a
transverse energy above about $15-17$~GeV in the early data taking periods and a 
high-level trigger also requiring \medium\ electron identification criteria 
in later periods.
Low-\ET\ electrons from \Jpsi\ are selected by a mixture of low \ET\ 
threshold EM triggers
depending on the data taking period. 
All events must have at least one primary vertex formed by at least 3 tracks. 

Electrons are required to be within $|\eta|<2.47$ excluding the transition region 
of $1.37<|\eta|<1.52$ 
for central, and within $2.5<|\eta|<4.9$ for
forward candidates. 
Electrons from \Wboson\ and \Zboson\ (resp. \Jpsi)
decays must have $\ET>20$~GeV (resp. $\ET>5$~GeV).

For central--central \Zboson\ selection, the \medium\ identification cut is applied
for both electrons, and for the central--forward \Zboson\ selection,
a central \tight\ and a \forwardloose\ electron are required. 
To suppress the
larger background \tight--\tight\ pairs are selected for the \Jpsi\ analysis. 
For \Zboson\ and \Jpsi\
selections in the central region, only oppositely charged electrons are considered (no charge
information is available in the forward region).
The dielectron invariant mass should be in the range $80-100$~GeV for \Zee\ 
and $2.5-3.5$~GeV for \Jee\ candidates.

For the \Wboson\ selection, a \tight\ electron is
required with additional cuts applied on jet cleaning~\cite{JetCleaning2010conf}, missing transverse momentum
$E_\mathrm{T}^\mathrm{miss}>25$~GeV and transverse mass\footnote{The transverse mass is defined as 
$$\MT = \sqrt{ 2 E_\mathrm{T}^{e} E_\mathrm{T}^\mathrm{miss} 
                      (1 - \cos(\phi^e - \phi^\mathrm{miss})) },$$
where $E_\mathrm{T}^{e}$ is the electron transverse energy, $E_\mathrm{T}^\mathrm{miss}$ is the missing transverse momentum,		      
$\phi^e$ is the electron direction and $\phi^\mathrm{miss}$ is the direction of $E_\mathrm{T}^\mathrm{miss}$ in $\phi$.
} $\MT>40$~GeV.
\Zee\ events are suppressed by rejecting events containing a second \medium\ electron.

In total, about 10000 central--central \Zboson\ and
3100  central--forward \Zboson\ candidates 
are selected in the reconstructed dielectron mass range $\mee=80-100$~GeV.
The number of \Jpsi\ candidates is about 8500 in the mass range $\mee=2.5-3.5$~GeV.
The largest statistics, about 123000 candidates, comes from \Wboson\ decays.

The amount of background contamination is estimated from data to be 
about 1\% for the central--central electron pairs 
and 14\% for the central--forward electron pairs for the \Zee\ selection. 
It is significantly higher, 23\%, for the \Jee\ selection.
It amounts to 7\% for the \Wen\ selection.

\subsubsection{Energy-scale determination using dielectron decays of \Zboson\ and \Jpsi\ particles}
\label{sec:ZCalib}

Any residual miscalibration for a given region $i$ is parametrised by
\begin{equation}
E^\mathrm{meas}=E^\mathrm{true}(1+\alpha_i),
\label{eq:bias}
\end{equation}
where $E^\mathrm{true}$ is the true electron energy, 
$E^\mathrm{meas}$ is the energy measured by the calorimeter
after MC-based energy-scale correction,
and $\alpha_i$ measures the residual miscalibration.
The $\alpha$ energy-scale correction factors are determined 
by a fit minimizing the negative unbinned log-likelihood~\cite{cscbook}:
\begin{equation}
- \ln L_{\mathrm{tot}} = \sum_{i,j} \sum^{N^\mathrm{events}_{ij}}_{k=1}
-\ln L_{ij}\left(\frac{m_{k}}{1+\frac{\alpha_{i}+\alpha_{j}}{2}}\right),
\label{eq:likelihood}
\end{equation}
where the indices $i,j$ denote the regions considered for 
the calibration with one of the electrons from the \Zee\ decay 
being in region $i$ and the other in region $j$,  
$N^\mathrm{events}_{ij}$ is the total number of selected \Zee\ decays
with electrons in regions $i$ and $j$,
$m_k$ is the measured dielectron mass in a given decay, 
and $L_{ij}(m)$ is the probability density function (pdf) quantifying the
compatibility of an event with the \Zboson\ lineshape. 
This pdf template is obtained from {\tt PYTHIA} MC simulation 
and smoothed to get a continuous distribution. 
Since the experimental distribution of the dielectron invariant mass depends
strongly on the cluster $\eta$ of the two electrons, 
mainly due to the material in front of the calorimeter, 
the pdf is produced separately for different bins in $|\eta|$ of the two electron clusters.

\begin{figure*}
\begin{center}
\includegraphics[width=0.49\textwidth]{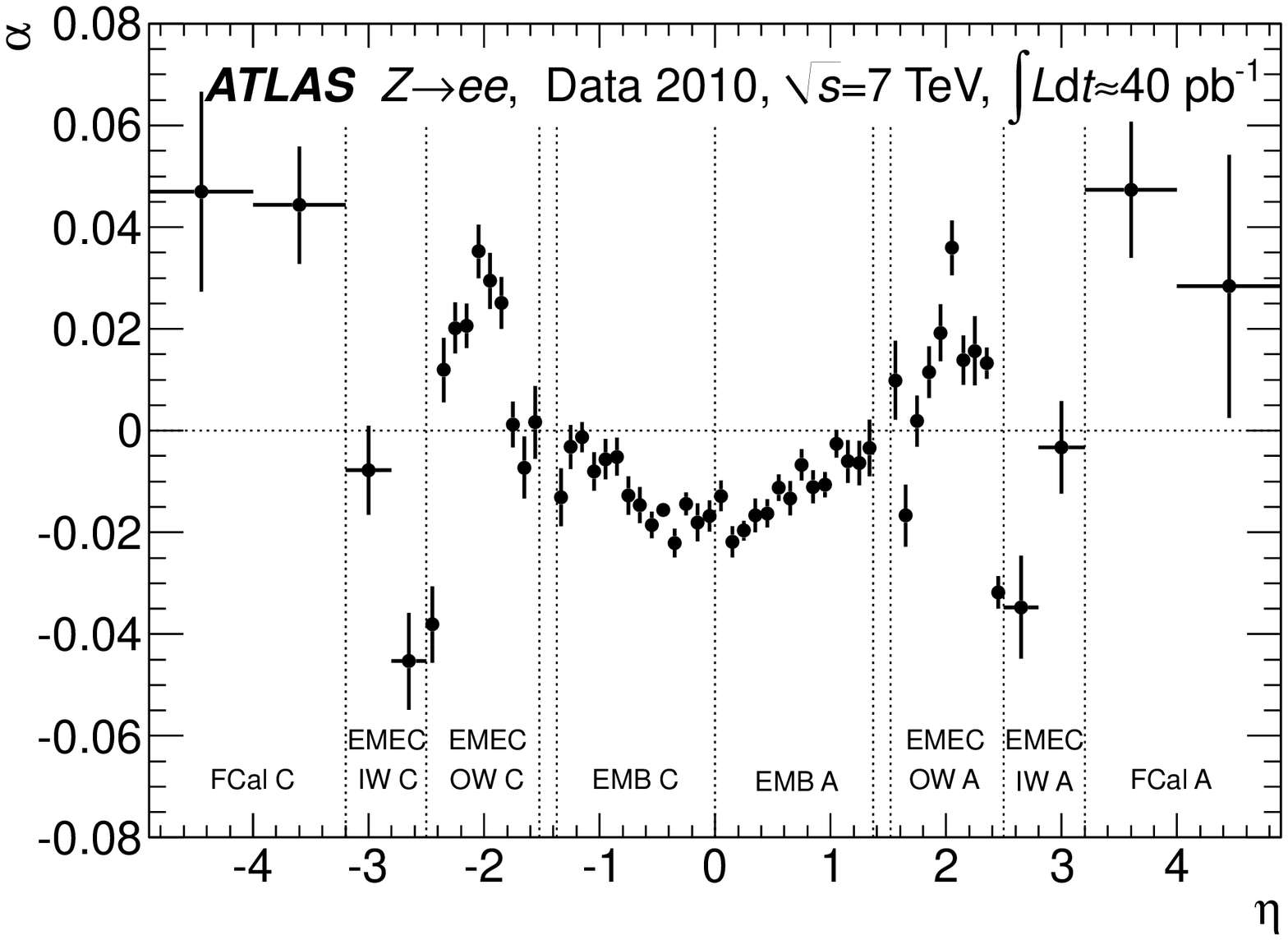}
\includegraphics[width=0.49\textwidth]{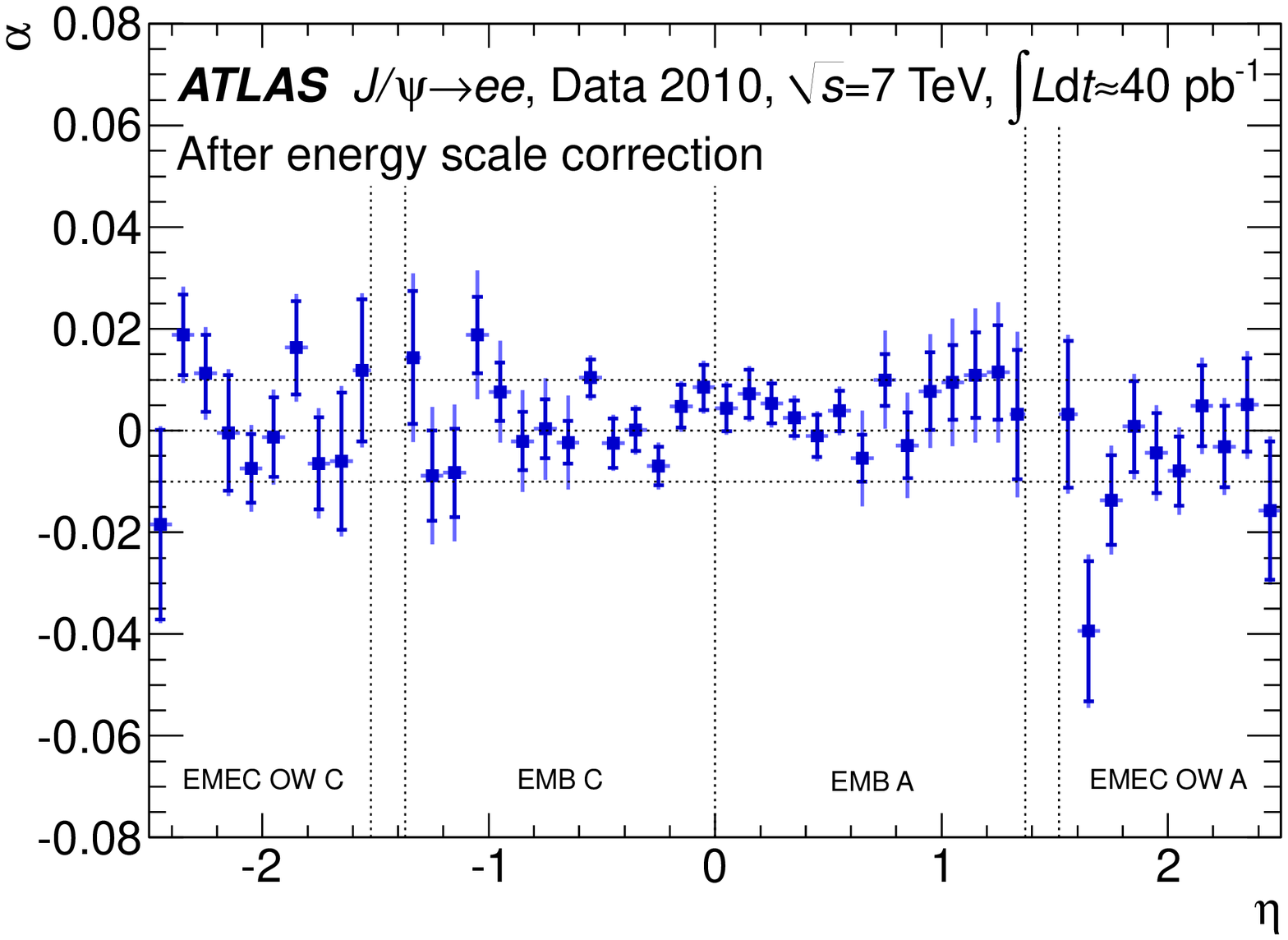}
\end{center}
\caption{The energy-scale correction factor $\alpha$ 
as a function of the pseudorapidity of the electron cluster 
derived from fits (left) to \Zee\ data and  
(right) to \Jee\ data. 
The uncertainties of the \Zee\ measurement are statistical only.
The \Jee\ measurement was made after the \Zee\ calibration had been applied.
Its results are given with statistical (inner error bars) and total (outer error bars) uncertainties.
The boundaries of the different detector parts  
defined in Section~\ref{sec:Detector} are indicated by dotted lines.}
\label{fig:Zscale}
\end{figure*}

The procedure described above was applied to the full 2010 dataset in 
58 $\eta$ bins over the full calorimeter coverage of $|\eta|<4.9$
and is considered as the {\it baseline calibration} method.  
The resulting $\alpha$ values
are shown on the left of Figure~\ref{fig:Zscale}. 
They are within $\pm$2\% in the barrel region and within $\pm$5\% in the forward regions.
The rapid variations with $\eta$ occur at the transitions between the different 
EM calorimeter systems as indicated in Figure~\ref{fig:Zscale}. 
The variations within a given calorimeter system
are due to several effects related to electronic calibration,
high-voltage corrections (in particular in the endcaps\footnote{
The form of the accordion in the \endcap\ varies as a function of the radius.
This implies a variation in the size of the LAr gap. Even though the HV is 
varied as a function of the radius to compensate this, the compensation is not
perfect and residual effects are present.}), 
additional material in front of the calorimeter,
differences in the calorimeter and presampler energy scales, and
differences in lateral leakage between data and MC.

The same procedure was applied using \Jee\ events to determine the electron energy scale.
The resulting $\alpha$ values are in good agreement with the \Zee\ measurement 
and the observed small differences are 
used in the following to estimate the uncertainty specific 
to low-\ET\ electrons.

\subsubsection{Systematic uncertainties}
\label{sec:ZCalibUncert}

\begin{figure*}
\begin{center}
\includegraphics[width=0.49\textwidth]{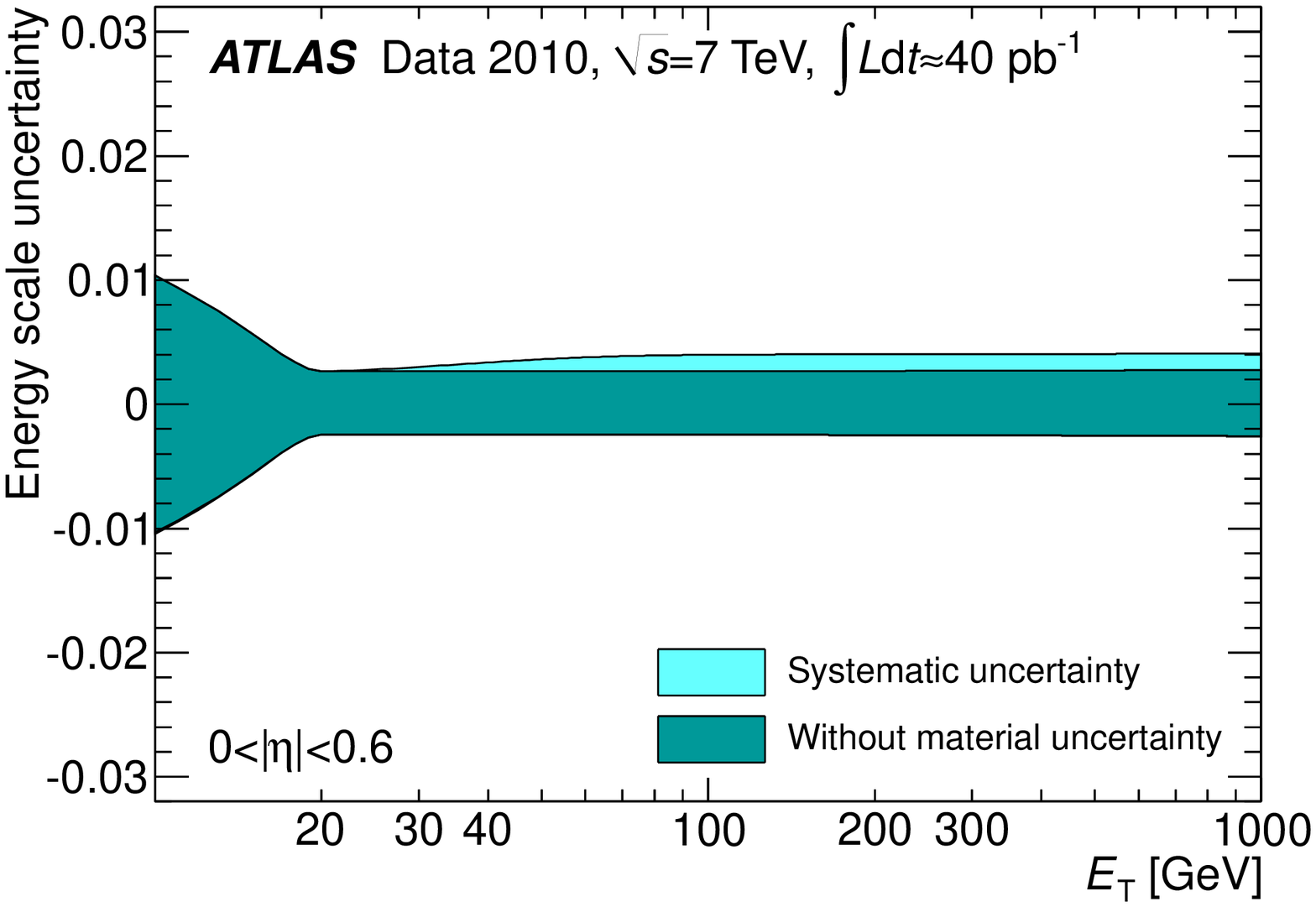} 
\includegraphics[width=0.49\textwidth]{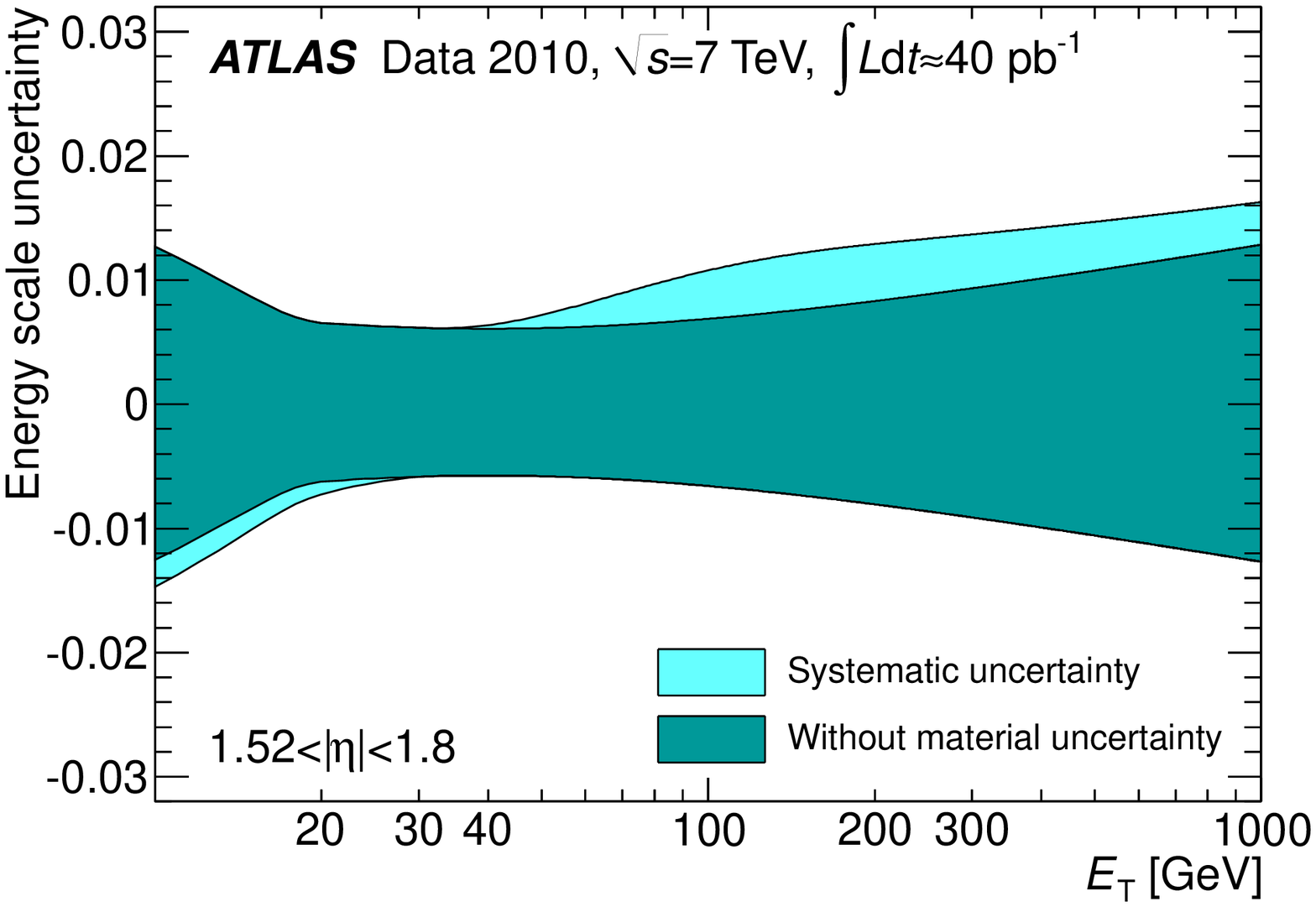} 
\end{center}
\caption{Total systematic uncertainty on the electron energy scale
(left) for the region $|\eta|<0.6$ which has the smallest uncertainty.  
and (right) for $1.52<|\eta|<1.8$ which has the largest uncertainty within the central region.
The uncertainty is also shown without the contribution due to the amount of additional
material in front of the EM calorimeters.
}
\label{fig:sys_ele_central}
\end{figure*}

The different sources of systematic uncertainties affecting the electron
energy-scale measurement are summarized in Table~\ref{tab:sys} and discussed below:
\begin{itemize}
\item{\bf Additional material}
The imperfect knowledge of the material in front of the EM calorimeter affects the electron
energy measurement since the deposited energy in any additional material is neither measured,
nor accounted for in the MC-based energy calibration. 
Nonetheless, if additional material were present in data, 
the $\alpha$ correction factors extracted from \Zee\ events would 
restore the electron energy scale on average. 
However, electrons from \Zboson\ decays have an \ET\ spectrum 
with a mean value around 40~GeV. 
For other values of \ET, a residual uncertainty arises due to the 
extrapolation of the calibration corrections,
as passive material affects lower-energy electrons more severely. 
This effect is estimated in two steps.
First the calibration procedure is applied on a \Zee\ MC sample 
produced, as explained in Section~\ref{sec:Samples},
with a dedicated geometry model with additional material in front of the calorimeters
using the nominal MC sample to provide the reference \Zboson\ lineshape, 
as performed on data.  
Then the non-linearity 
is measured using MC truth information
by comparing the most probable value of the
$E_\mathrm{reco}/E_\mathrm{truth}$ distributions between the nominal MC and 
the one with additional material
in bins of electron \ET.
The systematic uncertainty varies from $-$2\% to +1.2\%.  As expected and by construction, 
it vanishes for $\ET\sim 40$~GeV 
corresponding to the average electron \ET\ in the \Zee\ sample. 
This dominant uncertainty is therefore 
parametrised as a function of \ET\ for the different $\eta$ regions.

\item{\bf Low-\ET\ electrons} 
The energy-scale calibration results obtained for \Jee\ and \Zee\ decays can be compared.
As shown on the right of Figure~\ref{fig:Zscale}, the $\alpha$ correction factors extracted
using \Jee\ decays after applying the baseline calibration using \Zee\ decays are within 1\% of unity,
despite the very different 
\ET\ regimes of the two processes (the mean electron \ET\ in the \Jpsi\ selection is about 9 GeV).
This demonstrates the good linearity of the EM calorimeter and also that the amount of material before the
calorimeter is modelled with reasonable accuracy. Nonetheless, a 1\% additional uncertainty 
is added for electrons with $\ET=10$~GeV, decreasing linearly to 0\% for $\ET=20$~GeV.

Note, that the systematic uncertainties affecting 
the \Jee\ calibration are evaluated in the same manner as described here for the
\Zee\ analysis and are shown in Figure~\ref{fig:Zscale}. 
The dominant uncertainty comes from the imperfect knowledge of the material in front of the calorimeter
and varies between 0.2\% in the central barrel and 1\% close to the transition region between 
the barrel and \endcap\ calorimeters.

\item{\bf Presampler detector energy scale}
The sensitivity of the calibration to the measured presampler energy 
is significant because it is used to correct for energy lost upstream of the 
active EM calorimeter. 
Since the in-situ calibration only fixes one overall scale, 
it cannot correct for any difference between the
presampler detector and the EM calorimeter energy scales. By comparing the
energy deposited in the presampler by electrons from \Wen\ decays 
between data and MC simulation, one can
extract an upper limit\footnote{
As this limit is extracted from data--MC comparisons, 
it will include contributions from the uncertainty on the material
and therefore lead to some double-counting of this material uncertainty.
} on the presampler detector energy-scale uncertainty: it is about
5\% in the barrel and 10\% in the \endcap\ regions up to $|\eta|=1.8$.
The impact on the electron energy scale 
due to the uncertainty on the presampler energy scale depends on $\eta$ 
via the distribution of material in front of the calorimeter 
and on \ET, since the fraction of energy deposited in the presampler decreases as 
the electron energy increases. For very high-\ET\ electrons,
this uncertainty should decrease asymptotically to zero.
As for the material uncertainty, the 
$\alpha$ coefficients extracted from \Zee\ data correct the electron energy scale on average for
any bias on the presampler energy scale (giving by construction no bias at $\ET\sim 40$~GeV) 
but will not improve the response linearity in energy.
The largest uncertainty is 1.4\%, found
for the region $1.52<|\eta|<1.8$ and for $\ET=1$~TeV (due to the large extrapolation from $\ET=40$~GeV 
to this energy).

\item{\bf Calorimeter electronic calibration and cross-talk}
Cells belonging to different sampling layers in the EM calorimeters 
may have slightly different energy scales due to
cross-talk and uncertainties arising from an imperfect electronic calibration. 
The uncertainties on the energy scale
relative to the middle layer for cells in the strip and back layers of the calorimeter 
are estimated to be 1\%
and 2\%, respectively~\cite{Banfi:731840,Collard:1058294}. Using the same method as discussed above 
for the presampler detector energy scale,
the uncertainty on the strip layer energy scale is found to be 0.1\% for all $\eta$ and $\ET$,
while it is negligible 
on the back layer energy scale 
(as the energy deposited there is small).

\item{\bf Non-linearities in the readout electronics}
The readout electronics provide a linear response to typically 0.1\%~\cite{Abreu:2010zzc}.
This is taken as a systematic uncertainty on the extrapolation of the electron energy scale 
extracted from \Zee\ events to higher energies.

\item{\bf Requirements on calorimeter operating conditions} 
To check the possible bias due to these requirements, 
a tighter veto was applied on electrons falling close to dead regions 
and electrons in regions with non-nominal high voltage were excluded. 
No significant effect is observed for the barrel and endcap calorimeters, 
while differences of $0.6-0.8$\% are
seen in the forward region. 

\item{\bf Background and fit range} The effect of the background, predominantly from jets, on 
the extracted $\alpha$ values was studied by
tightening the electron selection thereby decreasing the amount of background 
significantly.
In addition, the fit range was also changed from $80-100$~GeV to $75-105$~GeV and $85-95$~GeV. 
The resulting uncertainty due to the electron selection is
+0.1\% in the barrel region
and reaches +1\% in the forward region, while due to the fit range it is 
0.1\% in the barrel region and grows to 0.6\% in the forward region.
These uncertainties are treated as uncorrelated.

\item{\bf Pile-up} The effect of 
pile-up is studied by determining the $\alpha$ coefficients as a function of
the number of reconstructed primary vertices (from 1 to 4). 
The average $\langle\alpha\rangle$ increases very slightly with the number of primary vertices and 
a systematic uncertainty of 0.1\% is assigned.

\item{\bf Possible bias of the method} 
The bias of the method is assessed by repeating the fit procedure on simulated
data, resulting in a systematic uncertainty of 0.1\% (0.2\%) in the central
(forward) region.
Moreover, the results of alternative fit methods were compared on data 
and agree within $0.1-0.5$\% ($0.8-1.0$\%).
This is added as an additional uncertainty due to possible biases of the method.

\item{\bf Theoretical inputs} In the extraction of the $\alpha$ coefficients from the data, 
the MC simulation, which uses a certain model of the Z lineshape, serves as a reference. 
Uncertainties related to the imperfect physics modelling of QED final state radiation, of the parton density 
functions in the proton, and of the underlying event are found to be negligible.
\end{itemize}
To summarize, the overall systematic uncertainty on the electron energy scale is 
a function of \ET\ and $\eta$. It is illustrated in
Figure~\ref{fig:sys_ele_central} for two $\eta$-regions. 
For central electrons with $|\eta|<2.47$, 
the uncertainty varies from 0.3\% to 1.6\%. 
The systematic uncertainties are smallest for $\ET=40$~GeV, typically below 0.4\%. 
Below $\ET=20$~GeV, the uncertainty grows linearly with decreasing \ET\ and slightly
exceeds 1\% at $\ET=10$~GeV.
For forward electrons with $2.5<|\eta|<4.9$, the uncertainties are larger and vary between
2\% and 3\%. 

\begin{table*}
\caption{Systematic uncertainties (in \%) on the electron energy scale
in different detector regions.}
\label{tab:sys} 
\begin{center}
\begin{tabular}{|l||c|c|c|}
\hline
                          & Barrel & \Endcap\ & Forward \\
\hline
\hline
Additional material & \multicolumn{3}{c|}{\ET- and $\eta$-dependent, from $-$2\% to +1.2\% } \\
\hline
Low-\ET\ region & \multicolumn{3}{c|}{\ET-dependent, from 1\% at 10 GeV to 0\% at 20 GeV} \\
\hline
Presampler energy scale & \multicolumn{2}{c|}{\ET- and $\eta$-dependent, $0-1.4$\%} &  \\
\hline
Strip layer energy scale &  0.1   & 0.1   &  0.1      \\
\hline
Electronic non-linearity  &  0.1   & 0.1   &  0.1      \\
\hline
Object quality requirements  & $<$0.1     &   $<$0.1  & 0.6$-$0.8 \\
\hline
Background and fit range   &  0.1   & 0.3  &  1.2  \\
\hline
Pile-up                   &  0.1   & 0.1   &  0.1      \\
\hline
Bias of method            &  0.1   & 0.1$-$0.5   & 0.8$-$1.0        \\
\hline
\end{tabular}
\end{center}
\end{table*}

%\subsubsection{\texorpdfstring{Energy-scale determination using $E/p$ measurements}
%{Energy-scale determination using E/p distributions}}
\subsubsection{Energy-scale determination using $E/p$ measurements}
\label{sec:EoP}

A complementary in-situ calibration method compares the energy $E$ measured by 
the electromagnetic calorimeter to the momentum $p$ measured by the inner detector.
It allows to take advantage of the larger statistics of \Wen\ decays.

\begin{figure*}
\begin{center}
\includegraphics[width=0.49\textwidth]{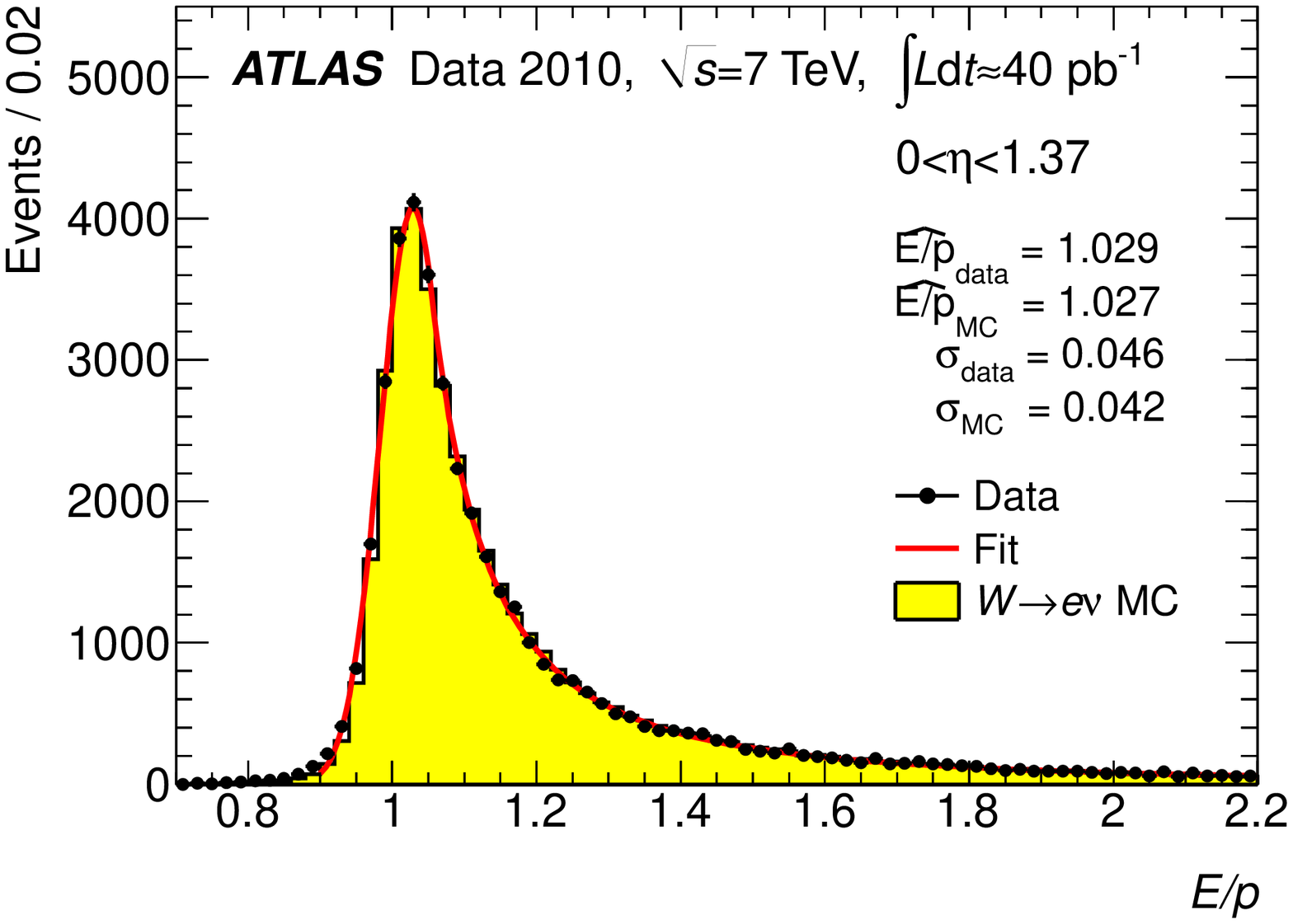}
\includegraphics[width=0.49\textwidth]{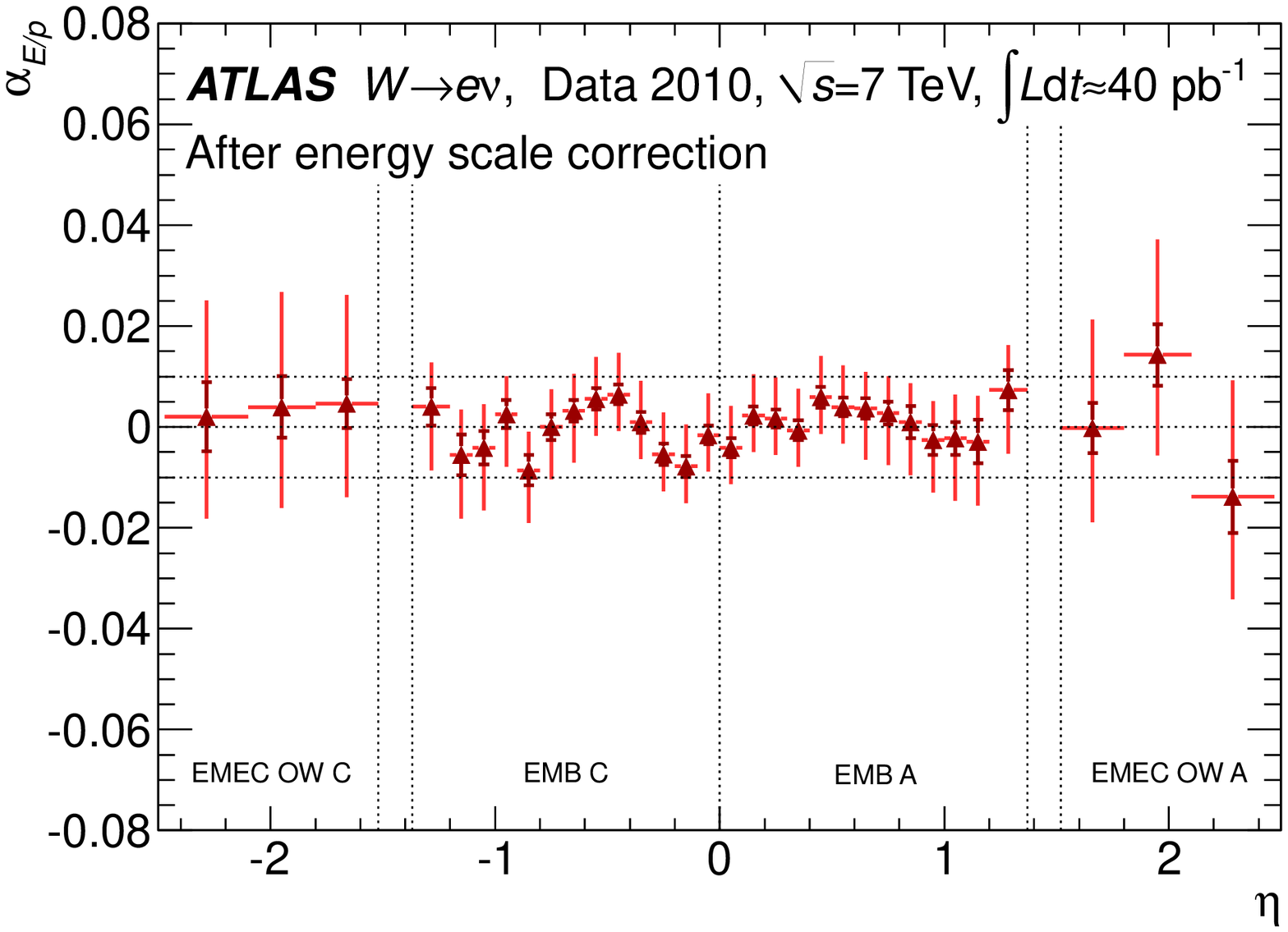}
\end{center}
\caption{(left) $E/p$ distributions of electrons and positrons from 
\Wen\ decays for $0. < \eta < 1.37$ in data (full circles with statistical error bars) 
and \Wen\ MC (filled histogram).
The result of the fit with a Crystal Ball function to the data is also shown (full line). 
The most probable value ($\widehat{E/p}$) and the Gaussian width ($\sigma$) of the fitted Crystal Ball function
are given both for the data and the signal MC.
(right) The $\alpha_{E/p}$ energy-scale correction factors 
derived from fits to $E/p$ distributions of \Wen\ electron and positron data, after the baseline
calibration had been applied. 
The inner error bars show the statistical uncertainty, 
while the outer error bars indicate the total
uncertainty.
The boundaries of the different detector parts defined in Section~\ref{sec:Detector}
are indicated by dotted lines.}
\label{fig:eop}
\end{figure*}

The ratio $E/p$ is shown on the left of Figure~\ref{fig:eop}
for electrons selected in the barrel EM calorimeter in \Wen\ events.
$E/p$ is close to unity, with a significant tail
at large values due to Bremsstrahlung occurring in the
inner detector.
The core of the distribution can be described by a Gaussian whose width
corresponds to the measurement error due to both the EM cluster energy 
and the track curvature resolutions.

The unbinned $E/p$ 
distributions are fitted by a Crystal Ball function~\cite{CrystalBall1,CrystalBall2}
and the most probable value, $\widehat{E/p}$, is extracted. 
The fit range, $0.9 < E/p < 2.2$, was chosen to be fully contained 
within the \ET- and $\eta$-dependent lower (0.7$-$0.8) and upper ($2.5-5.0$) 
cuts applied in the \tight\ electron selection.
The correction factors $\alpha_{E/p}$ are then derived by
\begin{equation}
 \widehat{E/p}_\mathrm{data}  = \widehat{E/p}_\mathrm{MC}(1+\alpha_{E/p}).
\label{eq:EoP}
\end{equation}
On the right of Figure~\ref{fig:eop} the $\eta$ dependence of 
the $\alpha_{E/p}$ coefficients measured
using electrons and positrons from \Wen\ decays are 
shown after the baseline calibration had been applied.
As expected, $\alpha_{E/p}\approx 0$ within about 1\%. 
The fluctuations are larger in the endcaps, where the statistics are poorer. 

The dominant systematic uncertainties on the measured $\alpha_{E/p}$ values arise from 
the fit procedure, (0.1-0.9)\%, 
the description of the material in front of the EM calorimeter, (0.3-0.9)\%, 
the background contamination in the selected electron sample, (0.2-1)\%, and
the track momentum measurement in the inner detector, (0.6-1.5)\%. 
The total uncertainty increases with $\eta$ and 
amounts to about 1\% in the barrel and 2\% in the endcaps.

The determination of the electron energy scale using the $E/p$ distributions measured in \Wen\
decays agrees, within its larger systematic uncertainties, with the baseline method using the 
invariant mass distribution in \Zee\ events, as shown on the right of Figure~\ref{fig:eop}.

\subsubsection{Energy response uniformity and linearity}
\label{sec:CalibUniformity}

\begin{figure*}
\begin{center}
\includegraphics[width=0.49\textwidth]{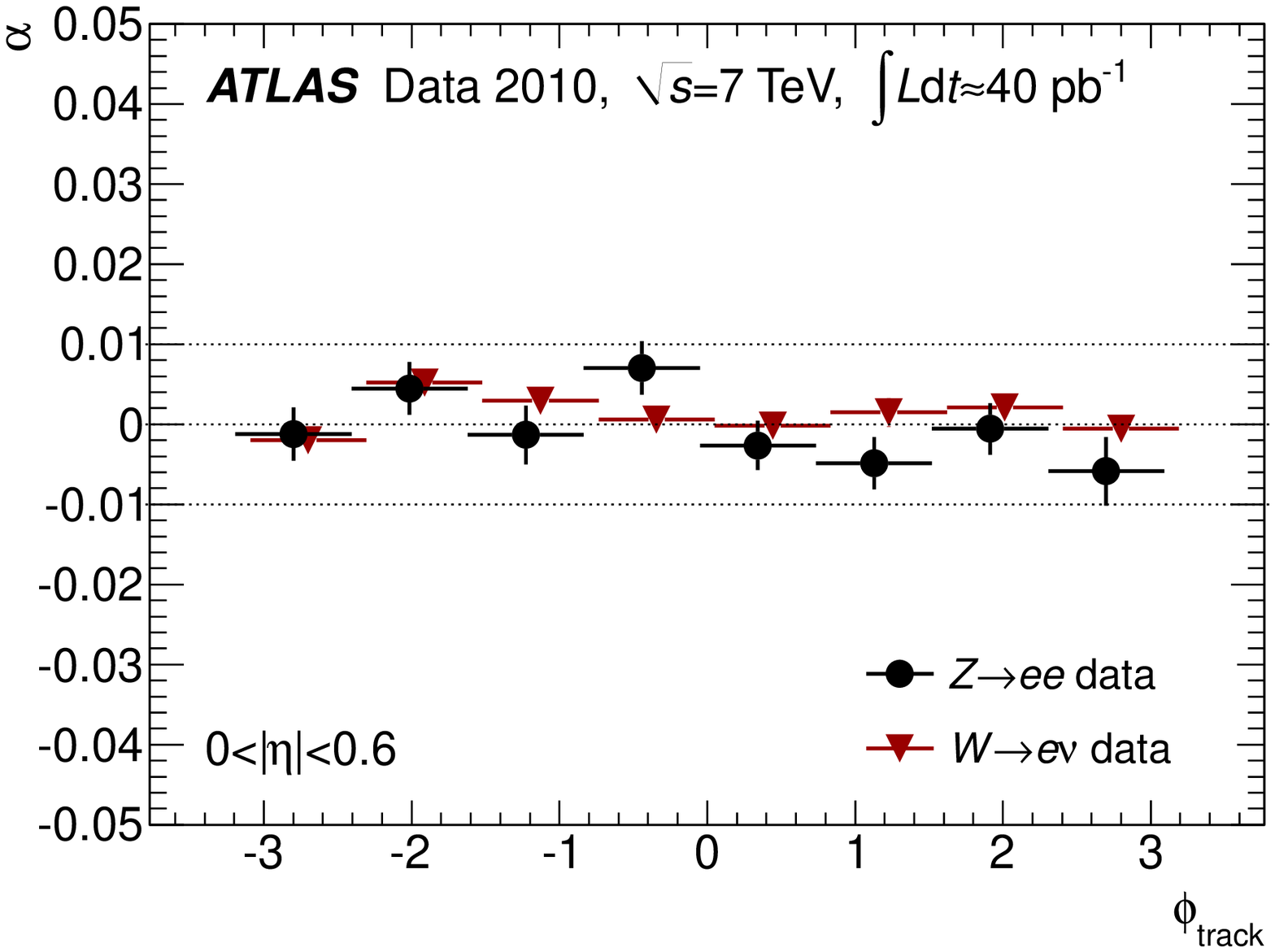}
\includegraphics[width=0.49\textwidth]{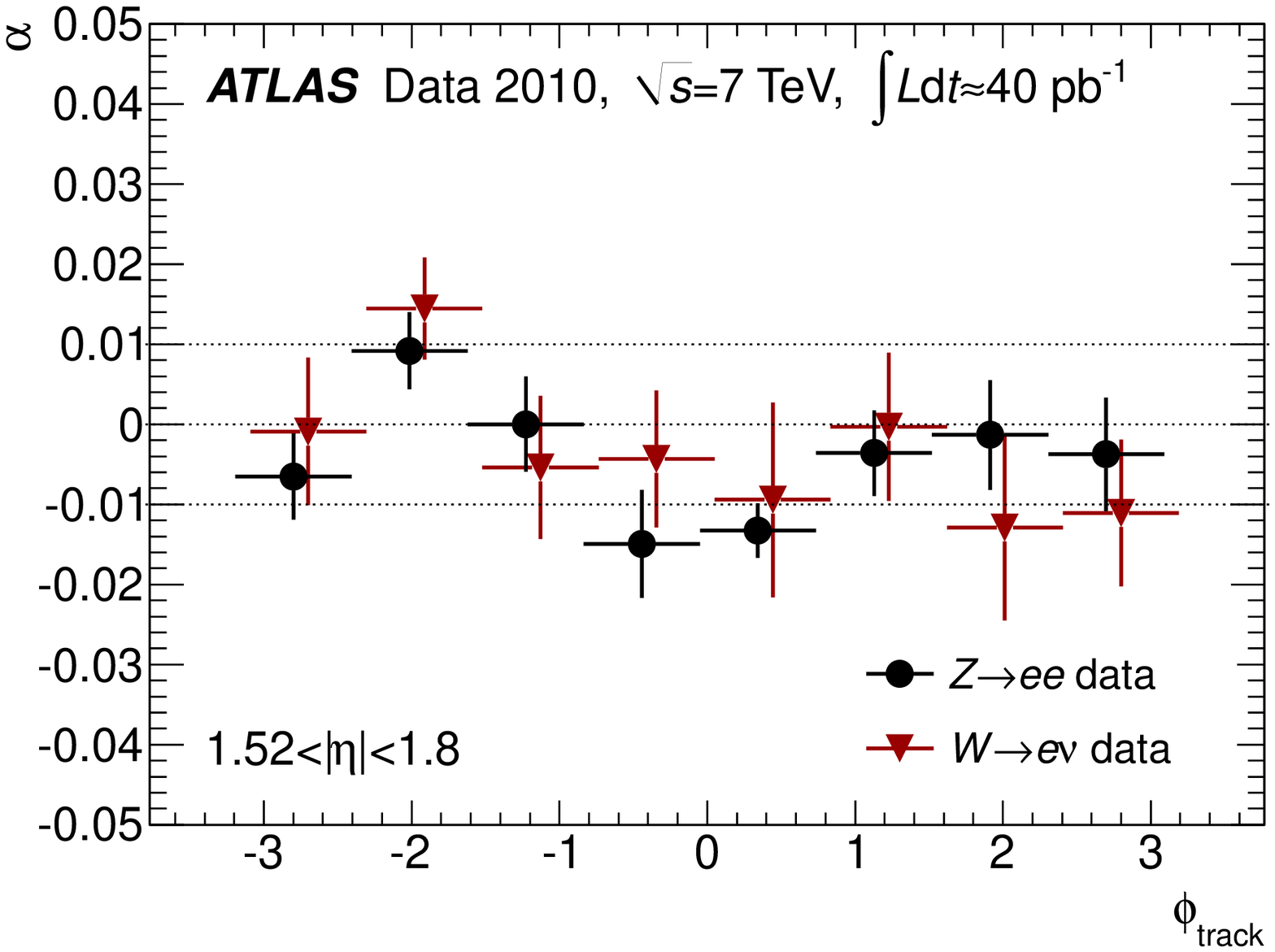}
\end{center}
\caption{The $\alpha$ energy-scale correction factor
as a function of the electron track $\phi$  
for (left) $|\eta|<0.6$ and (right) $1.52<|\eta|<1.8$
determined by the baseline calibration using \Zee\ decays (circles)
and by the $E/p$ method using \Wen\ decays (triangles).
Errors are statistical only.}
\label{fig:figeopphi}
\end{figure*}

The azimuthal uniformity of the calorimeter response is studied
using both the dielectron invariant mass distributions of \Zee\ events 
and the $E/p$ distributions of \Wen\ events,
after applying the $\eta$-dependent baseline calibration.
The results are shown in Figure~\ref{fig:figeopphi} for two $\eta$ regions.
They demonstrate a $\phi$ non-uniformity of less than about 1$\%$.

\begin{figure*}
\begin{center}
\includegraphics[width=0.49\textwidth]{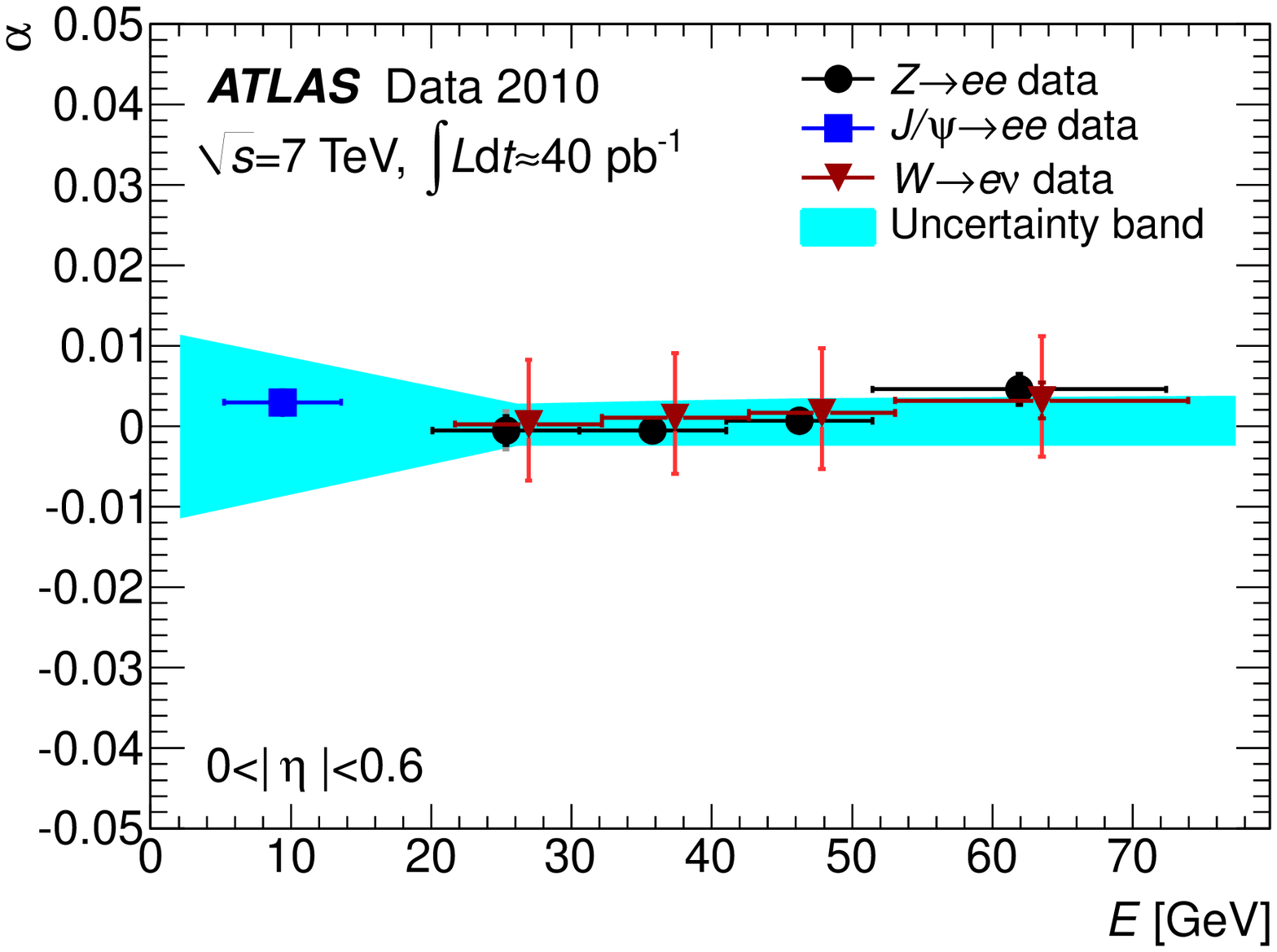}
\includegraphics[width=0.49\textwidth]{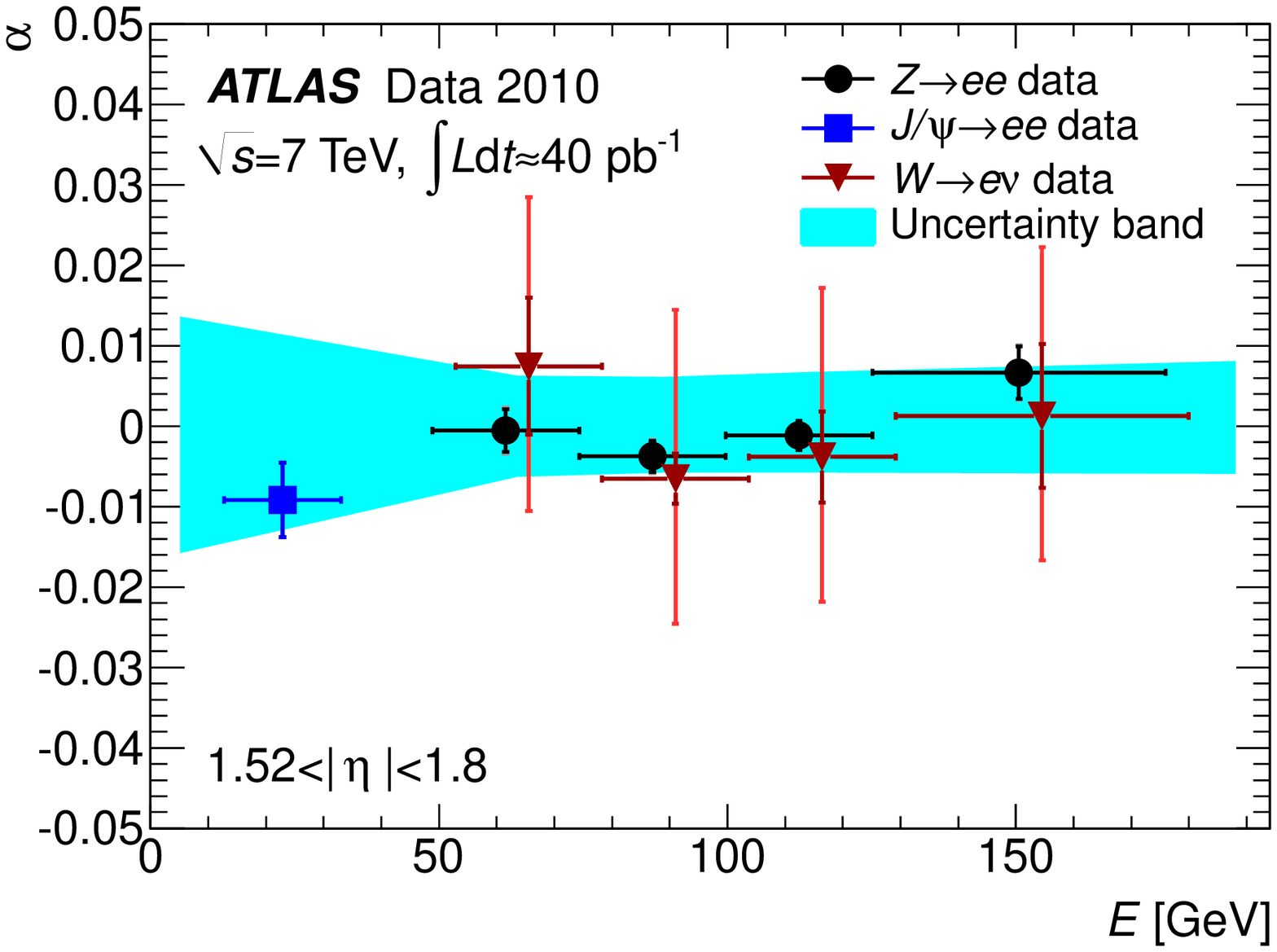}
\end{center}
\caption{The $\alpha$ energy-scale correction factor
as a function of the electron energy for (left) $|\eta|<0.6$ and (right) $1.52<|\eta|<1.8$
determined by the baseline calibration method using \Zee\ (circles) and \Jee\ (square) decays
and by the $E/p$ method described in Subsection~\ref{sec:EoP}
using \Wen\ decays (triangles).
For the \Zee\ data points, the inner error bar represents the statistical uncertainty
and the outer gives the combined error when bin migration effects are also included.
The error on the \Jee\ measurements are statistical only. 
The band represents the systematic errors on the energy scale
for the baseline calibration method as discussed in Table~\ref{tab:sys}.
For the $E/p$ method, the inner error bar represents the statistical and the outer
the total uncertainty.}
\label{fig:figlinear}
\end{figure*}

The linearity of the calorimeter response is studied, after applying the
$\eta$-dependent baseline calibration,
by determining the $\alpha$ coefficients in bins of electron energy.
The \Zee\ results are complemented at low energy by a \Jee\ calibration point
as shown in Figure~\ref{fig:figlinear} for two regions: on the left
the region $|\eta|<0.8$ which has the smallest uncertainties, and 
on the right the region $1.52<|\eta|<1.8$ which is affected by the largest material 
uncertainties.
Compared to \ET\ independent calibration, calibration factors
obtained as a function of \ET\ are more sensitive to
the description
of the energy resolution. This effect was estimated by varying the energy
resolution in MC simulation within its uncertainty and was
found to be about 0.1\% in the central region and up to 0.8\% in
the forward region.
All measurements are found to be within the uncertainty bands
assigned to the electron energy scale.
For the central region, the results are cross-checked with the $E/p$ method using 
\Wen\ events, averaged over the electron charge. 
Within the larger systematic uncertainties of the \Wen\ measurement,
the linearity measurements agree well with the \Zee\ data.

\subsection{Electron energy resolution}
\label{sec:CalibPerf}
The fractional energy resolution in the calorimeter is pa\-ram\-e\-tri\-sed as
\begin{equation}
\frac{\sigma_{E}}{E} = \frac{a}{\sqrt{E}} \oplus \frac{b}{E} \oplus c.
\label{eq:Eresolution}
\end{equation}
Here $a$, $b$ and $c$ are $\eta$-dependent parameters: 
$a$ is the sampling term, $b$ is the noise term and $c$ is the constant term. 

Great care was taken during the construction of the calorimeter to minimise all sources of
energy response 
non-uniformity, since any non-uniformity has a direct impact
on the constant term of the energy resolution.
The construction tolerances and the electronic 
calibration system ensure that the calorimeter response
is locally uniform, with a local constant term below 0.5\%~\cite{Aharrouche:2007nk} over regions
of typical size $\Delta\eta\times\Delta\phi=0.2\times0.4$. These regions are expected to be
intercalibrated in situ to $0.5\%$ achieving a global constant 
term\footnote{The long-range constant term is the residual
miscalibration between the different calorimeter regions, and
the global constant term is the quadratic sum of the
local and long-range constant terms.}
around 0.7\% for the EM calorimeter, which is well within the requirement
driven by physics needs, for example the $H\rightarrow\gamma\gamma$
sensitivity~\cite{atlastdr}.

To extract the energy resolution function from data, 
more statistics are needed than available in the 2010 data sample.
Therefore, only the constant term is determined here from a simultaneous analysis
of the measured and predicted dielectron invariant mass resolution from \Zee\
decays,
taking the sampling and noise terms from MC simulation. 

\begin{figure*}
\begin{center}
\includegraphics[width=0.485\textwidth]{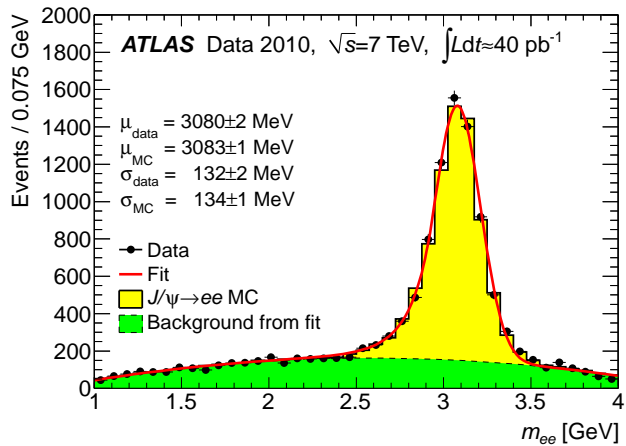}
\end{center}
\caption{Reconstructed dielectron mass distribution for \Jee\ decays, as measured
after applying the baseline \Zee\ calibration.
The data (full circles with statistical error bars) are compared to the sum of the MC signal 
(light filled histogram) and the background contribution (darker filled histogram) 
modelled by a Chebyshev polynomial. The mean ($\mu$) and the Gaussian width ($\sigma$) of the fitted 
Crystal Ball function are given both for data and MC.
}
\label{fig:mass_jpsi}
\end{figure*}

As shown in Figure~\ref{fig:mass_jpsi}, the measured dielectron mass
distribution of electrons coming from \Jee\ decays is in good agreement with the MC
prediction (both for the mean and the width). Since the electron
energy resolution at these low energies is dominated by the contribution 
from the sampling term, it is assumed 
that the term $a$ is well described, within 
a 10\% uncertainty,
as a function of $\eta$ by the MC simulation.
The noise term has a significant contribution only 
at low energies. Moreover, its effect on
the measurement of the constant term 
cancels out to first order, since the noise description in the MC simulation
is derived
from calibration data runs. The above assumptions lead to the formula:
\begin{equation}\label{eqt:cst}
c_\mathrm{data} = \sqrt{2\cdot\left(\left(\frac{\sigma}{m_Z}\right)_\mathrm{data}^2
-\left(\frac{\sigma}{m_Z}\right)_\mathrm{MC}^2\right) 
+ c_\mathrm{MC}^2} \phantom{i}_{,}
\end{equation}
where $c_{MC}$ is the constant term of about 0.5\% in the MC simulation.
The parameter $c_\mathrm{data}$ is an effective constant term which includes 
both the calorimeter constant term and the effect of inhomogeneities due to 
possible additional material. $m_Z$ denotes the \Zboson\ mass~\cite{Zedometry}, and 
$\sigma$ is the Gaussian component of the experimental resolution.

\begin{figure*}
\begin{center} 
\includegraphics[width=0.49\textwidth]{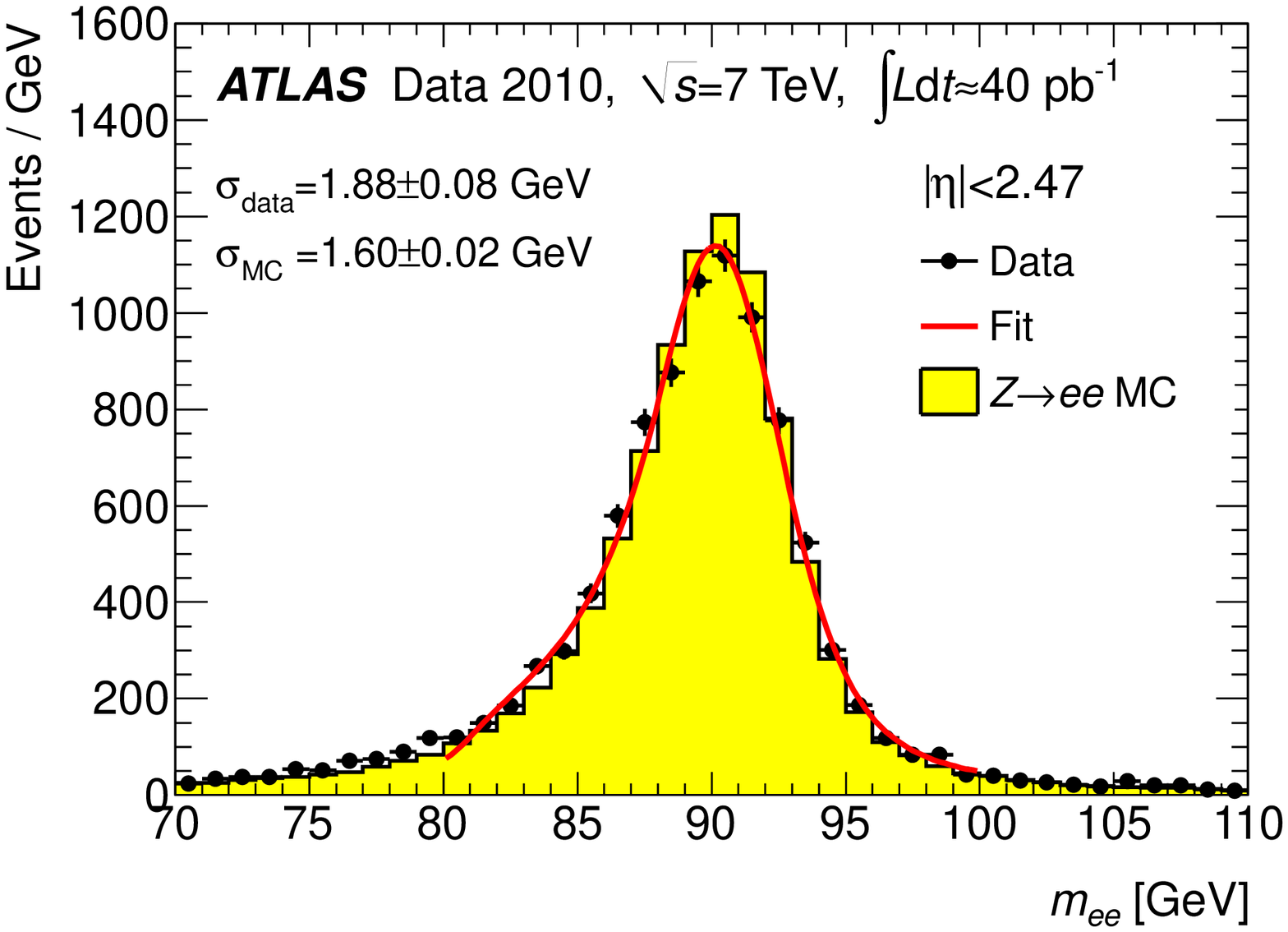}
\includegraphics[width=0.49\textwidth]{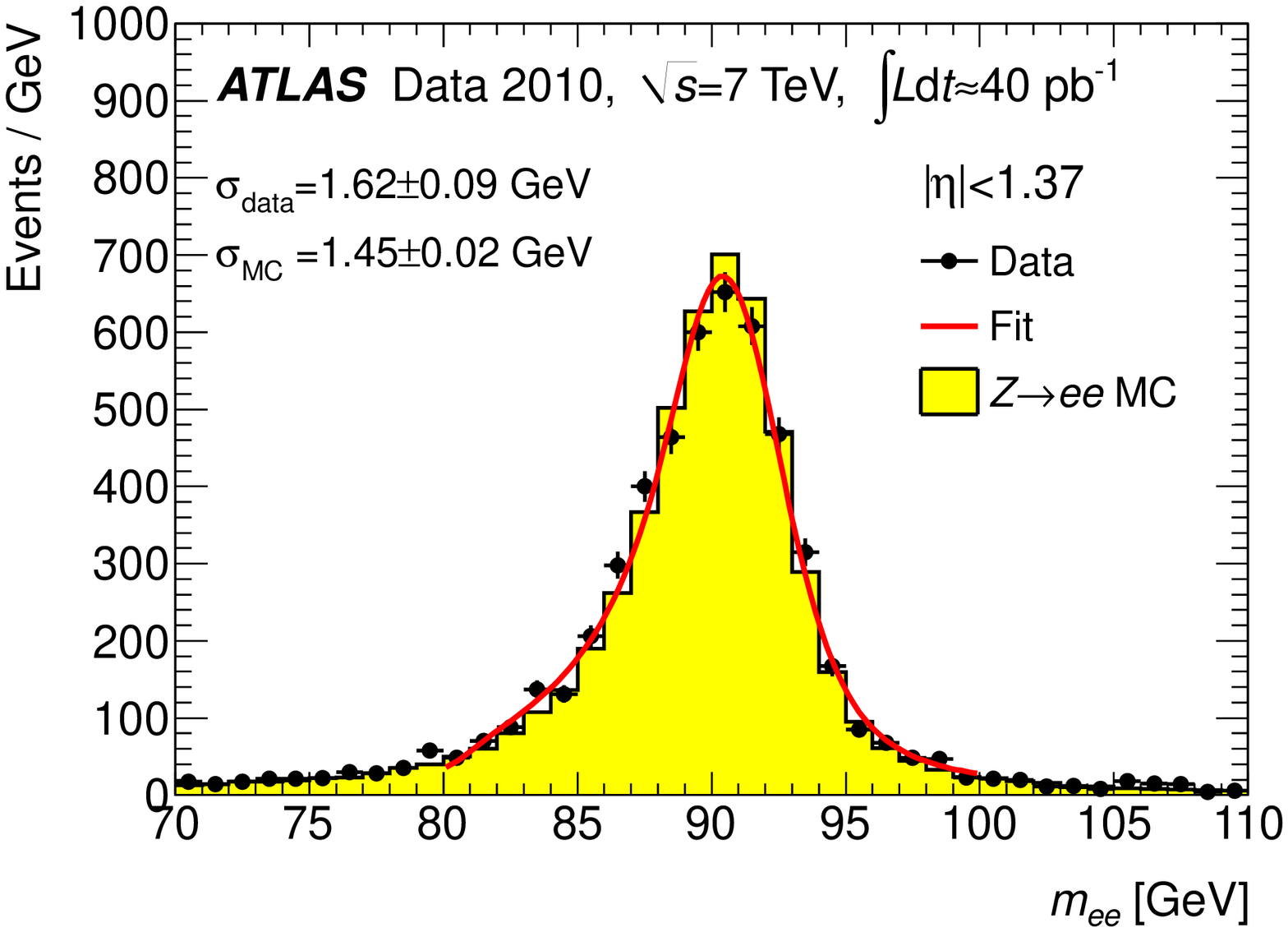} 

\includegraphics[width=0.49\textwidth]{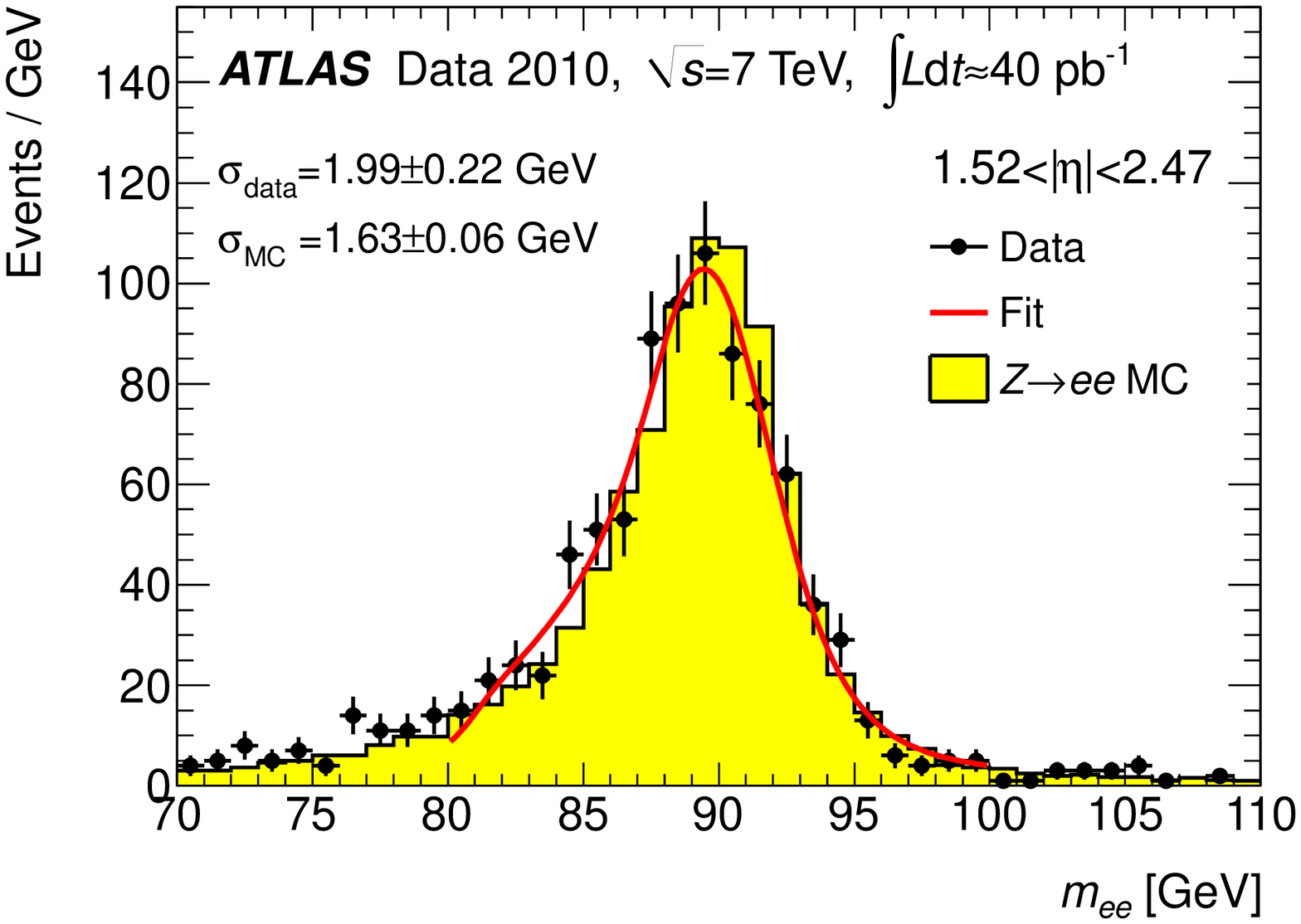} 
\includegraphics[width=0.49\textwidth]{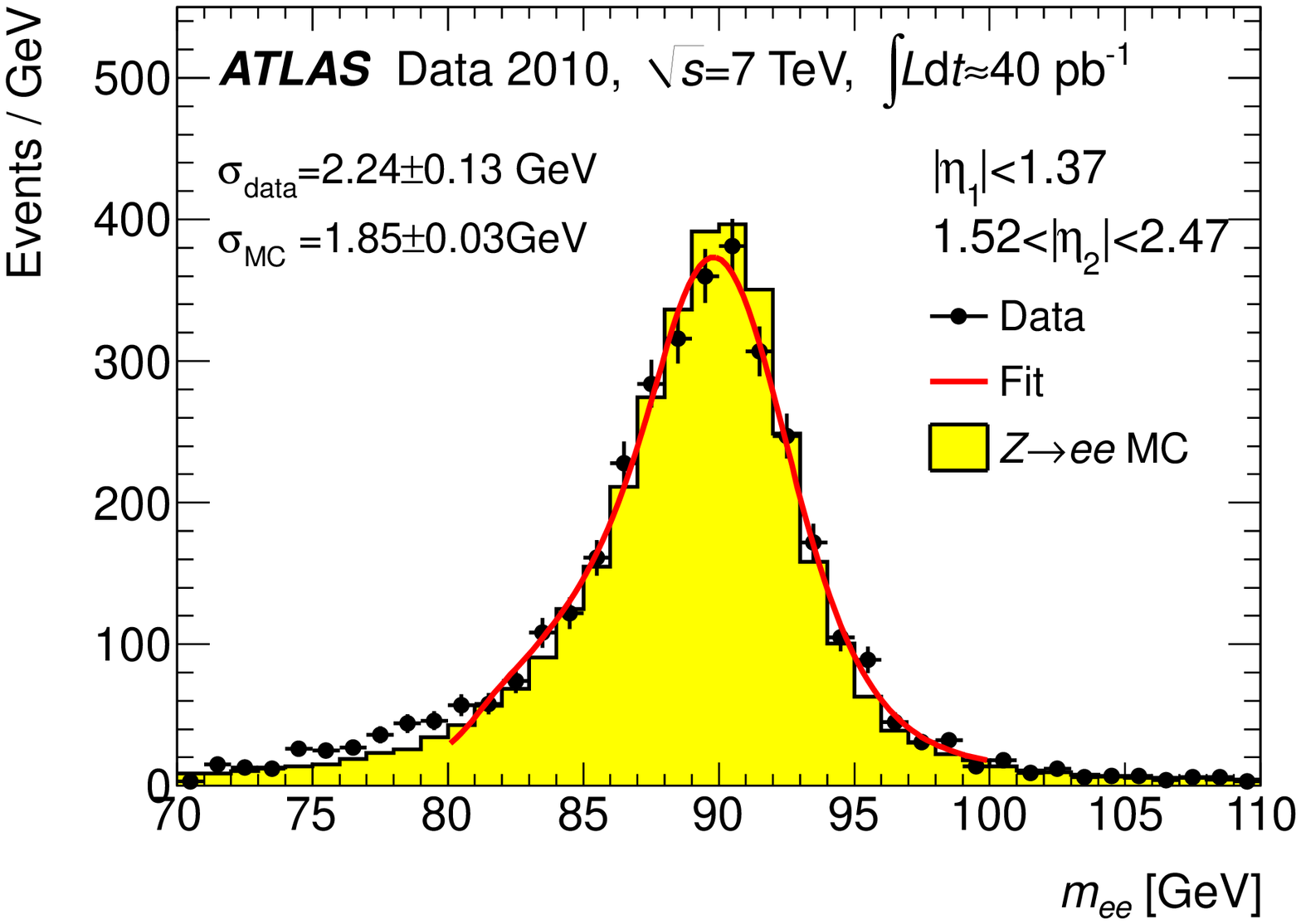} 

\includegraphics[width=0.49\textwidth]{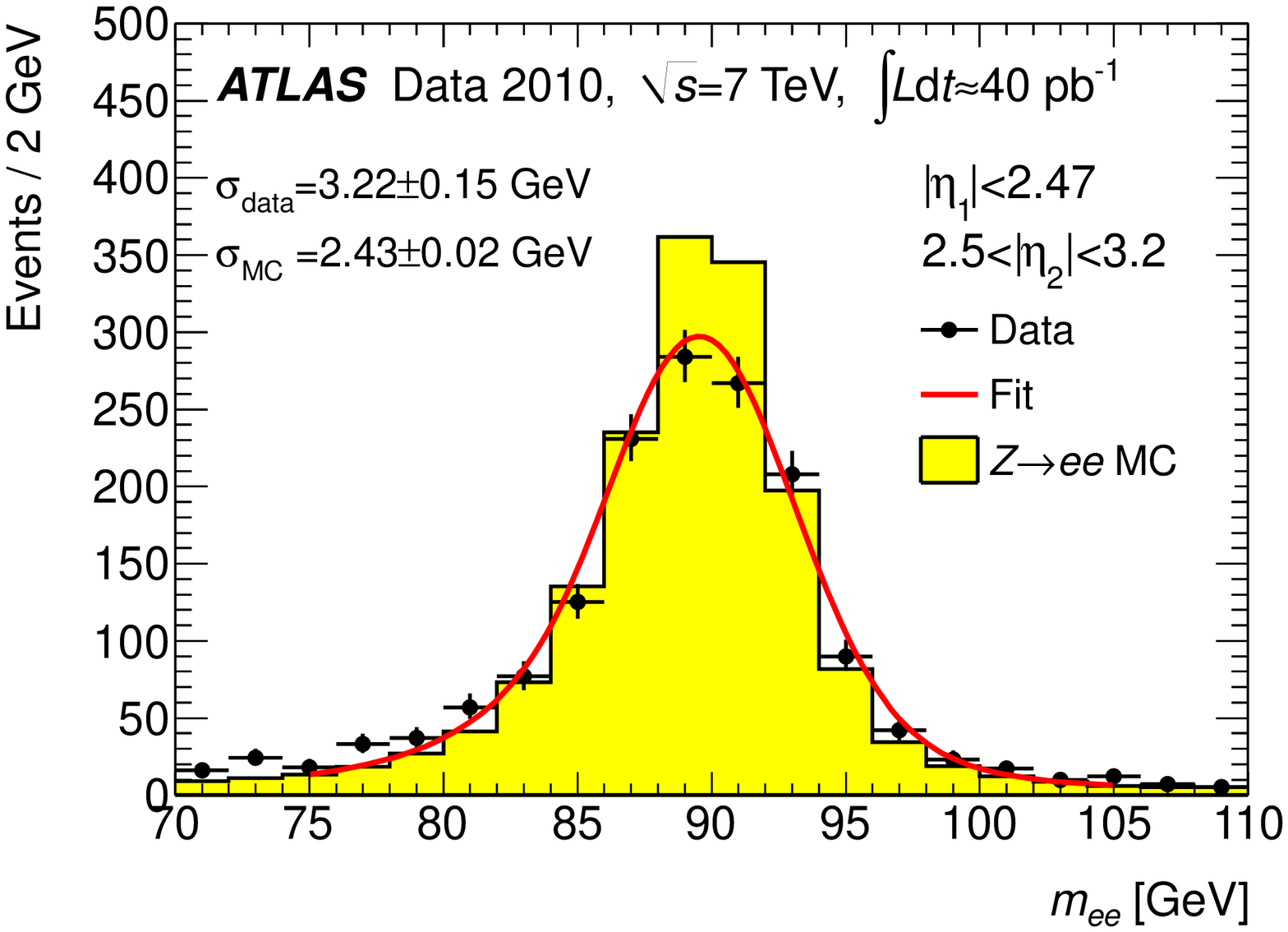}
\includegraphics[width=0.49\textwidth]{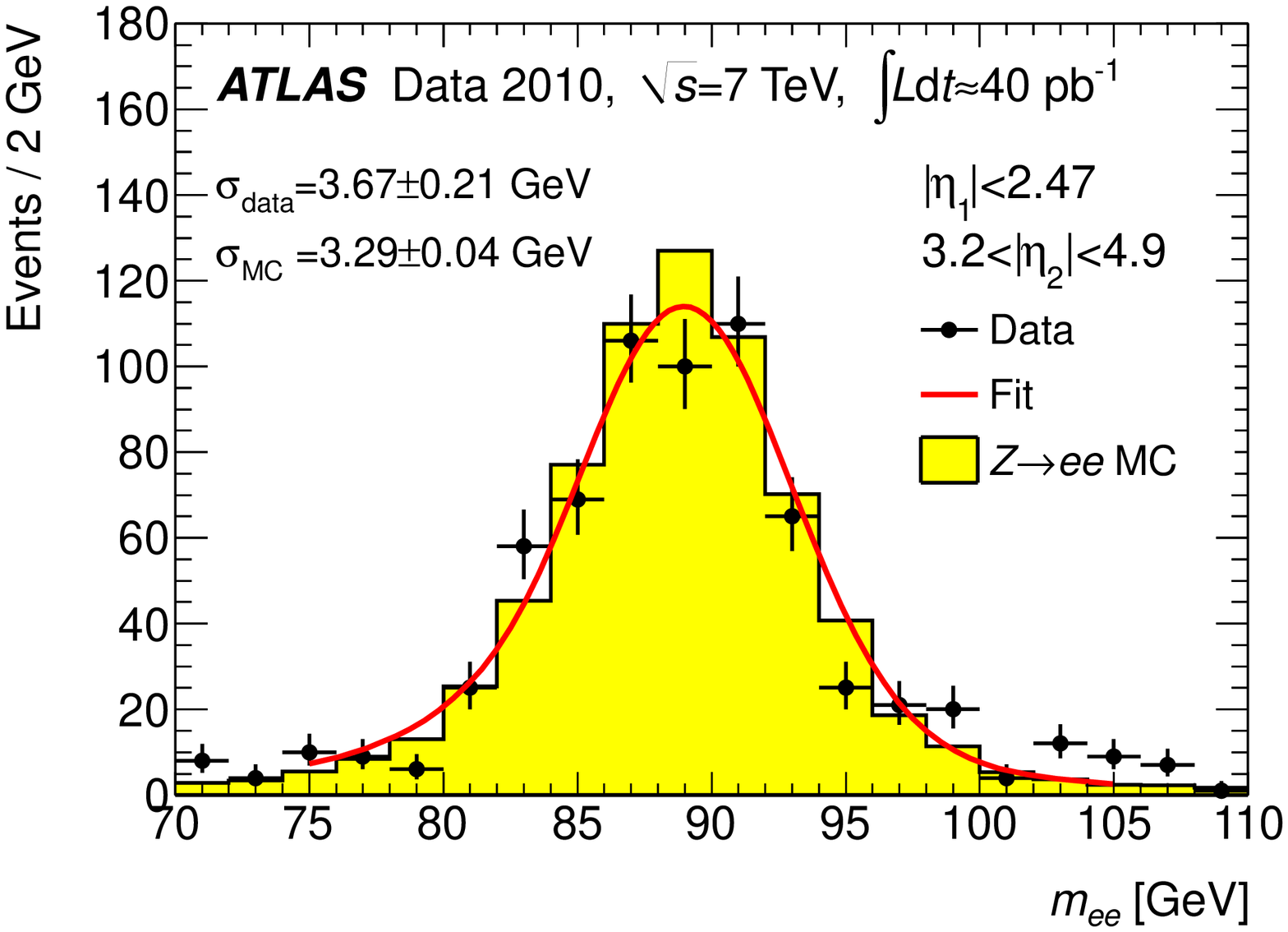}   
\end{center} 
\caption{Reconstructed dielectron mass distributions for \Zee\ decays for
different pseudorapidity regions
after applying the baseline \Zee\ calibration. 
The transition region $1.37<|\eta|<1.52$ is excluded.
The data (full circles with statistical error bars) are compared to the signal MC expectation
(filled histogram). The fits of a Breit-Wigner
convolved with a Crystal Ball function are shown (full lines).
The Gaussian width ($\sigma$) of the Crystal Ball function
is given both for data and MC simulation.} 
\label{fig:mass_Z} 
\end{figure*}

The resolutions are derived from fits to the invariant mass distributions using a Breit-Wigner
convolved with a Crystal Ball function in the mass range $80-100$~GeV for central-central events and in the
mass range $75-105$~GeV for central-forward events.  The Breit-Wigner width is fixed to the measured 
\Zboson\ width~\cite{Zedometry}, and the experimental resolution is described by the Crystal Ball function. 
Figure~\ref{fig:mass_Z} shows the invariant mass distributions of the selected \Zee\ decays:
the measured Gaussian components of the experimental resolution are always slightly worse than 
those predicted by MC, with the smallest deviation observed for barrel--barrel events 
(top left) and the largest one for central--EMEC-IW events (bottom left).

 \begin{table*}
\caption{Measured effective constant term $c_\mathrm{data}$ (see Eq.~\ref{eqt:cst}) from the observed width of the \Zee\ 
peak for different calorimeter $\eta$ regions.}  
\label{tab:cst_term}
\begin{center}
 \begin{tabular}{|l|l||l|} 
 \hline   
 Sub-system & $\eta$-range & Effective constant term, $c_\mathrm{data}$ \\ 
 \hline 
 \hline 
 EMB     & $|\eta|<1.37$ & 1.2\% $\pm$ 0.1\% (stat) $ ^{+\ 0.5\%}_{-\ 0.6\%}$ (syst)      \\ 
 EMEC-OW & $1.52<|\eta|<2.47$ &  1.8\% $\pm$ 0.4\% (stat) $\pm$ 0.4\% (syst)    \\ 
 EMEC-IW & $2.5<|\eta|<3.2$ &  3.3\% $\pm$ 0.2\% (stat) $\pm$ 1.1\% (syst)  \\ 
 FCal    & $3.2<|\eta|<4.9$ &  2.5\% $\pm$ 0.4\% (stat) $ ^{+\ 1.0\%}_{-\ 1.5\%}$ (syst) \\
 \hline 
 \end{tabular} 
\end{center}
\end{table*}

In central--forward events the two electrons belong to different detector regions.
Therefore, when extracting the constant term in the forward region,
a smearing is applied to the central electrons 
using the results of the barrel--barrel and \endcap--\endcap\ measurements.

The results obtained for the effective constant term are shown in Table~\ref{tab:cst_term}. 
Several sources of systematic uncertainties are investigated.
The dominant uncertainty is due to the uncertainty on the sampling term,
as the constant term was extracted assuming that the sampling term is correctly reproduced by the simulation. 
To assign a systematic uncertainty due to this assumption, the simulation was modified by increasing 
the sampling term by 10\%. The difference in the measured constant term is found to be
about 0.4\% for the EM calorimeter and 1\% for the forward calorimeter. 
The uncertainty due to the fit procedure was estimated by varying the fit range.
The uncertainty due to pile-up was investigated by comparing simulated MC 
samples with and without pile-up and was found to be negligible.

\section{Efficiency measurements} 
\label{sec:Efficiencies}
In this section, the measurements of electron 
selection efficiencies are presented using the tag-and-probe method~\cite{TPMethod-CDF,TPMethod-D0}.
\Zee\ events provide a clean environment to study all components of the electron
selection efficiency discussed in this paper. In certain cases, such as
identification or trigger efficiency measurements, the statistical power of the
results is improved using \Wen\ decays, as well. To extend the reach towards
lower transverse energies, \Jee\ decays are also used to measure the electron
identification efficiency. However the available statistics of \Jee\ events
after the trigger requirements in the 2010 data sample
are limited and do not allow a precise separation of the isolated signal component
from b-hadron decays and from background processes.

\subsection{Methodology} 
A measured electron spectrum needs to be corrected for
efficiencies related to the electron selection in order to
derive cross-sections of observed physics processes 
or limits on new physics.  
This correction factor is defined as the product 
of different efficiency terms. 
For the case of a single electron in the final state one can write:
\begin{equation}
\label{eqn:CWApprox}
C = \epsilon_{event} \cdot \alpha_{reco} \cdot \epsilon_{ID} 
    \cdot \epsilon_{trig} \cdot \epsilon_{isol}.
\end{equation}
Here $\epsilon_{event}$ denotes the efficiency of the event preselection cuts, 
such as primary vertex requirements and event cleaning.
$\alpha_{reco}$ accounts for the basic reconstruction efficiency  
to find an electromagnetic cluster 
and to match it loosely to a reconstructed charged particle track
in the fiducial region of the detector and also for any 
kinematic and geometrical cuts on the reconstructed object itself.
$\epsilon_{ID}$ denotes the efficiency of the identification cuts 
relative to reconstructed electron objects.
$\epsilon_{trig}$ stands for the
the trigger efficiency 
with respect to all reconstructed and identified electron candidates.
$\epsilon_{isol}$ is the efficiency of any isolation requirement, if applied,
limiting the presence of other particles (tracks, energy deposits) close to the
identified electron candidate.

In this paper, three of the above terms are studied: 
the dominant term of $\alpha_{reco}$ that accounts for the efficiency to loosely 
match a reconstructed track fulfilling basic quality criteria to a 
reconstructed cluster, %$\epsilon_{reco}$, 
the identification efficiency $\epsilon_{ID}$,
and the trigger efficiency $\epsilon_{trig}$ for the most important single electron
triggers used in physics analyses based on 2010 data.

Note that the above decomposition is particularly useful as it allows the use of data-driven
measurements of the above independent efficiency terms, such as the
ones presented in this paper using the tag-and-probe (\TandP) technique,
in physics analyses, and therefore limits the reliance on MC simulation.
This is usually done by correcting the MC predicted values of the
above efficiency terms for a given physics process in bins (of typically \ET\ and $\eta$) 
by the measured ratios of the data to MC efficiencies in the
\TandP\ sample in the same bins. The range of validity of this method
depends on the kinematic parameters of the electrons used in the physics
analysis itself and on more implicit observables such
as the amount of jet activity in the events considered in the analysis with respect
to that observed in the \TandP\ sample. 

The \TandP\ method 
aims to select a clean and unbiased sample of electrons, called {\it probe} electrons, 
using selection cuts, called {\it tag} requirements, primarily on other objects in the event.
The efficiency of any selection cut can then be measured by applying it to the sample of probe electrons. 
In the following, a well-identified electron is used 
as the tag in the \Zee\ and \Jee\ measurements 
and high missing transverse momentum is used in the \Wen\ 
measurements.

For most efficiency measurements presented here, 
the contamination of the probe sample by background (for example
hadrons faking electrons, or electrons from heavy flavour decays, or electrons
from photon conversions) 
requires the use of some background estimation technique (usually 
a side-band or a template fit method).
The number of electron candidates is then independently estimated both 
at the probe level 
and at the level where the probe passes the cut of interest.
The efficiency is then equivalent to the fraction of probe candidates 
passing the cut of interest.

Depending on the background subtraction method, 
different formulae for computing the statistical 
uncertainties on the efficiency measurements have 
been used as discussed in~Ref.~\cite{CDFErrors}. 
These formulae are approximate but generally conservative. 
When no background subtraction is necessary, the simple binomial formula is
replaced by a Bayesian evaluation of the uncertainty.

The statistics available with the full 2010~dataset are not sufficient 
to measure any of the critical
efficiency components as a function of two parameters, 
so the measurements are performed separately in bins of~$\eta$ and~\ET\
of the probe.
The bins in $\eta$ are adapted to the detector geometry, 
while the \ET-binning corresponds to the 
optimization bins of the electron identification cuts.

\subsection{Electron identification efficiency in the central region}
\label{sec:IDPerf}
The measurements of the efficiency of electron identification with the predefined
sets of requirements, called \medium\ and \tight\ and described in Table~\ref{tab:IDcuts},  
were performed on three complementary samples of \Wen, \Zee\ and \Jee\ events. 
While the electrons from \Wen\ and \Zee\ decays are typically well-isolated,
the \Jee\ signal is a mix of isolated and non-isolated electrons.
Both prompt ($pp \rightarrow J/\psi X$) and non-prompt 
($b \rightarrow J/\psi X'$) production contribute, and in the latter case 
the electrons from the \Jee\ decay are typically accompanied by other particles from the 
decay of the b-hadron. This, coupled with the higher background levels in the 
low-\ET\ region, makes the \Jpsi\ analysis more demanding.
The measurements cover the central region of the EM calorimeter 
within the tracking acceptance, $|\eta| < 2.47$, and the electron transverse
energy range $\ET=4-50$~GeV. Electrons in the forward region, $2.5 < |\eta| < 4.9$,
are discussed in Subsection~\ref{sec:FwdPerf}.

\subsubsection{Probe selection}
\label{sec:IDProbeSelection}

The three data samples were obtained using a variety of triggers:
\begin{enumerate} 

\item \Wen\ decays are collected using a set of \MET~triggers. These triggers
had an increasing \MET~threshold from approximately 20~GeV initially at low
luminosity to 40~GeV at the highest luminosities obtained in~2010. The total number
of unbiased electron probes in this sample after background
subtraction amounts to about 27500. 

\item \Zee\ decays are obtained using a set of single inclusive electron
triggers with an \ET~threshold of~15~GeV. The total number of unbiased
electron probes in this sample is about 14500 after background subtraction. 

\item \Jee\ decays are selected using a set of low-\ET\  single electron
triggers with thresholds between 5~and~10~GeV. Towards the end of 2010, 
these triggers had to be heavily prescaled and a different trigger was
used, requiring an electromagnetic cluster
with~$\ET > 4$~GeV in addition to the single electron trigger. 
The total number of unbiased electron probes in this sample
amounts to about 6000 after background subtraction. As already noted, 
they are a mix of isolated
and non-isolated electrons from prompt and non-prompt \Jpsi\ decays, respectively,
with their fractions
depending on the transverse energy bin. 

\end{enumerate}

Only events passing data-quality criteria, in
particular concerning the inner detector and the calorimeters,
are considered.
At least one reconstructed primary vertex with at least three tracks should be
present in the event. Additional
cuts were applied to minimise the impact of beam backgrounds and to remove electron
candidates pointing to problematic regions of the calorimeter readout as discussed
in Subsection~\ref{sec:OQ}. 

Unbiased samples of electron probes, with minimal
background under the signal, were obtained by applying stringent cuts to the trigger
object in the event (a neutrino in the case of \Wen\ decays 
and one of the two electrons in the case of \Zee\ and \Jee\ decays), which is thus the \tag,
and by selecting the electron probe following very loose requirements
on the EM calorimeter cluster and the matching track: 
\begin{itemize}
\item In the case of \Wen\ decays, simple kinematic requirements 
were made: $\MET > 25$~GeV and $\MT > 40$~GeV. 
For the fake electron background 
from multijet events, there is
usually a strong correlation in the transverse plane
between the azimuthal angle of the \MET~vector and 
that of one of the highest \ET\ reconstructed jets.
Thus a large rejection against 
fake electrons from hadrons or photon conversions can be 
obtained by requiring \MET~isolation: the difference between the azimuthal angles 
of the missing transverse momentum and any jet having $\ET>10$~GeV
was required to be $\Delta \phi > 2.5$ for the baseline analysis.
This $\Delta \phi$ threshold was varied between~0.7 and~2.5 to assess
the sensitivity of the measurements to the level of background under the \Wen\
signal.
\item In the case of \Zee\ (resp.~\Jee) decays, the tag electron was required to 
have $\ET > 20$~(resp.~5)~GeV, to match the corresponding trigger object, 
and to pass the \tight\ electron identification requirements. 
The identification requirements were varied between the \medium\ and \tight\ 
selections to evaluate the sensitivity of the measurements to the level of 
background under the \Zee\ and \Jee\ signal. The probe electron was required 
to be of opposite charge to the tag electron. In the \Jee\ selection, 
to address the case of high-\ET\ electrons that would often produce close-by 
EM~showers in the calorimeter, the distance in~$\Delta R$ between the 
two electron clusters was required to be larger than~0.1. 
All tag--probe pairs passing the cuts were considered.
\item The probe electron was required to have $|\eta|<2.47$, and $\ET>15$~GeV for
\Wen, $\ET>15$~GeV for \Zee, and $\ET>4$~GeV for \Jee\ decays.
\item To reject beam-halo muons producing high-energy bremsstrahlung 
clusters in the EM~calorimeter in the data sample collected by 
\MET~triggers for the \Wen\ channel, 
certain track quality requirements have to be applied on the electron probes: 
the electron tracks should have at least one pixel hit
and a total of at least seven silicon (pixel plus SCT) hits. 
These cuts have been applied in all three selections, \Wen, \Zee\ and \Jee.
Their efficiency is measured separately using \Zee\ events
as described in~Subsection~\ref{sec:RecoPerf}.
\end{itemize}

\begin{figure*}
\begin{center}                               
\includegraphics[width=0.49\textwidth]{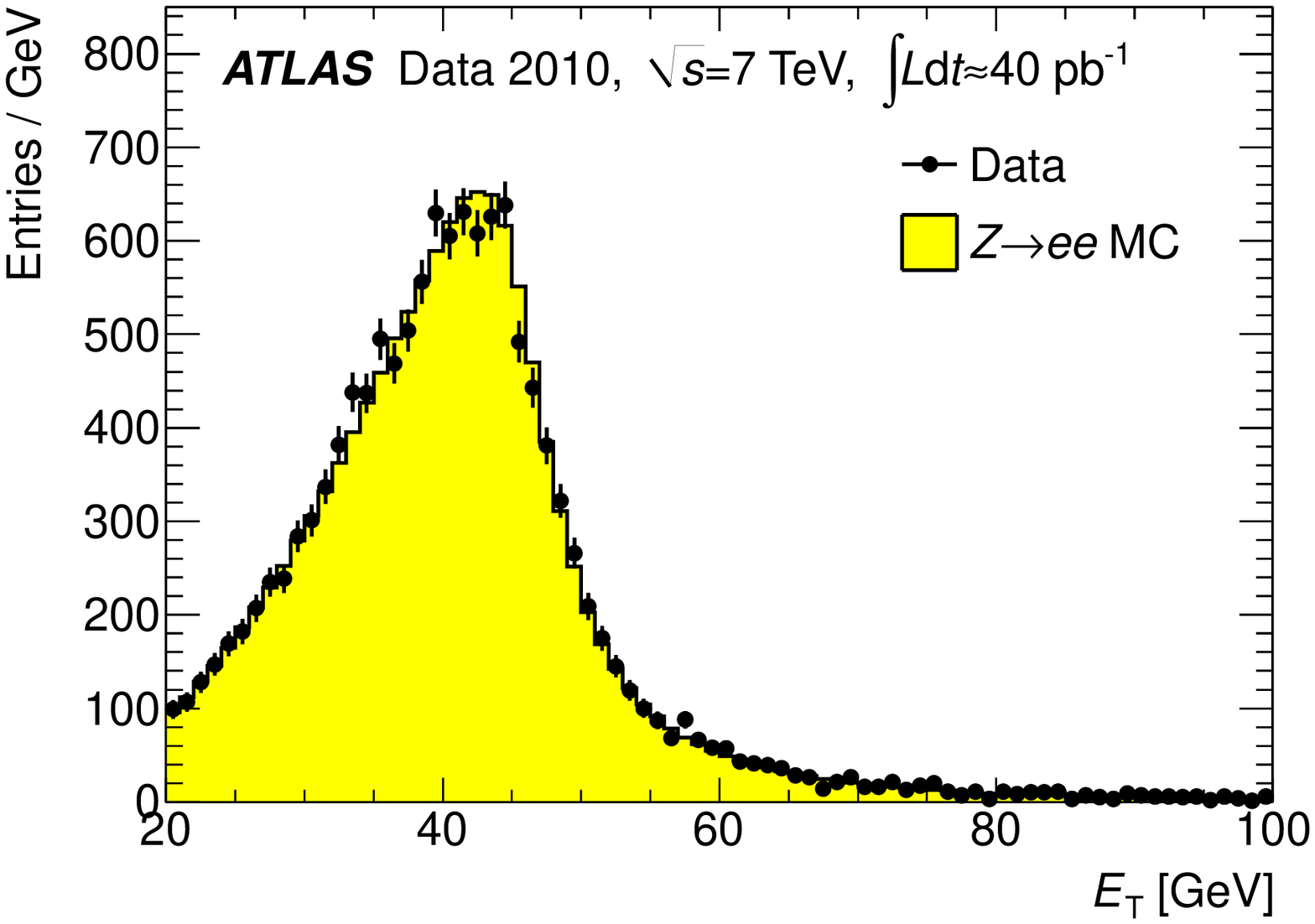}
\includegraphics[width=0.49\textwidth]{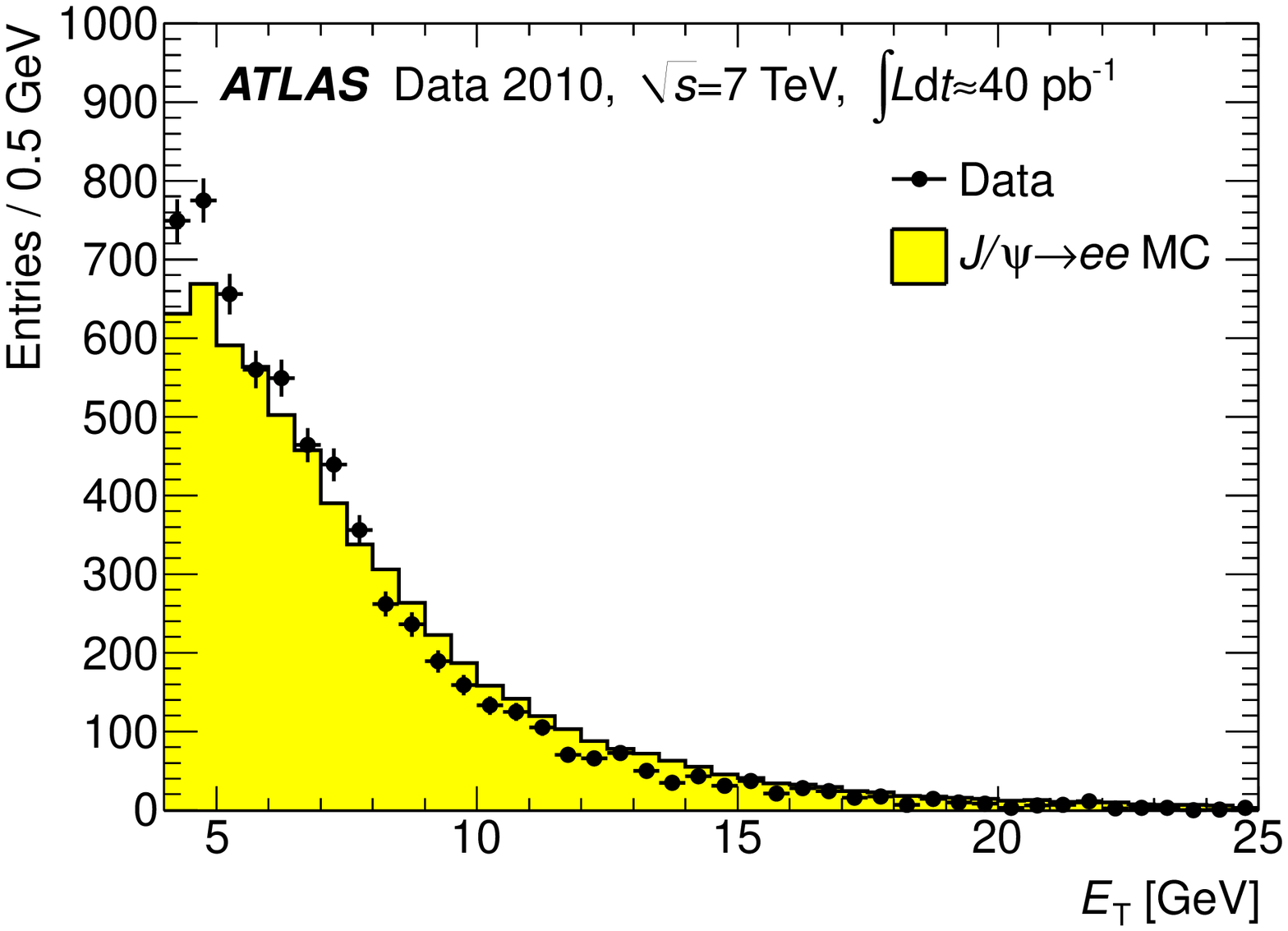}
   
\includegraphics[width=0.49\textwidth]{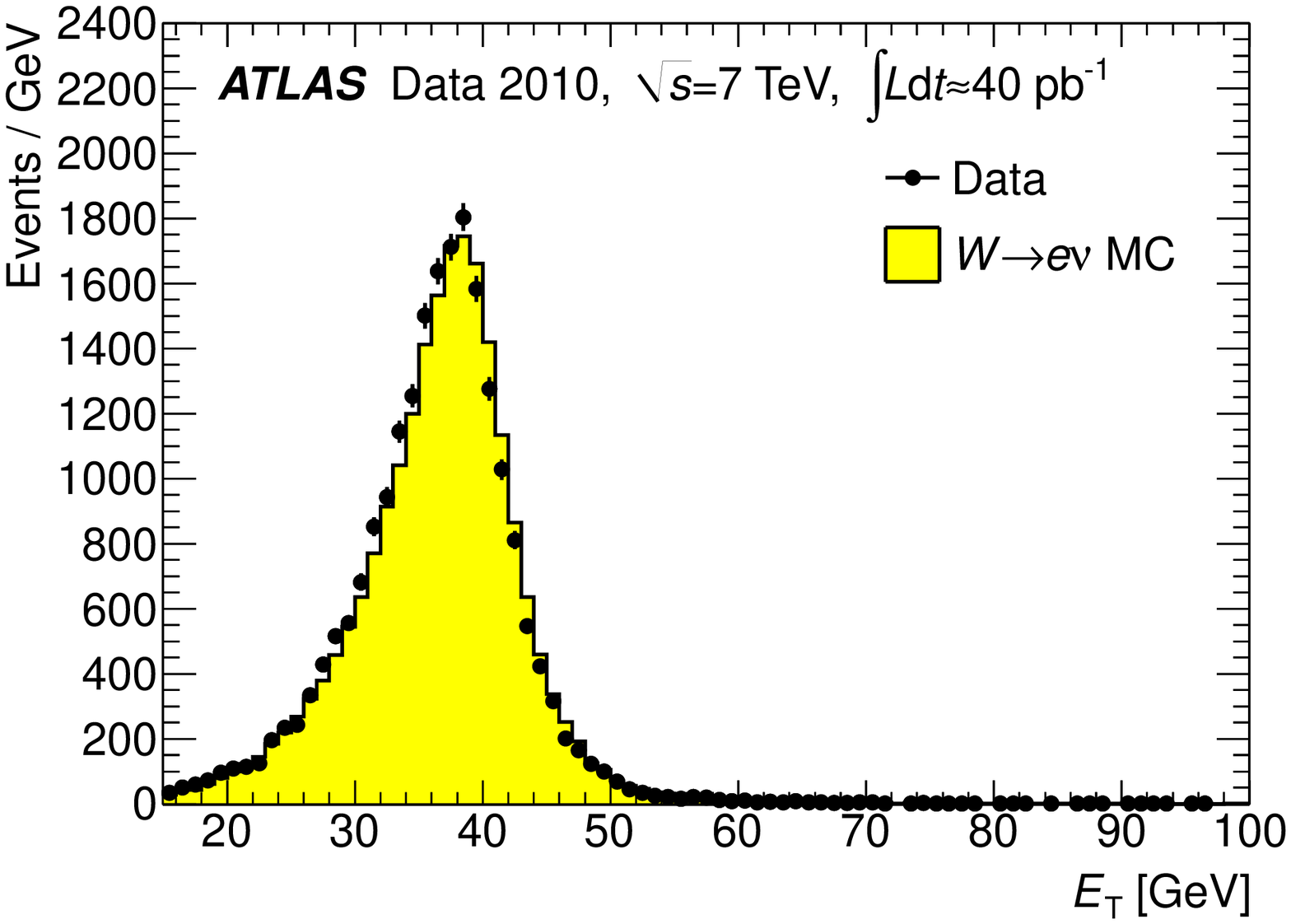}
\includegraphics[width=0.49\textwidth]{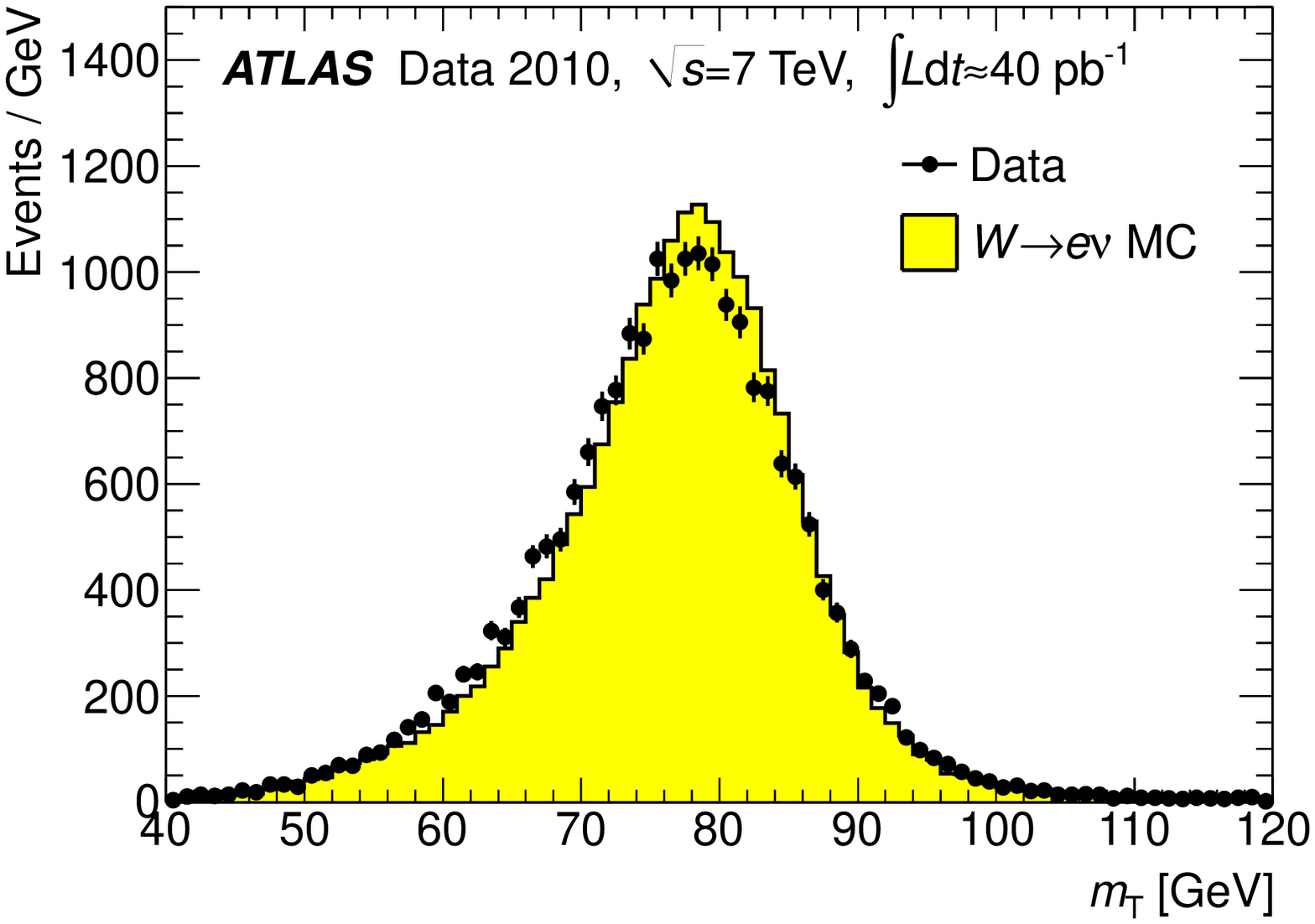}
\end{center}
\caption{Transverse energy spectra, compared between data and MC, for the selected
electron probes passing \tight\ identification cuts for the (top left) \Zee, (top right) 
\Jee, and (bottom left) \Wen\ channels, together with 
(bottom right) the transverse mass distribution for the \Wen\ channel.
The data points are plotted as full circles with statistical error bars, 
and the MC prediction, normalised to the number of data entries,
as a filled histogram. 
}
\label{WZJpsi_kinematics}
\end{figure*}

The same procedure is applied to the MC~simulation, with in addition a 
reweighting of the MC to reproduce the pile-up observed in data as well as 
the proper mixture of the various triggers. Figure~\ref{WZJpsi_kinematics} shows the
transverse energy distributions of the probes for each of
the three channels and, for completeness since the \Wen\ channel relies on an
orthogonal trigger based on~\MET, the transverse mass distribution for the \Wen\
selected probes. In order to compare these distributions to those expected from a
signal~MC, \tight\ identification cuts have been applied to the probes resulting
in very high purity in the case of the \Wen\ and \Zee\ channels. In the case of the
\Jee\ channel however, some background remains even at this stage, as can be seen
from the excess of probes in data compared to~MC at low~\ET. The small differences seen 
between data and MC distributions in the \Wen\ measurement arise primarily 
from the imperfections
of the modelling of the \MET-triggers in simulation.

\subsubsection{Background subtraction}

The next step in the analysis is to use a discriminating variable to estimate 
the signal and background contributions in each
\ET~or~$\eta$~bin. This variable should
ideally be uncorrelated to the electron identification variables.

\paragraph{Dielectron mass for the \Zee\ and \Jee\ channels}

The reconstructed dielectron mass is the most efficient discriminating
variable to estimate the signal and background contributions in 
the selected sample of electron probes from \Zee\ and \Jee\ decays. 
The signal integration ranges, 
typically $80<\mee<100$~GeV for the \Zee\ channel
and $2.8<\mee<3.2$~GeV for the \Jee\ channel, were
chosen to balance the possible bias of the
efficiency measurement and the systematic uncertainty on the background
subtraction. 

\begin{figure*}
\begin{center}                               
\includegraphics[width=0.49\textwidth]{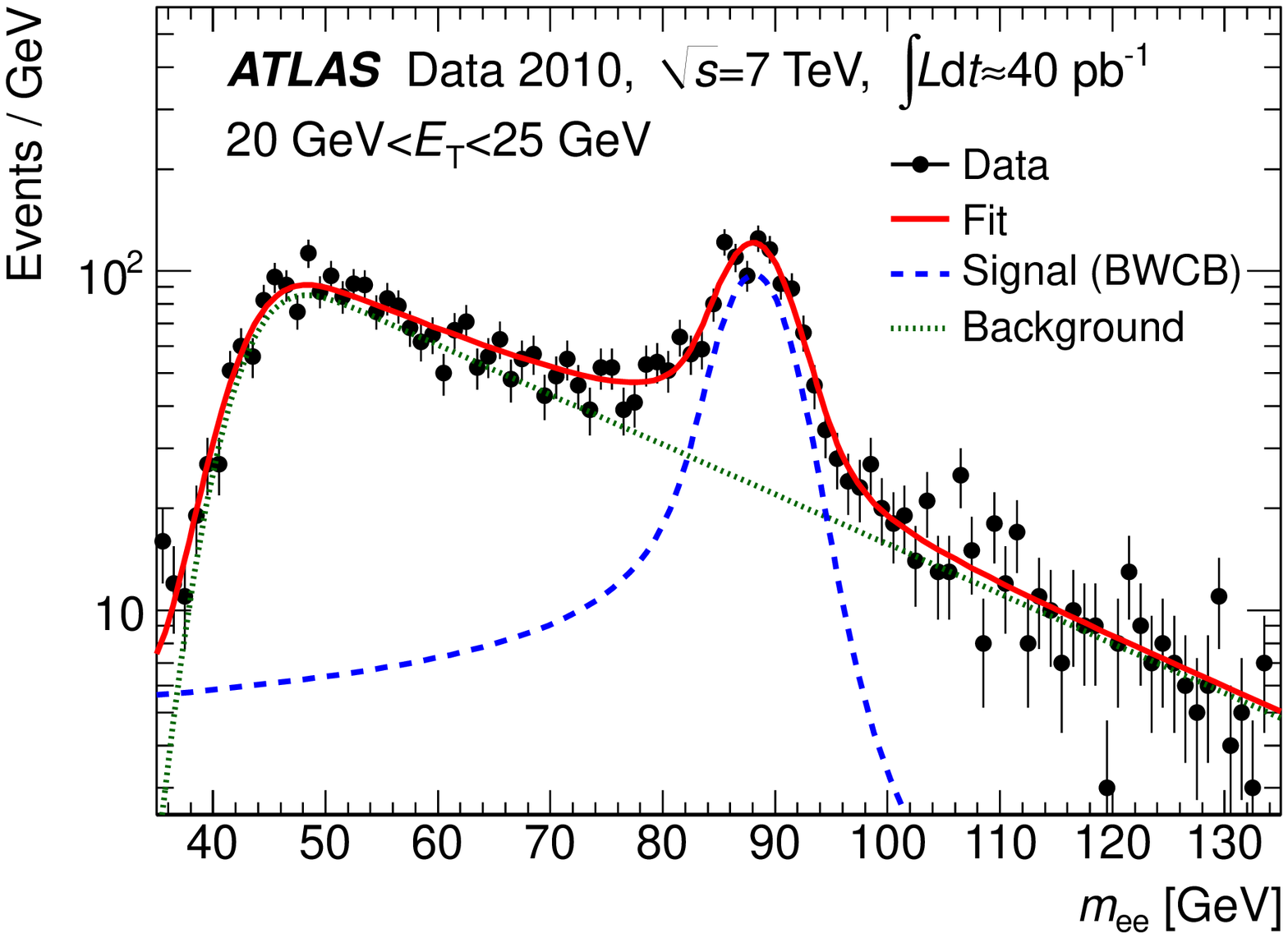}
\includegraphics[width=0.49\textwidth]{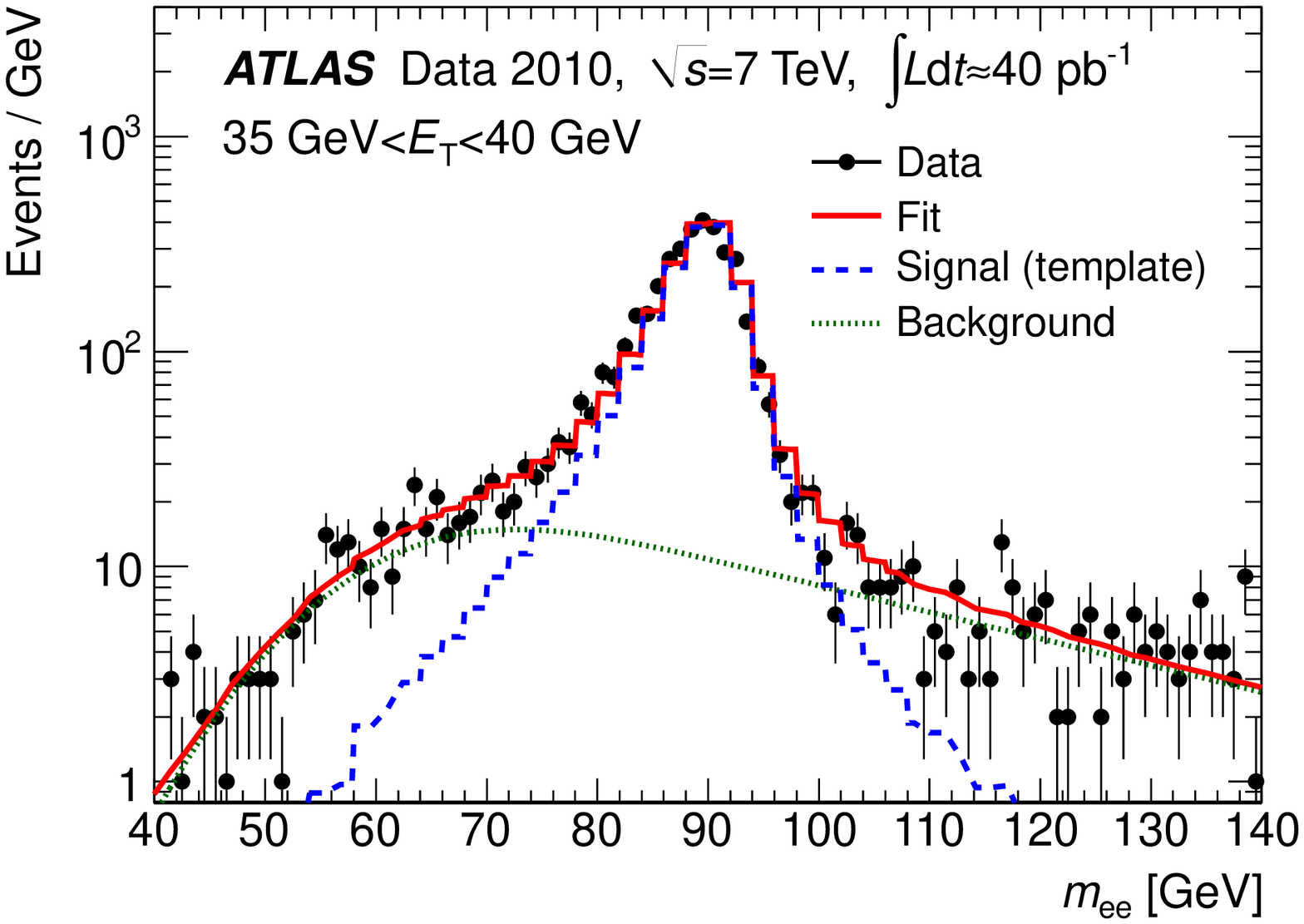}
\end{center}
\caption{The distributions of the dielectron invariant mass of \Zee\ candidate events, 
before applying electron identification cuts on the probe electron,
in the \ET-range (left) $20-25$~GeV and (right) $35-40$~GeV.
The data distribution (full circles with statistical error bars) is fitted with the sum (full line) of 
a signal component (dashed line) modelled by a Breit-Wigner convolved with a Crystal Ball 
function (BWCB) on the left or by a MC template on the right,  
and a background component (dotted line) chosen here as an exponential 
decay function convolved with a Gaussian.}
\label{Zee_massfits}
\end{figure*}

\begin{figure*}
\begin{center}                               
\includegraphics[width=0.49\textwidth]{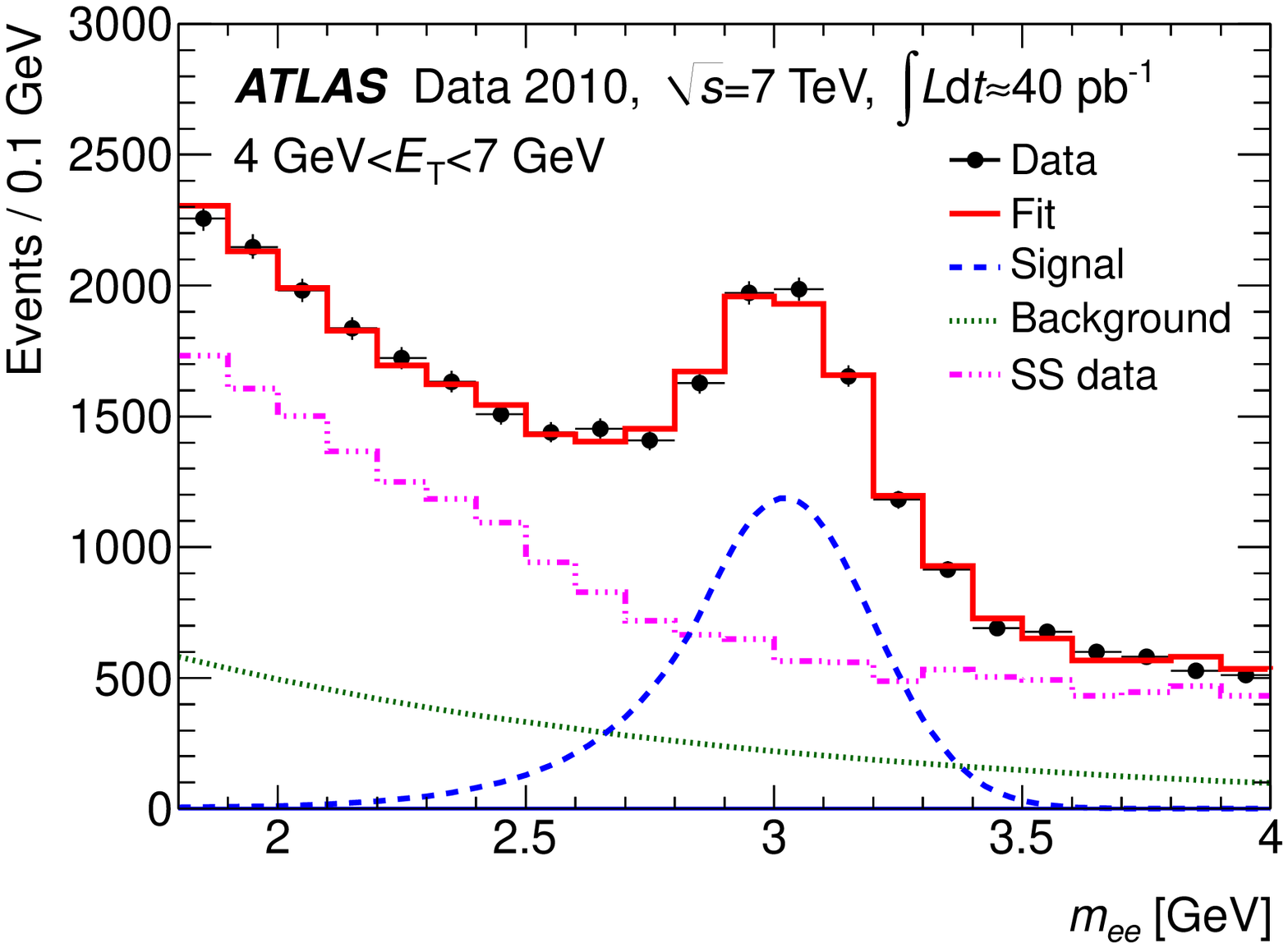}
\includegraphics[width=0.49\textwidth]{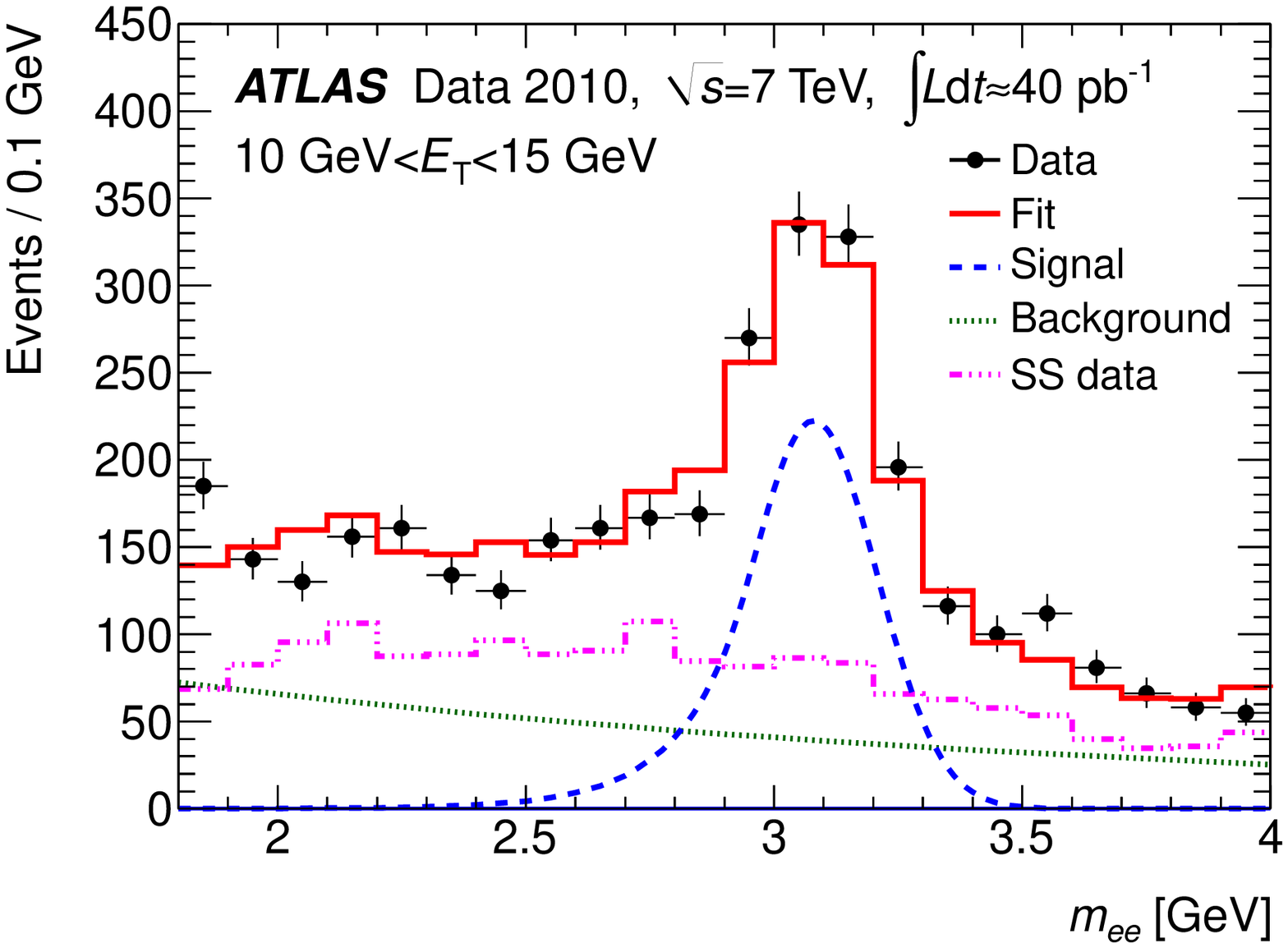}
\end{center}
\caption{The distributions of the dielectron invariant mass of \Jee\ candidate events, 
before applying electron identification cuts on the probe electron,
in the \ET-range (left) $4-7$~GeV and (right) $10-15$~GeV.
The data distribution (full circles with statistical error bars) is fitted with the sum (full
line) of 
a signal component (dashed line) described by a Crystal Ball function 
and two background components, one taken from same-sign pairs
in the data (dash-dotted line) and 
the remaining background modelled by an exponential function 
(dotted line).} 
\label{Jpsiee_massfits}
\end{figure*}

In the \Zee\ channel, which has more events and lower background
contamination, the efficiency measurements in $\eta$-bins (for
transverse energies $20<\ET<50$~GeV) were performed with a simple same-sign background subtraction. 
For both channels, the shape of the background under the dielectron mass peak depends
strongly on the \ET-bin due to kinematic threshold effects. 
Therefore for the measurements in \ET-bins (integrated over
$|\eta| < 2.47$ and excluding the overlap region $1.37 < |\eta| < 1.52$), 
the background subtraction is performed as follows.
\begin{itemize}

\item In the \Zee\ channel, a two-component fit with a signal
contribution plus a background contribution is performed in each bin to 
the \mee\ distribution over typical fit mass ranges of $40 < \mee < 160$~GeV. 
The signal contribution is modelled either by a Breit-Wigner distribution convolved with a
parametrisation of the low-mass tail, arising mostly from material effects, by a Crystal Ball 
function, or by a template obtained from \Zee\ MC simulation.
For the background contribution a variety of fit
functions were considered. In the \Zee\ measurement, an exponential and a
single-sided exponential convolved with a Gaussian are used. 

\item In the case of the \Jee\ selection, where the background contamination is highest, 
the amount and shape of the background vary significantly with the
\ET~of the probe, and depend strongly on the selection criteria applied to the probe.
Therefore, the fit described above, and
applied typically over $1.8 < \mee < 4$~GeV, contains a third component, 
which is based on the spectrum of same-sign
pairs in the data. Use of the same-sign sample has the advantage that it describes the shape
of a large fraction of the background (random combinations of fake or real
electrons), in particular in the signal region. The remaining background is
modelled on each side of the signal region by an exponential, a Landau
function or a Chebyshev polynomial.
\end{itemize}
Examples of the fit results are shown in~Figure~\ref{Zee_massfits} for the \Zee\
and in Figure~\ref{Jpsiee_massfits} for the \Jee\ measurement.

\paragraph{Calorimeter isolation for the \Wen\ channel}

The \Wen\  sample is selected with very stringent \MET\
requirements. There is only a limited choice of observables
to discriminate the isolated electron
signal from the residual background from jets. 
One suitable observable, which is nevertheless slightly correlated
with some of the electron identification variables, is the energy isolation measured in
the calorimeter. This isolation variable, denoted hereafter \Isol, 
is computed
over a cone of half-angle $\Delta R = 0.4$ as follows. The transverse energies of all EM~and
hadronic calorimeter cells are summed except for those which are in the~$5\times 7$
EM~calorimeter cells in $\Delta \eta \times \Delta \phi$ space around the cluster
barycentre. 
This sum is normalised to the transverse energy of the EM~cluster to yield \Isol. For
isolated electrons, the \Isol\ distribution is expected to peak at values close to zero, with
a width determined by the combination of electronic noise, shower leakage, 
underlying event and
pile-up contributions. For the background from jets, a much wider distribution is
expected reaching values well beyond unity. The signal region is defined by
requiring that the calorimeter isolation be below a certain threshold,
typically~$0.4$. 
The residual background in the signal region is estimated 
using template distributions derived from data by requiring that
the electron probes fail certain electron identification cuts. 
The obtained templates are normalized in the background region, 
above the chosen isolation threshold, to the number of selected 
electron probes.

\begin{figure*}
\begin{center}                               
\includegraphics[width=0.49\textwidth]{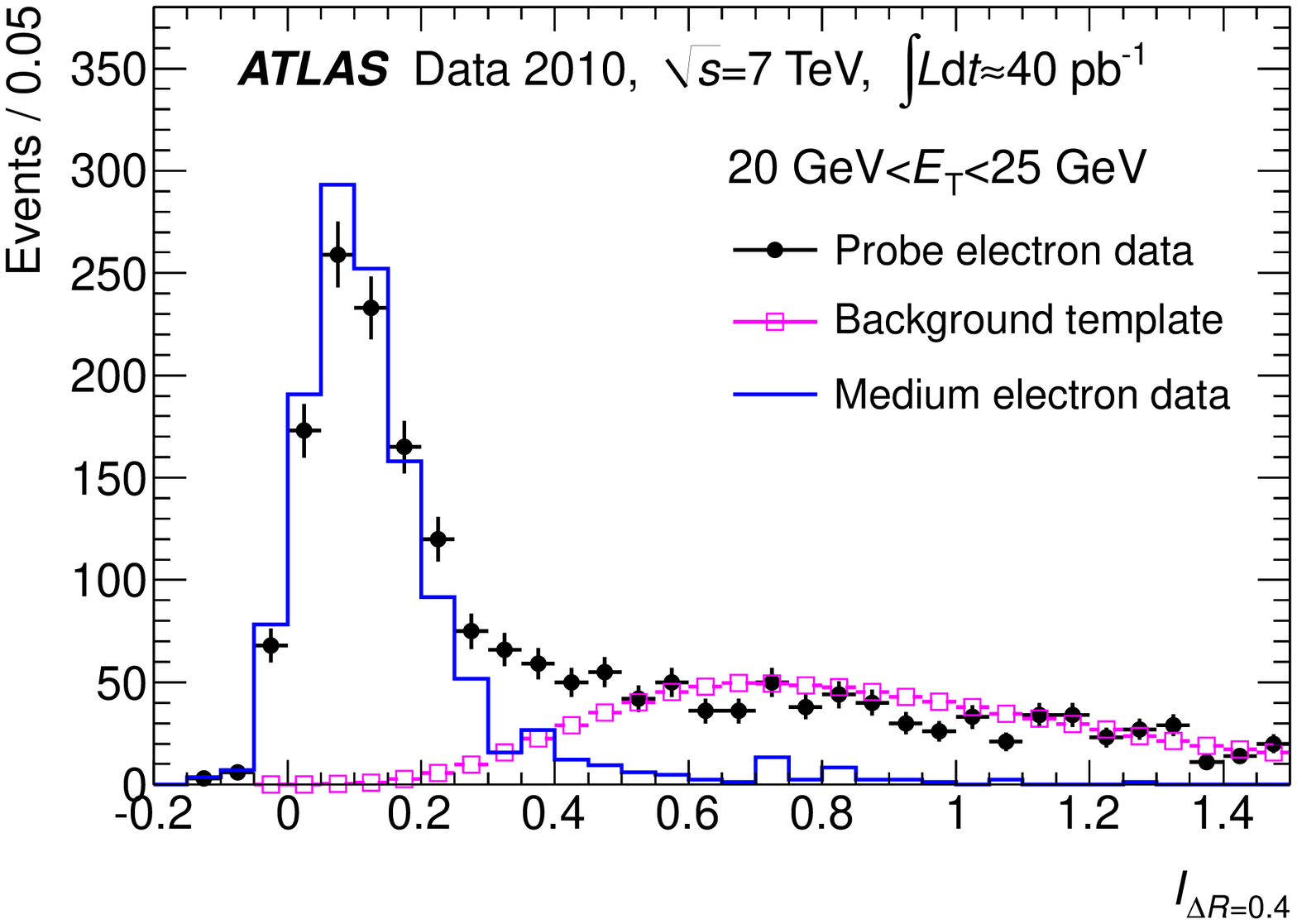}
\includegraphics[width=0.49\textwidth]{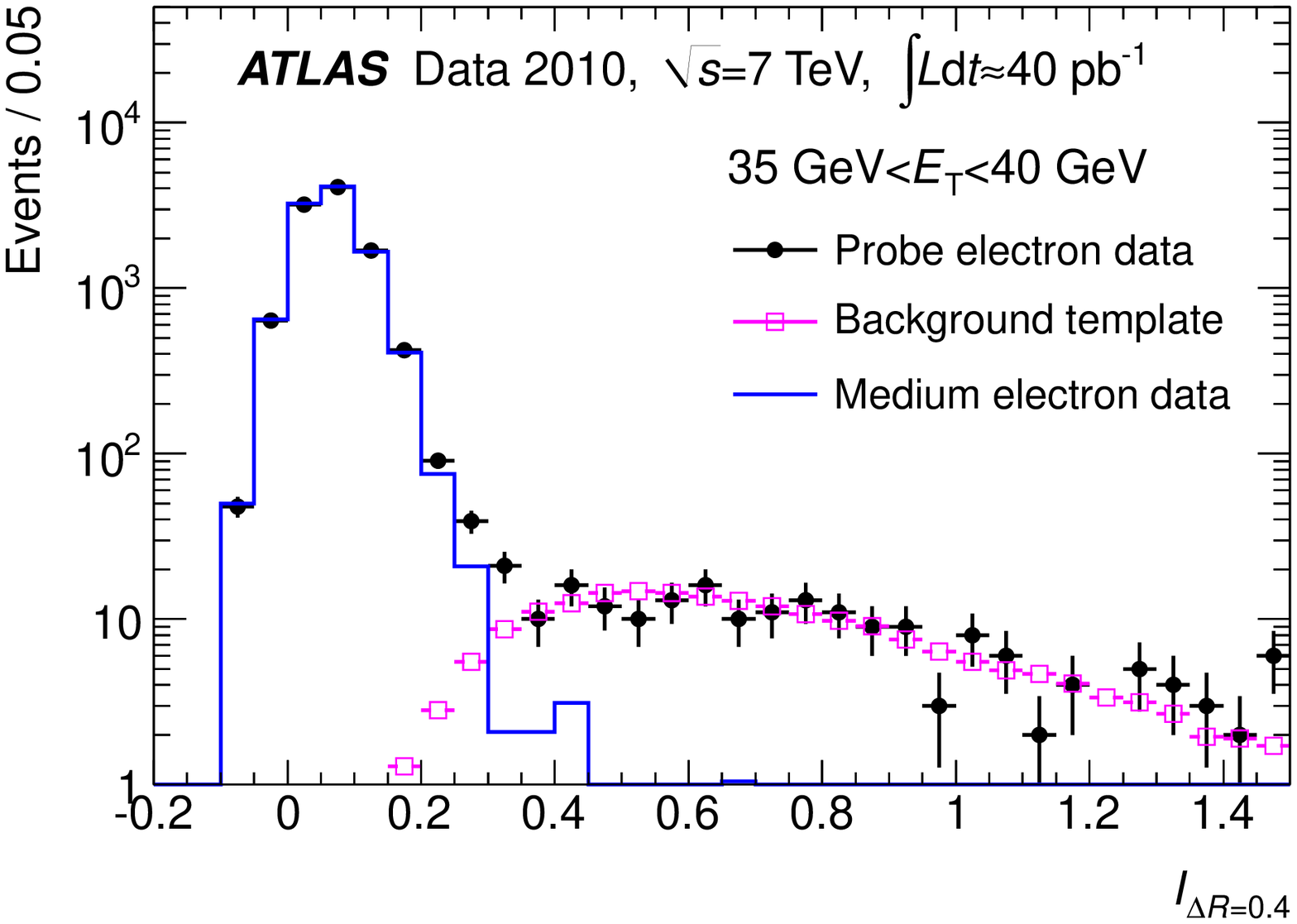}
\end{center}
\caption{The distributions of the calorimeter isolation variable, 
\Isol\ for the \Wen\ data sample for (left) 
$20 < \ET < 25$~GeV and (right)  $35 < \ET < 40$~GeV. 
The full circles with statistical error bars correspond to the probe electrons
before applying any identification cuts.
The open squares show the corresponding 
background template, derived from data,
normalised to the probe electron data in the region $\Isol>0.4$.  
To illustrate the expected shape of the \Wen\ signal,
the distributions obtained for electron probes passing the \medium\
identification cuts and 
normalised to the calculated \Wen\ signal are shown by full histograms.}
\label{Fig:simpleSubtract_1}
\end{figure*}

Figure~\ref{Fig:simpleSubtract_1} shows the \Isol\ distribution 
for the data in
two regions of phase space: a low-\ET\ bin, $20 < \ET < 25$~GeV, where the 
background contribution is high, and the \ET~bin, $35 < \ET < 40$~GeV, which has
the largest fraction of the signal statistics and a very high
signal-to-background ratio.

\paragraph{Samples obtained after background subtraction}

Once the background subtraction procedure has been well defined,
the next step in the process of measuring the
efficiencies of the electron identification criteria (relative to electron
reconstruction with additional track silicon hit requirements, as described above) 
is to define the total
numbers of signal probes before and after applying the identification cuts, together
with their statistical and systematic uncertainties.  
The ratios of these two numbers
in each \ET-bin or $\eta$-bin are the efficiencies measured in data.

\begin{table*}
\caption{Numbers of signal and background probes and signal-over-background
ratios~($S/B$), in different \ET~ranges, for the \Wen, \Zee, and \Jee\ channels. The
errors are statistical only.} 
\label{SignalProbeStats}
\begin{center}
{\footnotesize
\begin{tabular}{|l||c|c|c||c|c||c|c|}
\hline
& \multicolumn{3}{|c||}{ \Wen\ } & \multicolumn{2}{|c||} { \Zee\ } 
                       & \multicolumn{2}{|c|} { \Jee\ } \\
\hline
    \ET\ [GeV]  & $15-20$ & $20-25$ & $35-40$ & $20-25$ & $40-45$ & $4-7$  & $15-20$  \\
\hline\hline
Signal&$455\pm 20$&$1040\pm 30$&$10090\pm 100$&$870\pm 40$&$3710\pm 60$&$3900\pm 90$&$155\pm 15$ \\
\hline
Background&$60\pm 10$&$140\pm 20$&$35\pm 6$&$460\pm 20$&$160\pm 20$&$3330\pm 190$&$120\pm 20$ \\
\hline
$S/B$&$7.3\pm 1.0$&$7.3\pm 1.1$&$290\pm 50$&$1.9\pm 0.1$&$24\pm 3$&$1.2\pm 0.1$&$1.3\pm 0.3$ \\
\hline
\end{tabular}
}
\end{center}
\end{table*}

Table~\ref{SignalProbeStats} shows several examples of the numbers of signal and
background probes and of the corresponding signal-to-background ratios~($S/B$) for
the three channels and for selected \ET-bins. The $S/B$~ratios were found to be
fairly uniform as a function of~$\eta$ for a given channel and \ET-bin. In contrast,
as expected, the $S/B$~ratios improve considerably for high-\ET\ electrons from
\Wen\ and \Zee\ decay. The $S/B$~ratios for the \Wen\ channel are
considerably higher than for the \Zee\ channel partly due to the higher \Wboson\
cross-section and partly because of the more stringent kinematic
cuts applied to the neutrino tag (high \MET\ and \MET\ isolation) than to the electron
tag in the \Zee\ case. Such stringent kinematic cuts were not applied
to the \Zee\ channel because of limited statistics. At the much lower
\ET-values covered by the \Jee\ channel, the $S/B$~ratios are of order unity before
applying any electron identification cuts and therefore the systematic uncertainties
from the background subtraction procedure will be larger than for the \Wen\
and \Zee\ channels, as shown in~Subsection~\ref{systID}.

\subsubsection{Systematic uncertainties}
\label{systID}

The dominant systematic uncertainties on the efficiency measurements described above
are linked to the background subtraction from the probe samples, especially
before applying the electron identification cuts. The background level under the
signal was varied substantially to verify the stability of the
background subtraction procedure, mostly by varying the cuts applied to the tag
component of the event. Furthermore, the background subtraction method itself
was also varied.
The following sources of systematic uncertainties were considered:
\begin{itemize}
 
 \item{\bf Background level}
 The tag requirements 
 (such as the electron identification level, \medium\ or \tight,  for
 \Zee\ and \Jee, and the \MET\ and electron isolation, for \Wen\ and \Zee,
 respectively), were varied to
 induce variations of the background level under the signal.

 \item{\bf Discriminating variable used in the background estimation}
 Several analysis choices were varied to estimate the uncertainty due to the discriminating variable chosen
 (calorimeter isolation for \Wen\ and invariant mass for \Zee\ and \Jee):
 the size of the signal window; 
 the definition of the side-band region 
 used for background subtraction for the
 $\eta$-dependent efficiencies in the \Zee\  channel;
 the signal and background models (functions or templates) 
 used in the fits 
 for the \ET-dependent efficiencies in
 the \Zee\  and \Jee\ channels;
 the definition of the isolation variable and the normalization region for the background
 template distributions for the \Wen\ channel.

\item{\bf Possible bias related to the method of the background subtraction}
The possible bias from the correlations between the discriminating variable and the
efficiencies themselves in the case of calorimeter isolation for the \Wen\ channel was studied
by changing the selection used when producing
the background isolation template (trigger stream, selection cuts).

Also, wherever feasible the possible bias of the efficiency extraction method 
(in particular the background subtraction) was also studied 
by repeating the measurements on simulated data and comparing the results to the MC
truth. Typically, these {\it closure tests} were performed by 
mixing a high-statistics simulated signal sample and a background contribution
with the background shape taken from a control region in data. 
The signal-to-background ratios were estimated from data and varied within
reasonable limits. 
Any observed bias (defined as the difference of the measured and the true MC value 
in the test) was taken as an additional systematic uncertainty.

\end{itemize}

All combinations of the above variations were used to extract the efficiency, 
yielding about a hundred distinct measurements for each channel and for each
kinematic bin.
Given
the complexity of the background subtraction procedure and the variety of kinematic
configurations studied, no single preferred method for background subtraction could
be defined. The central value of the measured efficiency was therefore defined as
the mean of the distribution of all the efficiency values obtained through these
variations and the systematic uncertainty was defined as the root mean square of the
distributions. The statistical error is the mean of the statistical errors 
of all measurements corresponding to these analysis configurations.

Other potential sources of uncertainty were also checked but
led to negligible contributions to the overall systematic uncertainty
on the measurements:
\begin{itemize}

 \item the impact of the energy-scale corrections discussed
  in~Subsection~\ref{sec:Calibration} of this paper;

 \item the charge-dependence of the efficiencies in the \Wen\ measurement;
 
 \item the time-dependence of the efficiencies in the \Wen\ measurement;
  
 \item the size of the dead regions in the EM~calorimeter;
 
 \item the amount of pile-up considered in the simulation.
\end{itemize}
 
When comparing the measured efficiencies with MC predictions,
uncertainties related to the composition of the \TandP\ sample 
potentially also need to be considered. 
In the case of the \Jee\ channel, the uncertainties on the 
{\bf fraction of non-prompt \Jpsi\ decays}~\cite{JpsimmPaper} 
in the probe sample, which depend both on the
kinematic bin and on the trigger conditions, are important.
The uncertainties linked to the trigger, reconstruction and identification
efficiencies of the non-prompt contribution are included.
The effect of the modelling of the 
{\bf mixture of triggers} used in the \Wen\ and \Jee\ channels
was also studied. It is negligible in the \Wen\ case.

Table~\ref{SystIDeff} illustrates the main components of the measurement uncertainties 
on the efficiency of the \tight\ electron identification cuts for a few typical \ET-bins 
and for each channel. These uncertainties are somewhat larger than those for the \medium\ cuts. 
The total uncertainties are computed as the quadratic sum of the
statistical and the total systematic uncertainties.
In the \Zee\ and \Jee\ measurements, the total systematic uncertainty is obtained by adding linearly the closure test biases
to the quadratic sum of all other components. 

\begin{table*}
\caption{Relative uncertainties (in~\%) on the measured efficiencies of the \tight\ electron
identification for \Wen, \Zee, and \Jee\ decays for a few typical \ET~bins (integrated over the full
$\eta$-range). 
For the \Jee\ channel, the uncertainties affecting the MC prediction for the efficiency are also given.}
\label{SystIDeff}
\begin{center}
\begin{tabular}{|l||c|c|c||c|c||c|c|}
\hline
& \multicolumn{3}{|c||}{ \Wen\ } & \multicolumn{2}{|c||} { \Zee\ } & \multicolumn{2}{|c|} { \Jee\ } \\
\hline
    \ET\ range (GeV)    & 15-20 & 20-25 & 35-40 & 20-25 & 40-45 & 4-7  & 15-20  \\
\hline\hline
Statistics              & 3.0 & 1.7 & 0.3 & 3.5 & 0.9 & 2.5 & 9.9 \\
\hline
Background level        & 1.2 & 1.3 & 0.3 & 4.4 & 0.9 & 2.2 & 3.1 \\
\hline
Discriminating variable & 4.8 & 1.9 & 0.3 & 3.3 & 1.5 & 4.9 & 9.6 \\
(nature, shape, range)  & & & & & & & \\
\hline
Possible bias of        & 3.7 & 0.6 & 0.1 & 1.7 & 1.8 & 3.6 & 3.1 \\
background subtraction  & & & & & & & \\
\hline\hline
Total                   & 7.1 & 3.1 & 0.5 & 8.0 & 3.6 & 9.3 & 16.5 \\
\hline
\hline
MC statistics           & & & & &                     & 0.2 & 0.8 \\
Non-prompt $J/\psi$     & & & & &                     & 5.2 & 7.7 \\
\hline
Trigger mixture         & & & & &                     & 5.1 & 2.4 \\
\hline\hline
MC total                & & & & &                     & 7.3 & 8.1 \\
\hline
\end{tabular}
\end{center}
\end{table*}

\subsubsection{Measured efficiencies}

\begin{figure*}
\begin{center}                               
\includegraphics[width=0.48\textwidth]{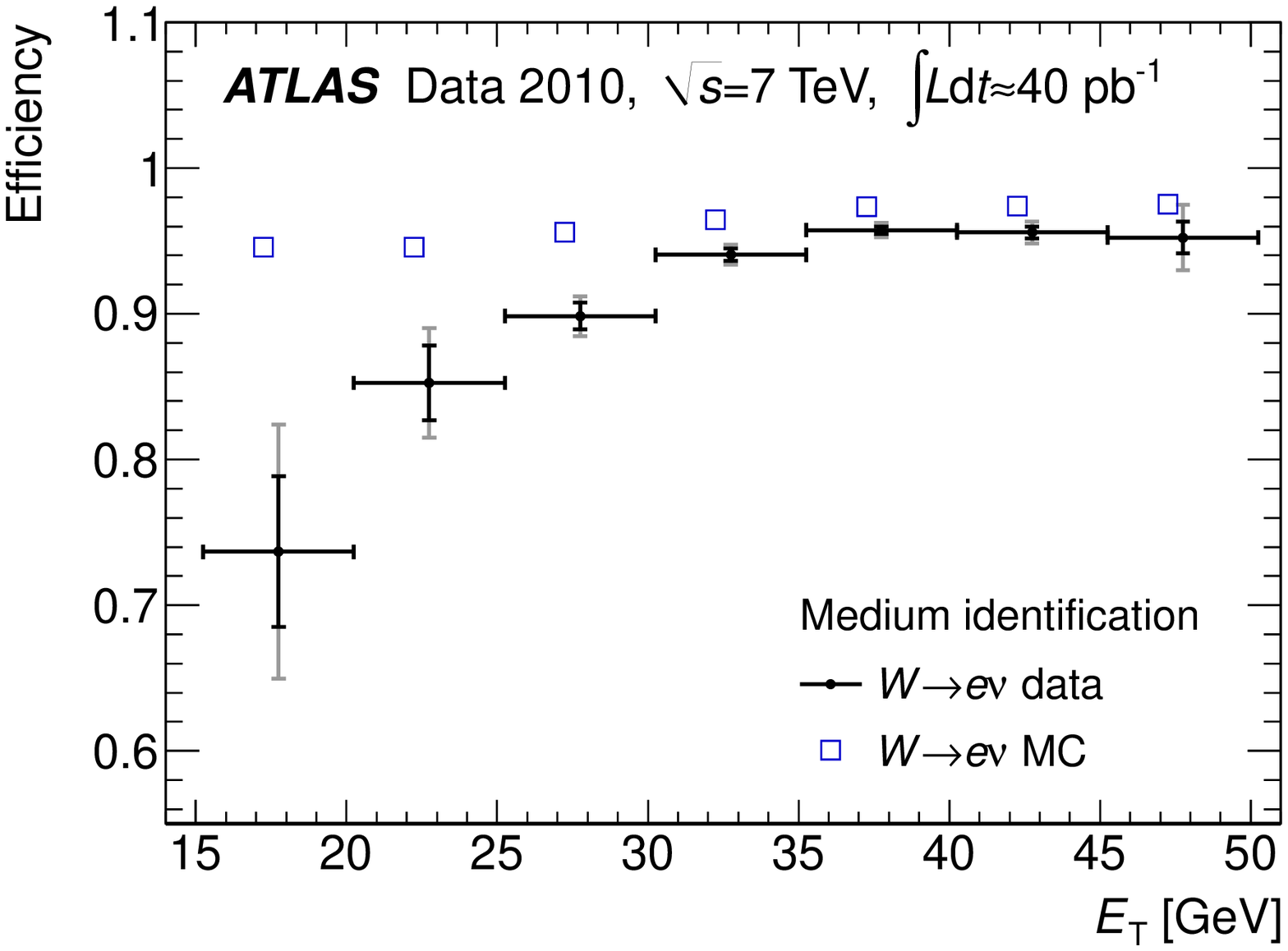}
\includegraphics[width=0.48\textwidth]{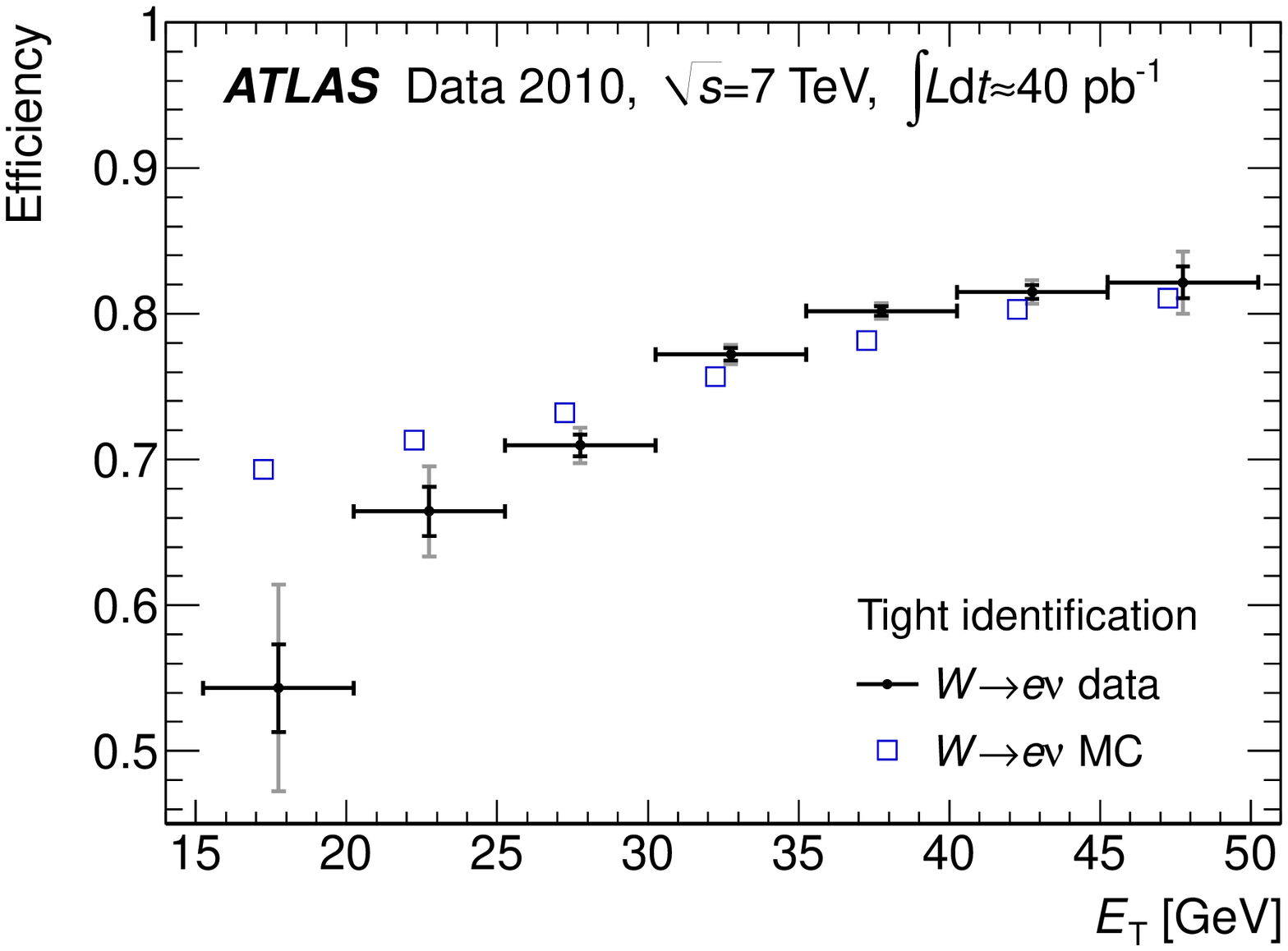}

\includegraphics[width=0.48\textwidth]{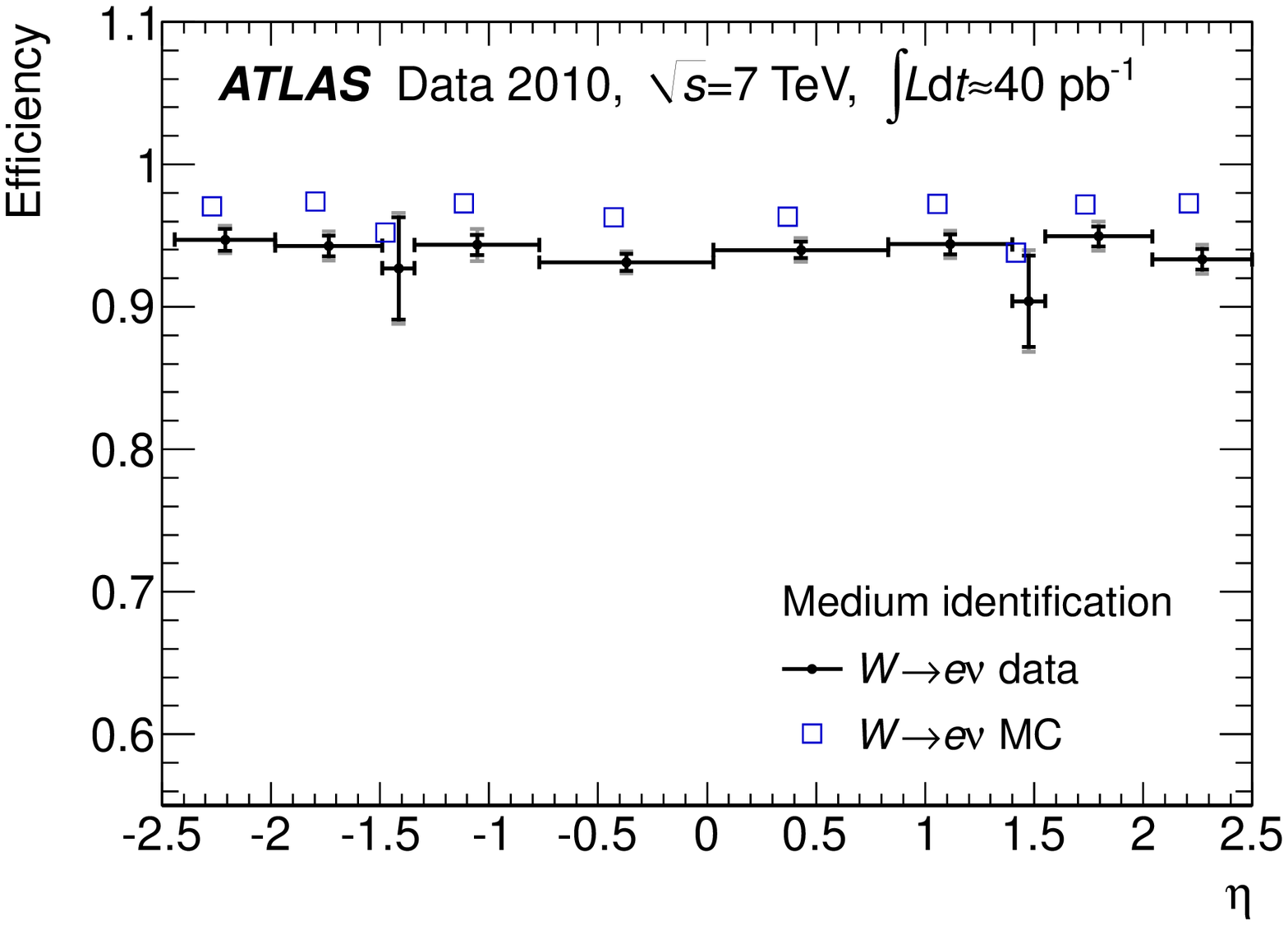}
\includegraphics[width=0.48\textwidth]{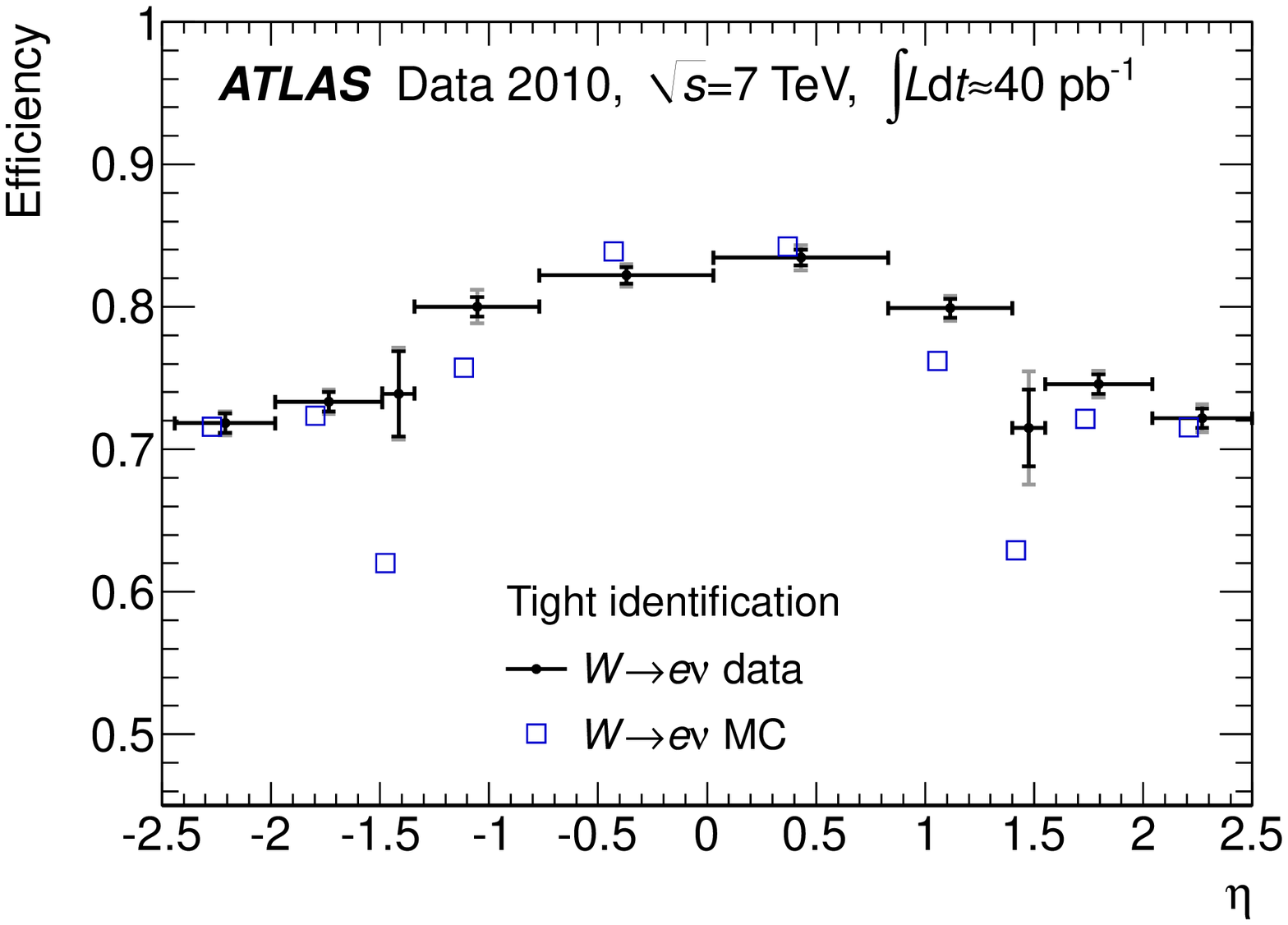}
\end{center}
\caption{Electron identification efficiencies measured from \Wen\ events and predicted by~MC for (left) \medium\ and (right) \tight\
identification as a function (top) of~\ET\ and integrated over~$|\eta|<2.47$ excluding the
transition region $1.37<|\eta|<1.52$ and (bottom) of~$\eta$ and integrated
over $20<\ET<50$~GeV. The results for the data are shown with their statistical (inner error bars)
and total (outer error bars) uncertainties. The statistical error on the MC efficiencies plotted as 
open squares is negligible. 
For clarity, the data and MC points are slightly displaced horizontally in opposite directions.}
\label{Weff}
\end{figure*}

\begin{figure*}
\begin{center}                               
\includegraphics[width=0.48\textwidth]{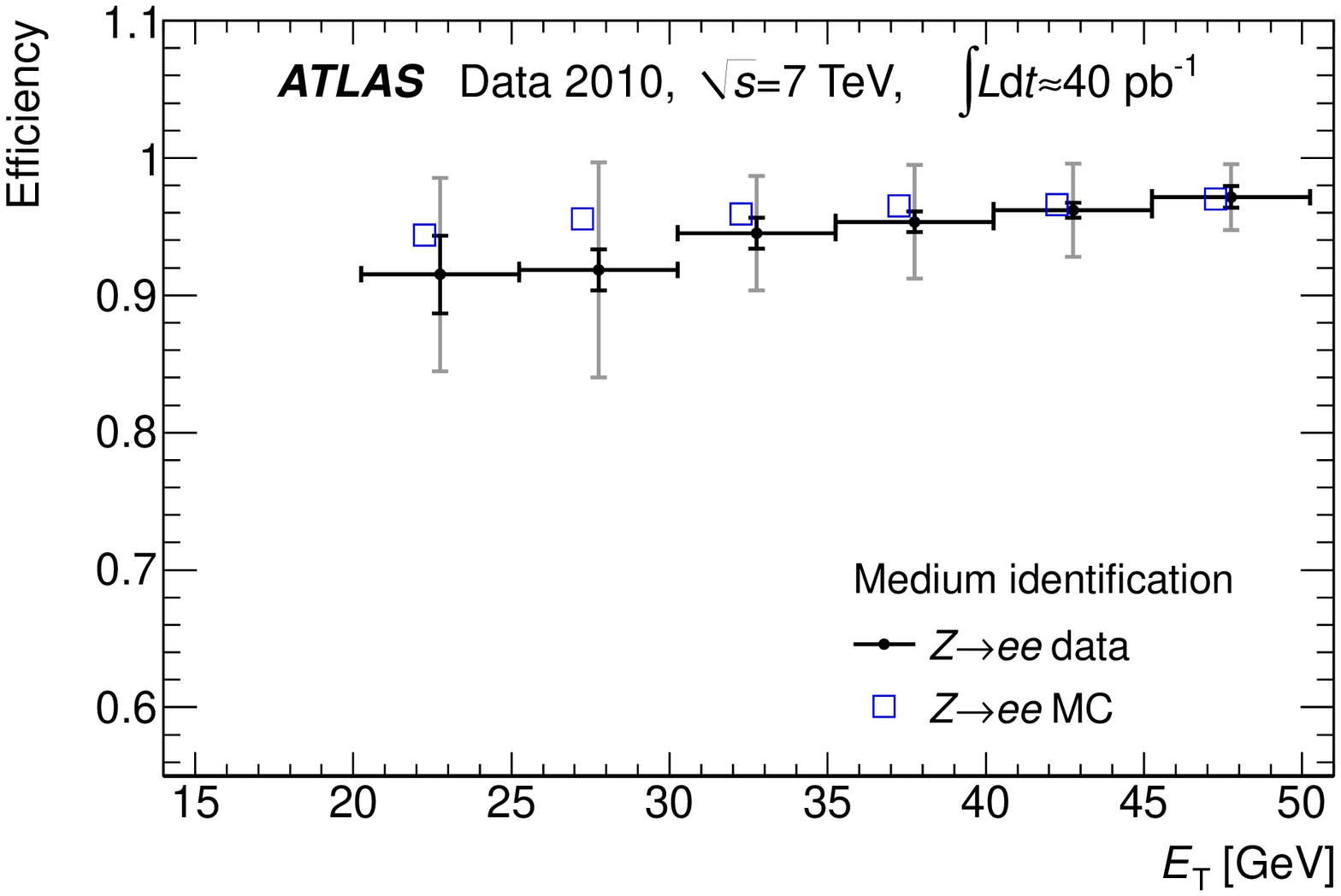}
\includegraphics[width=0.48\textwidth]{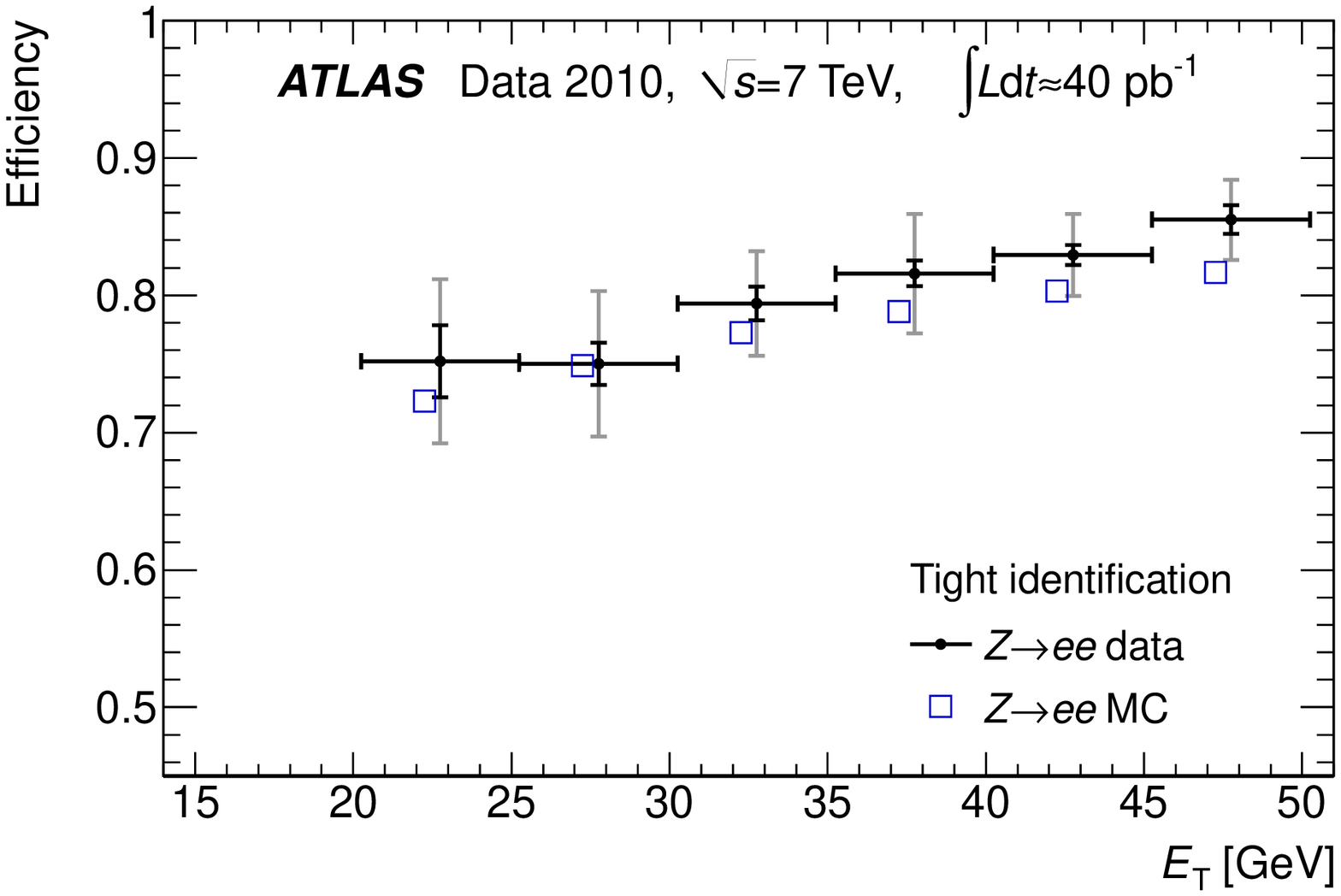}

\includegraphics[width=0.48\textwidth,angle=0]{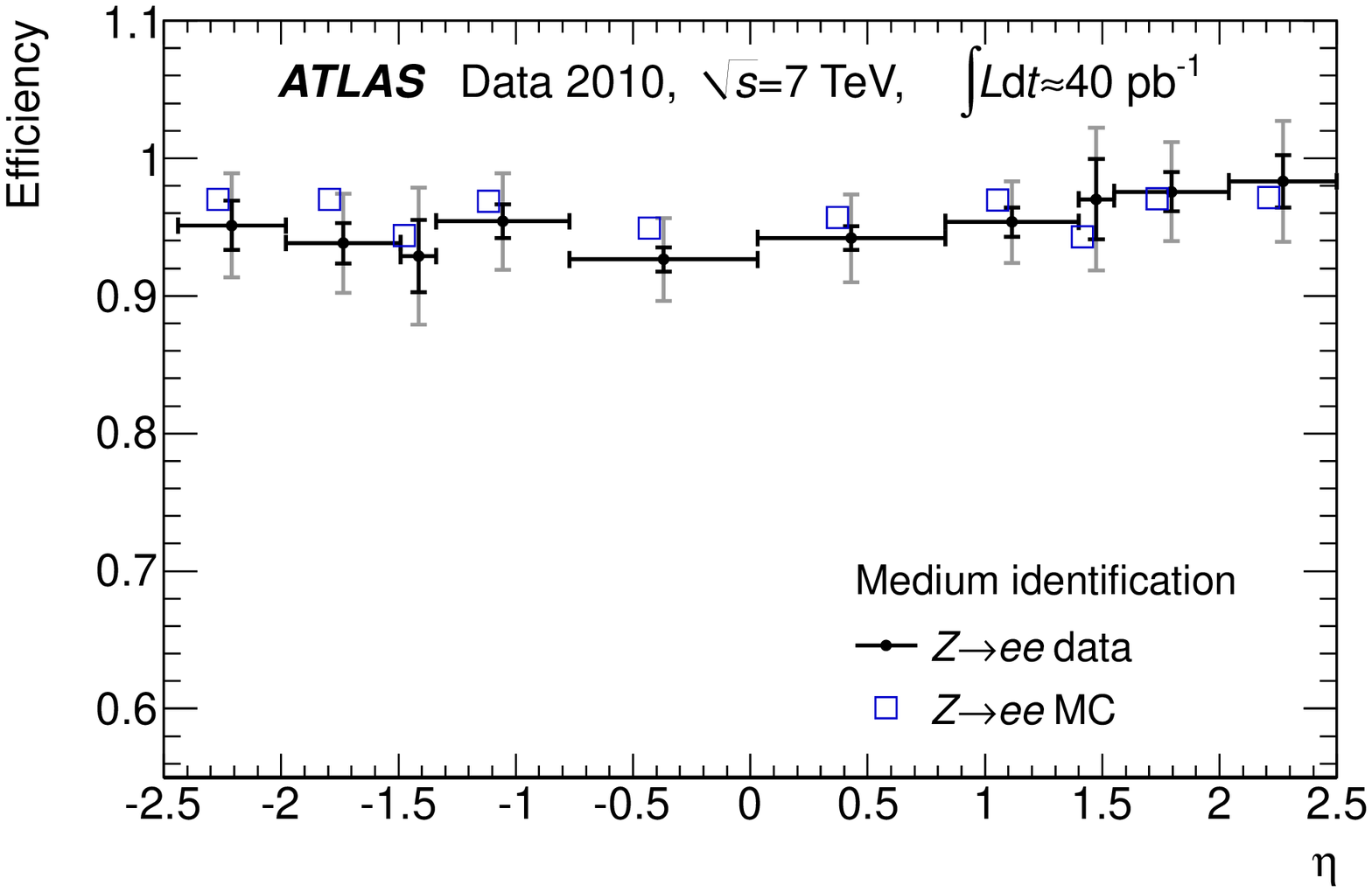}
\includegraphics[width=0.48\textwidth,angle=0]{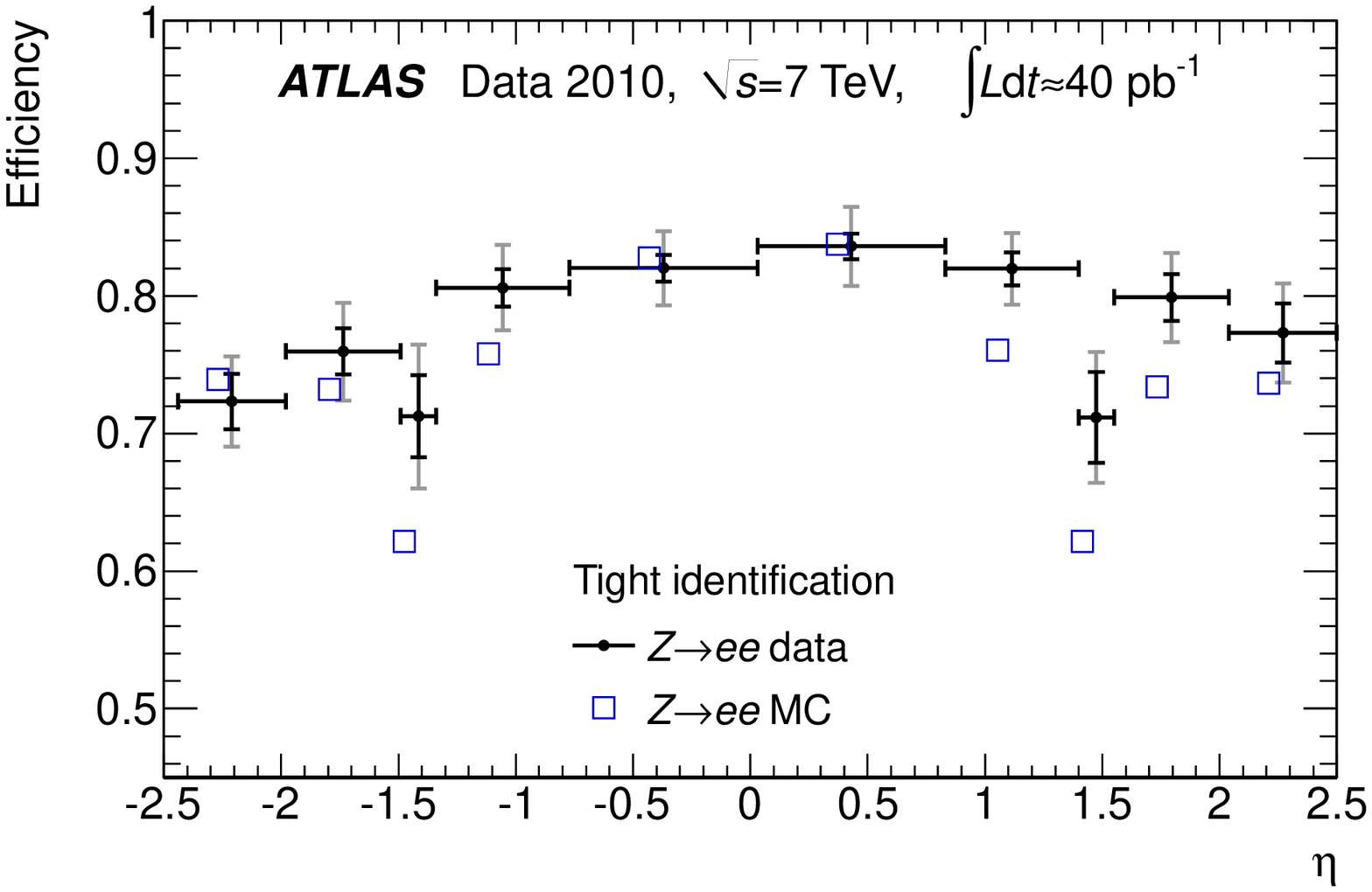}
\end{center}
\caption{Electron identification efficiencies measured from \Zee\ events and predicted by~MC for (left) \medium\ and (right) \tight\
identification as a function (top) of~\ET\ and integrated over $|\eta|<2.47$ excluding the
transition region $1.37<|\eta|<1.52$ and (bottom) of~$\eta$ and integrated
over $20<\ET<50$~GeV. The results for the data are shown with their statistical (inner error bars)
and total (outer error bars) uncertainties. The statistical error on the MC efficiencies plotted as
open squares is negligible. 
For clarity, the data and MC points are slightly displaced horizontally in opposite directions.}
\label{Zeff}
\end{figure*}

\begin{figure*}
\begin{center}                               
\includegraphics[width=0.49\textwidth]{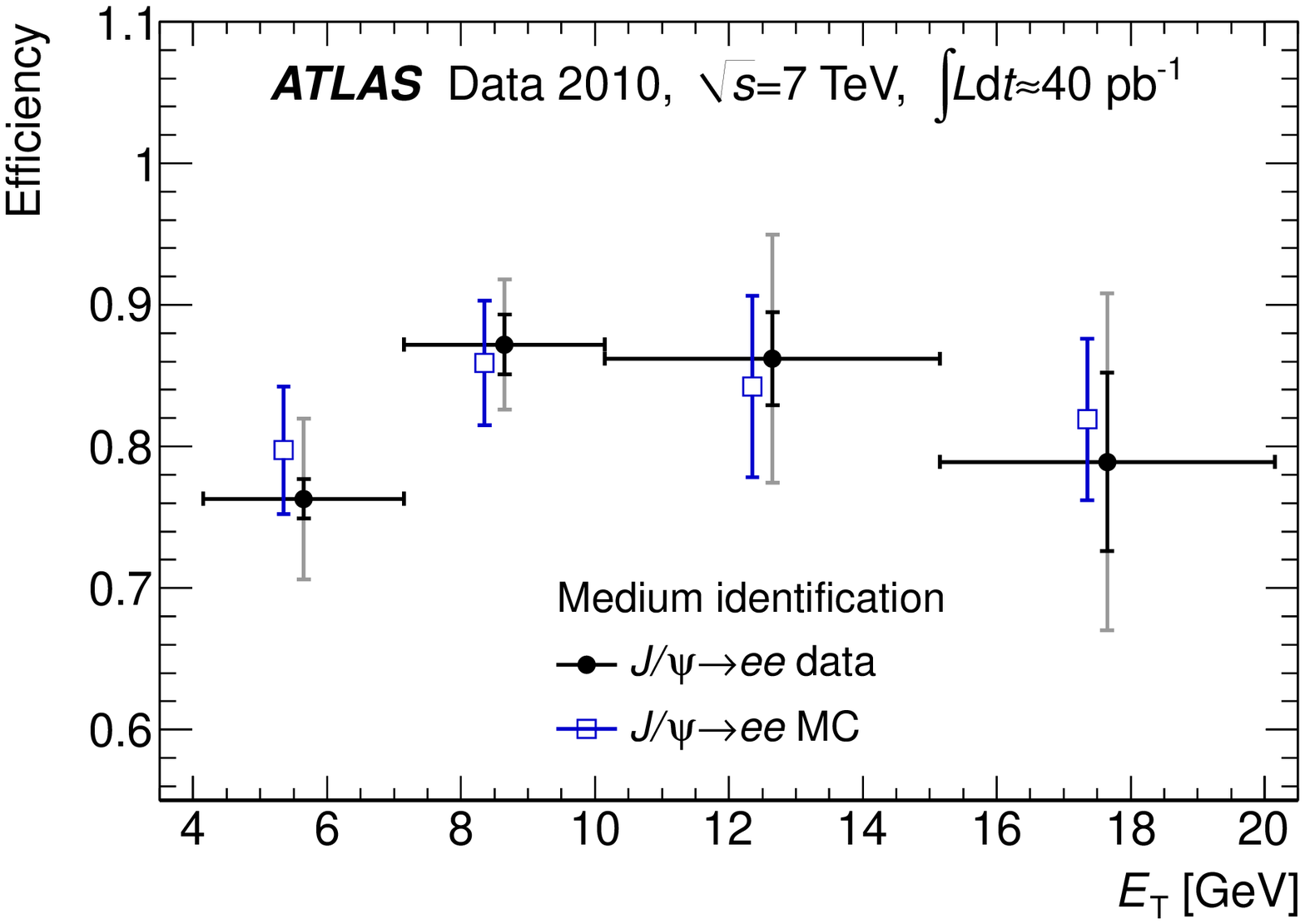}
\includegraphics[width=0.49\textwidth]{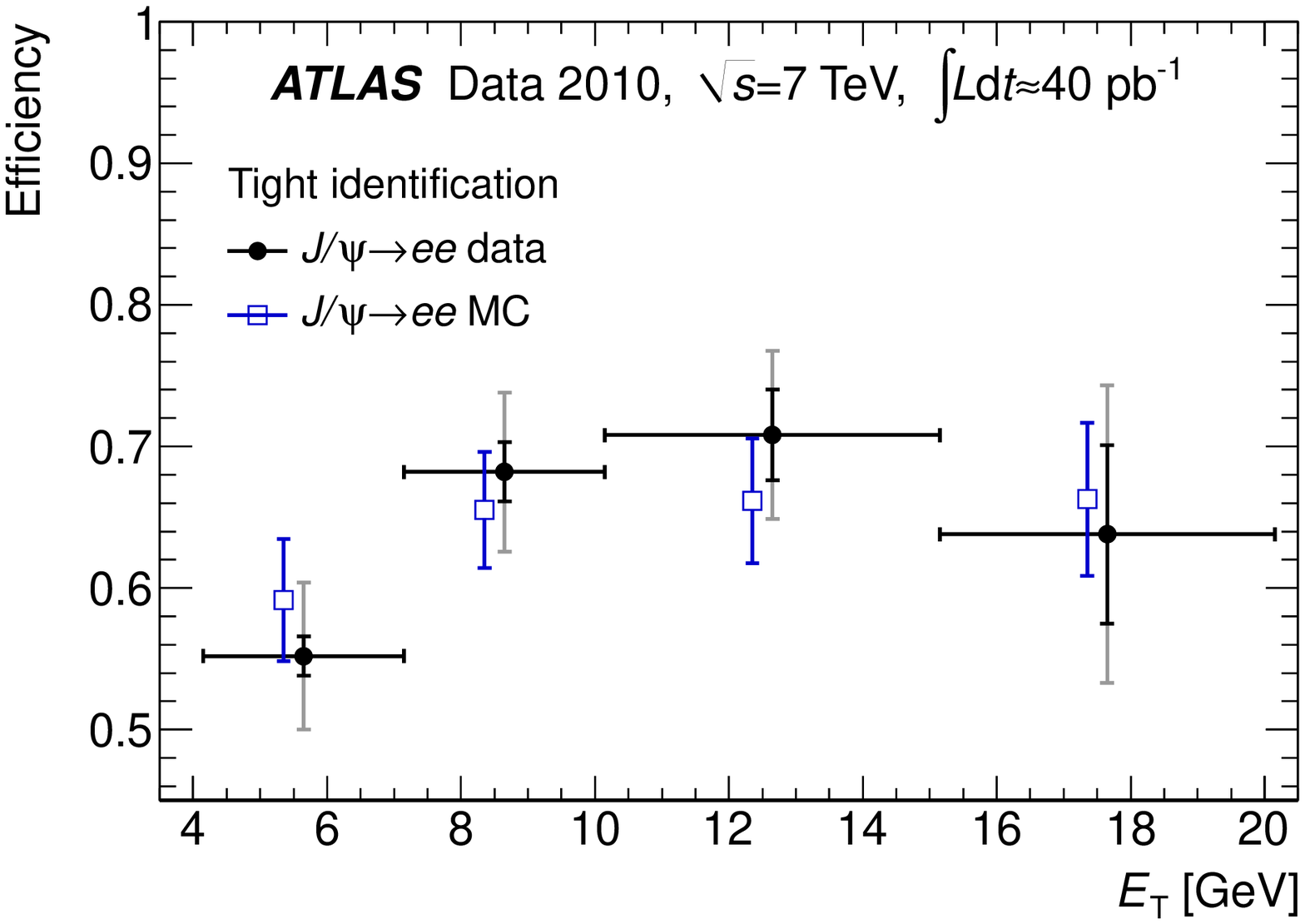}
\end{center}
\caption{Electron identification efficiencies measured from \Jee\ events and predicted by~MC for (left) \medium\ and (right) \tight\
identification as a function of \ET\ and integrated over $|\eta|<2.47$ excluding the
transition region $1.37<|\eta|<1.52$. The results for the data are shown with their statistical (inner error bars)
and total (outer error bars)
uncertainties. The MC~predictions are a weighted average of the efficiencies expected for prompt and non-prompt \Jpsi\
production as explained in the text. The total error on the MC efficiencies plotted as open squares is also shown. 
For clarity, the data and MC points are slightly displaced horizontally in opposite directions.}
\label{Jpsieff}
\end{figure*}

The efficiencies of the
\medium\ and \tight\ electron identification cuts 
as a function of~\ET\ and~$\eta$ are shown in 
Figures~\ref{Weff},~~\ref{Zeff} and~~\ref{Jpsieff},
respectively, for the \Wen, \Zee\ and \Jee\ channels. 
For the \Jee\ channel, only the
measurements in four bins of \ET\ are presented due to the limited
statistics, especially in the \endcap s. 
For the \Wen\ and \Zee\ channels, the measured efficiencies are compared
directly to those expected from the MC~simulations, whereas, for the \Jee\ channel,
the measured efficiencies are compared to a weighted average of the efficiencies
expected from prompt and non-prompt \Jpsi\ production. 
As the \tight\ cuts rely on tracking information, their performance is quite
sensitive to interactions of electrons in the inner detector material.
Their efficiency versus~$\eta$ is expected to be much less uniform than
that of the \medium\ cuts. 

The observed differences
between data and MC are discussed in terms of differences in electron
identification variables in Subsection~\ref{sec:Shapes}, in particular for the
calorimeter shower shapes (used in the \medium\ and \tight\ selections) and for the ratio  of
high-threshold transition radiation hits to all hits in the TRT detector (used in the
\tight\ selection).

Overall, the $\eta$ dependence of the identification efficiency is
in good agreement between data and~MC, with the most important deviations
seen around the transition region between the barrel and endcap calorimeters.
 Larger differences are seen 
as a function of~\ET, especially in the \Wen\ measurement, 
where the efficiency appears to decrease more at
low \ET\ for data than for~MC. More data are needed to properly understand this result.

The \ET-dependence of the efficiencies in the case of the \Jee\ measurements is in
good agreement between data and~MC. The shape can be attributed to the combination of
the increasing contribution of non-isolated electrons from non-prompt \Jpsi\
production (for which the efficiency decreases with \ET\ and 
is significantly lower at all \ET\ than for electrons from prompt \Jpsi\ production) and
to the rapidly improving efficiency for isolated electrons from prompt \Jpsi\
production as \ET~increases in this low-\ET\ range.   

\begin{figure*}
\begin{center}                               
\includegraphics[width=0.48\textwidth,angle=0]{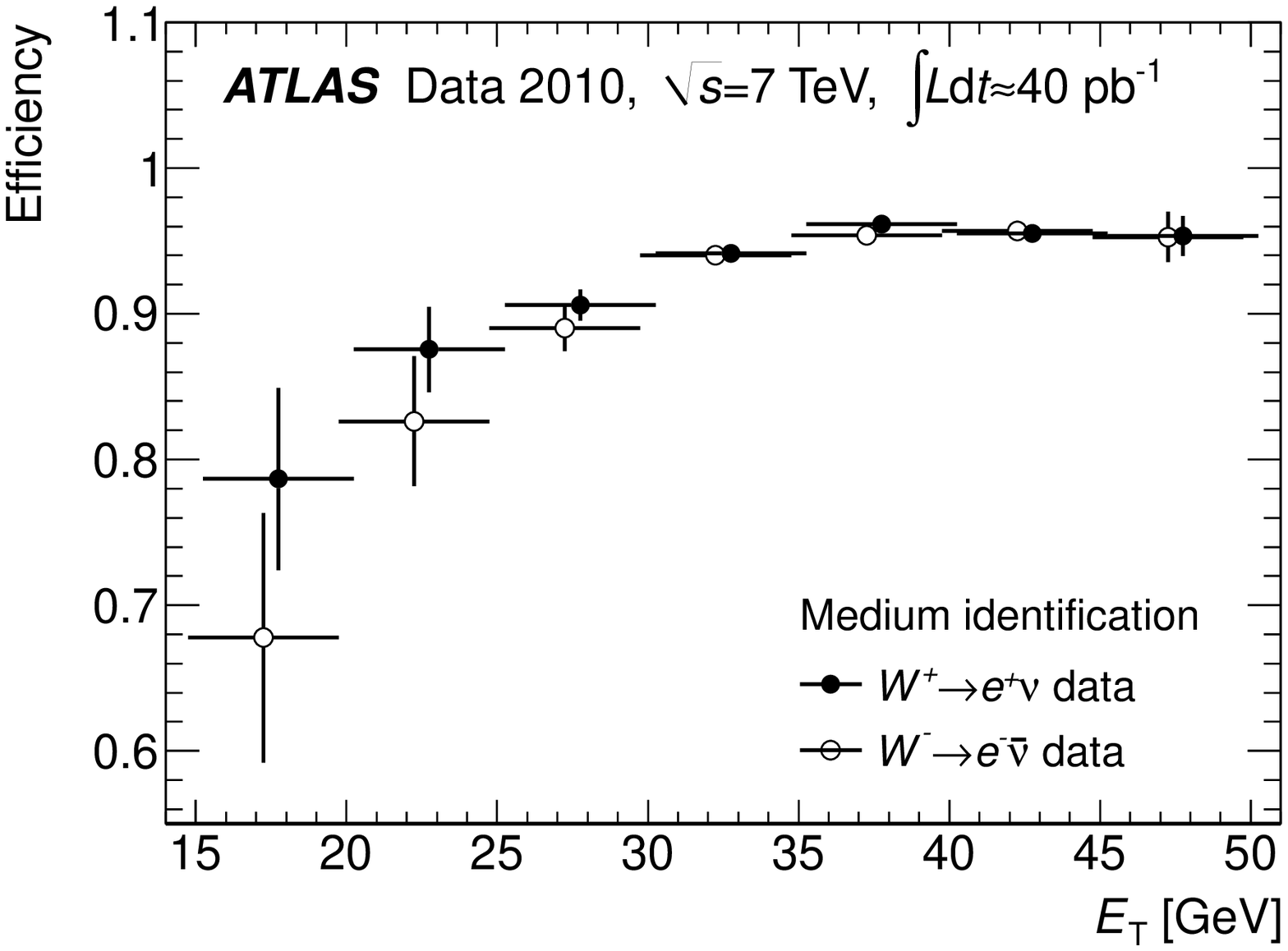}
\includegraphics[width=0.48\textwidth,angle=0]{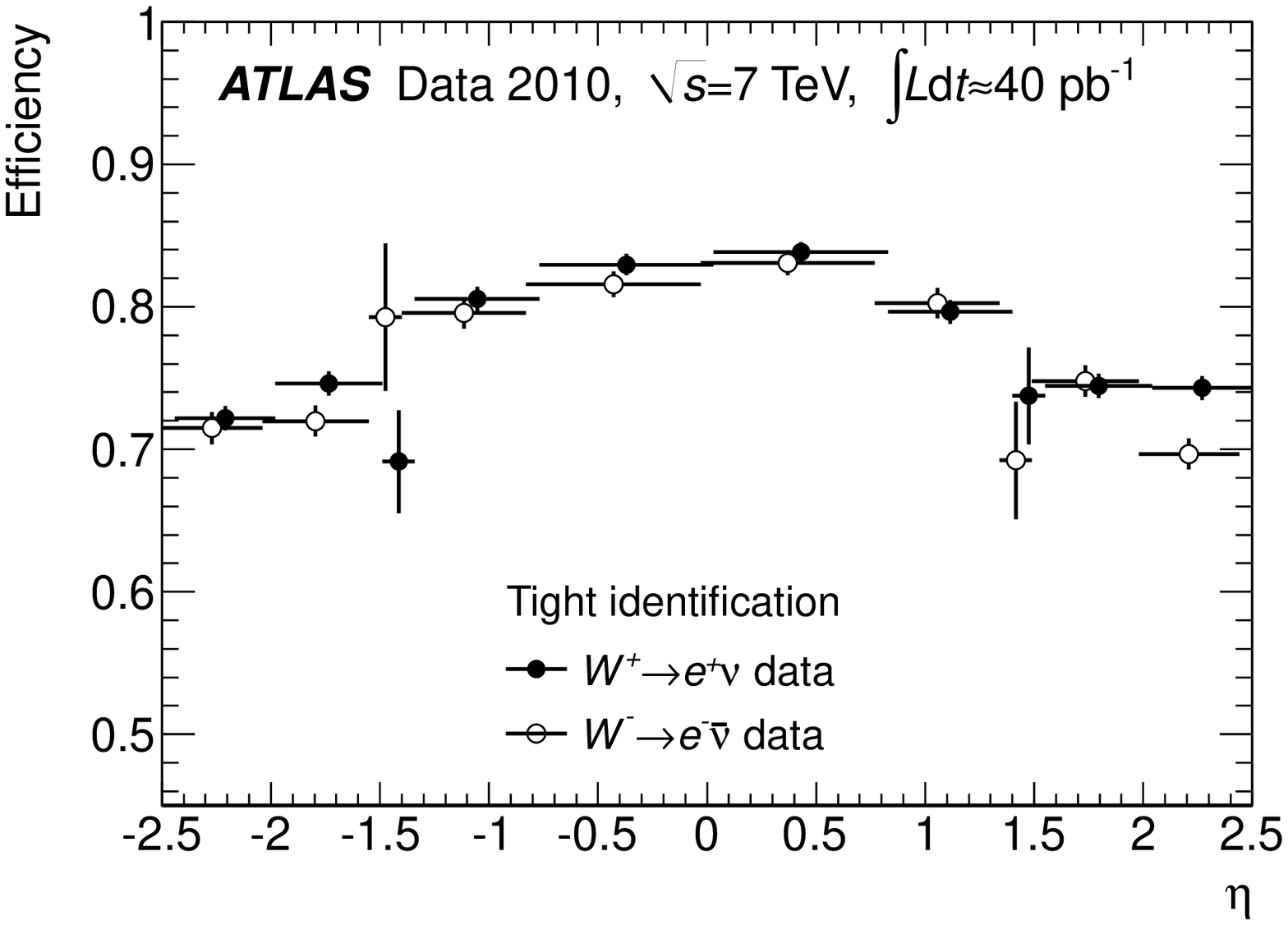}
\end{center}
\caption{Electron identification efficiencies measured separately for positrons (full circles) and electrons (open circles)
from \Wen\ events (left) for 
\medium\ identification as a function of~\ET\ and integrated over $|\eta|<2.47$ excluding the
transition region $1.37<|\eta|<1.52$ and (right) for \tight\
identification as a function of~$\eta$ and integrated over $20<\ET<50$~GeV. 
The results are shown with statistical uncertainties only.
For clarity, the electron and positron data points are slightly displaced horizontally in opposite directions.}
\label{Weffe+vse-}
\end{figure*}

To check the consistency of the measurements, the electron and positron identification
efficiency from the \Wen\ sample is compared, in  Figure~\ref{Weffe+vse-}, for \medium\ cuts as a
function of~\ET\ and for \tight\ cuts as a function of~$\eta$. Only statistical
uncertainties are shown. The systematic uncertainties are in general significantly larger
and correlated to some extent between the electron and positron measurements
in the same \ET- or $\eta$-bin.

The identification efficiency is expected to be higher for positrons than for electrons,
since there are about 40\% more positrons produced than electrons from \Wboson\ decays.
Although the charge misidentification probability due to material effects is itself charge independent,
the higher rate of $W^+\rightarrow e^+\nu$ will induce more charge-misidentified probes in the electron
sample than in the positron sample.
The lower identification efficiency of these charge-misidentified electrons and positrons,
also as a consequence of the material effects, leads to the expected difference in efficiency.
This difference is estimated in MC simulation to be as large as 3\%
at high $\eta$-values where the amount of material is larger.

Since the dominant systematic uncertainties on the measurement arise from background subtraction
and the number of events in the electron channel is smaller to start with,
somewhat higher total uncertainties are observed in the measurements for electrons
than for positrons. Small disagreements between data and MC
in some $\eta$-bins indicate that there might be some contribution
also from residual misalignment effects in the inner detector.
The discrepancies observed in these few bins have, therefore,
been added in quadrature to the total uncertainty for the charge-averaged measurements. 
It is
expected that, with more data and improved inner-detector alignment constants, these
discrepancies will be reduced.

\begin{figure*}
\begin{center}                               
\includegraphics[width=0.485\textwidth]{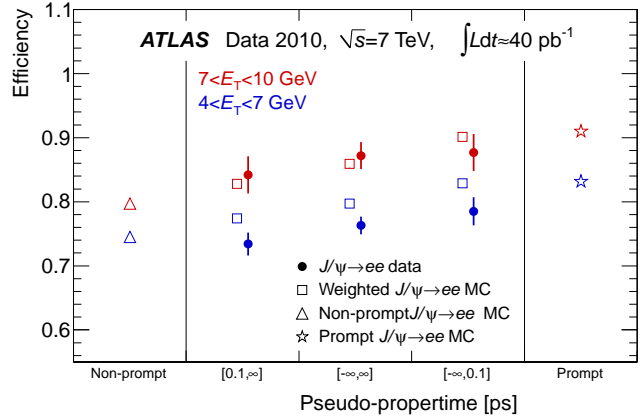}
\end{center}
\caption{Electron identification efficiencies measured from \Jee\ events and predicted by~MC for \medium\ identification
for two~\ET\ ranges: $4<\ET<7$~GeV (lower points) and $7<\ET<10$~GeV (higher points) for different ranges of
pseudo-proper time. The left-most open triangles show the MC efficiencies for a pure non-prompt \Jpsi\ sample,
while the right-most open stars show them  for a pure prompt \Jpsi\ sample integrated over
all pseudo-proper time values.
The MC~predictions plotted as open squares in the middle 
are weighted averages of the efficiency values expected for prompt and non-prompt
\Jpsi\ production as explained in the text. 
The results for the data are shown with statistical uncertainties only. 
For clarity, the data and MC points are slightly displaced horizontally in opposite directions.} 
\label{Jpsieff_lifetime}
\end{figure*}

The measurements for the \Jee\ channel have also been repeated for the \medium\ identification
criteria for different ranges of the measured pseudo-proper time, defined as
\begin{equation}
\tau_0 = \frac{L_{xy}\cdot m}{p_\mathrm{T}},
\label{eq:tau0}
\end{equation}
where $L_{xy}$ is the distance between the primary vertex 
and the extrapolated common vertex of the two electron candidates in the
transverse plane, $m$ is the reconstructed dielectron mass, and 
$p_\mathrm{T}$ is the reconstructed transverse momentum of the \Jpsi\
candidate.
Restricting the allowed pseudo-proper time to low (resp. high) values will improve the purity  
of the sample in terms of prompt (resp. non-prompt) \Jpsi\ decays.  
The results of these measurements for the two highest statistics \ET-bins are compared
in~Figure~\ref{Jpsieff_lifetime} to the MC~expectations for the weighted prompt plus non-prompt sample. 
The efficiencies expected for pure prompt and non-prompt \Jpsi\ production are
also shown. The efficiencies increase by several percent as the fraction of non-prompt decays decrease.
The data show the same trend but more statistics are needed to measure clearly the variation of
the efficiency with the fraction of decays from prompt \Jpsi\ production in the data, 
and ultimately to separate the prompt and non-prompt \Jpsi\ samples in the electron channel.

The \Wen\ and \Zee\ samples cover very similar \ET\ and $\eta$-ranges, 
but they are not identical, so the one-dimensional identification efficiencies 
presented here are not expected to be exactly equal for a given bin in each channel. 
The measured identification efficiencies, integrated over  
$\eta$ and for $20<\ET<50$~GeV, are given in Table~\ref{tab:IDWZEffi}.
Within their respective total uncertainties, 
the departures from the expected MC~efficiencies observed for \Wen\ and \Zee\ decays are 
compatible.

In contrast, the overlap between the \Wen\ and \Jee\ samples is limited to the \ET-range
between 15~and~20~GeV, a region in which both samples suffer from quite low statistics
and from large systematic uncertainties of about 10\%. 
Moreover, the \Jee\ efficiency is the weighted average of prompt and 
non-prompt \Jpsi\ decays, where only the former 
should be comparable to the electron efficiency obtained from \Wen\ decays. 
As the $\eta$-distributions of the two samples are not as similar as 
those of electrons from \Wen\ and \Zee\ decays,  
the measured and expected identification efficiencies and their ratios are
compared in the $15-20$~GeV \ET-bin in Table~\ref{tab:IDWJpsiEffi}, but only 
over a limited $\eta$-range, $|\eta|<0.8$. 
The MC efficiencies for \Wboson\ and prompt \Jpsi\ production agree within 
a few percent.
The measurement uncertainties are however still too large to draw 
firm conclusions.

\begin{table*}
\caption{\textit{Medium} and \tight\ identification efficiencies (in \%) measured in the \Wen\ and \Zee\ channels, 
integrated over $|\eta|<2.47$ excluding the transition region between barrel and 
\endcap\ EM calorimeters at $1.37<|\eta|<1.52$ and over~$20<\ET<50$~GeV. 
The measured data efficiencies are given together with the expected
efficiencies from MC~simulation and with their ratios. 
For the data measurements and for the ratios, the
first error corresponds to the statistical uncertainty and the second to the systematic
uncertainty. For the MC~expectations, the statistical uncertainties are negligible.}
\label{tab:IDWZEffi}
{\footnotesize
\begin{center}
\begin{tabular}{|l|l||l|l|l|}
\hline
Selection & Channel & Data [\%] & MC [\%] & Ratio \\
\hline\hline
{\it Medium} & \Wen & $ 94.1 \pm 0.2 \pm 0.6 $  & $ \ \ 96.9 $ & $ 0.971 \pm 0.002 \pm 0.007$ \\
& \Zee & $ 94.7 \pm 0.4 \pm 1.5 $  & $ \ \ 96.3 $ & $ 0.984 \pm 0.004 \pm 0.015$ \\
\hline
{\it Tight} & \Wen & $ 78.1 \pm 0.2 \pm 0.6 $  & $ \ \ 77.5 $ & $1.009 \pm 0.003 \pm 0.007$ \\
& \Zee & $ 80.7 \pm 0.5 \pm 1.5 $  & $ \ \ 78.5 $ & $1.028 \pm 0.006 \pm 0.016$ \\
\hline
\end{tabular}
\end{center}
}
\end{table*}

\begin{table*}
\caption{\textit{Medium} and \tight\ identification efficiencies (in \%) measured in the \Wen\ and \Jee\ channels,
integrated over $|\eta|<0.8$ and $15<\ET<20$~GeV. The measured data
efficiencies are given together with the expected efficiencies from MC~simulation and
their ratios. The MC~efficiencies for the \Jee\ channel are obtained as a weighted
average of the expected prompt and non-prompt components (see text). For completeness,
the expected MC efficiencies for a pure sample of \Jee\ decays from prompt
\Jpsi\ production are also given. For the data measurements and for the ratios, the first
error corresponds to the statistical uncertainty and the second to the systematic
uncertainty. For the MC~expectations, the statistical uncertainties are negligible.}
\label{tab:IDWJpsiEffi}
{\footnotesize
\begin{center}
\begin{tabular}{|l|l||l|l|l|l|}
\hline
Selection & Channel & Data [\%] & MC [\%] & Ratio & MC [\%]  \\
& &  &   &   & prompt \Jpsi  \\
\hline\hline
{\it Medium} & \Wen & $ 75.8 \pm 8.8 \pm \ \, 8.1 $  & $ \ \ 94.9 $ & $ 0.80 \pm 0.09 \pm 0.07$ &  \\
& \Jee & $ 80.0 \pm 7.3 \pm 10.2 $  & $ \ \ 81.9 $ & $ 0.98 \pm 0.09 \pm 0.14$ & \ \ \ \ \ 92.9 \\
\hline
{\it Tight} & \Wen & $ 61.9 \pm 6.0 \pm \ \, 7.0 $  & $ \ \ 78.3 $ & $0.79 \pm 0.08 \pm 0.09$ &  \\
& \Jee & $ 68.1 \pm 7.3 \pm \ \, 9.0 $  & $ \ \ 69.1 $ & $0.99 \pm 0.11 \pm 0.15$ & \ \ \ \ \ 78.3  \\
\hline
\end{tabular}
\end{center}
}
\end{table*}

\subsubsection{Electron identification variables}
\label{sec:Shapes}
The efficiencies measured in data and predicted by MC simulation presented in 
Figures~\ref{Weff},~\ref{Zeff}~and~\ref{Jpsieff} manifest 
some marked differences. These differences are related to
discrepancies in electron identification variables. In
this section, the distributions of calorimeter shower shapes and  
of the high threshold hit fraction in the
TRT are discussed.

\paragraph{Shower-shape distributions}

Lateral shower shapes in the EM~calorimeter (listed in Table~\ref{tab:IDcuts}) 
play a crucial role in \medium\
electron identification. They are extracted by the \TandP\ method using \Zee\ events
in bins of the probe \ET,
with \tag\ requirements and \probe\ definition as described in Section~\ref{sec:IDPerf}. 
The residual background, which could distort the measured distributions, 
is removed on a statistical basis using the technique of Ref.~\cite{sPlot:2005}. 
The method assigns a weight to each event based on a likelihood
fit to the \mee\ distribution in the range of $40-180$~GeV. 
These weights are then used to build the shape distributions.   
In order to obtain unbiased results, the correlations
between the discriminating variable (\mee)
and the extracted variables (shower shapes) need to be negligible.
This was verified using MC simulation.

To obtain bin-by-bin systematic uncertainties on the extracted electron shower shapes,
different models for the signal and background dielectron-mass distributions were investigated as in the 
efficiency measurement.  

The background subtraction method was validated by a closure
test performed on MC events by applying the
same procedure as used for the data.
For some distributions, the observed bias is of
the same order as the systematic uncertainty due to the choice of fit functions.  
The total bin-by-bin systematic uncertainties are calculated as the % quadratic
sum of these two  uncertainties and are \ET\ dependent.
They amount to $1-5$\% in the \ET-bin $25-30$~GeV,
and to $1-3$\% in the bin $40-50$~GeV, depending on the shape
variable.
With the 2010 dataset, the total uncertainty is dominated by the statistical uncertainty.

\begin{figure*}
\begin{center}
\includegraphics*[width=0.48\textwidth]{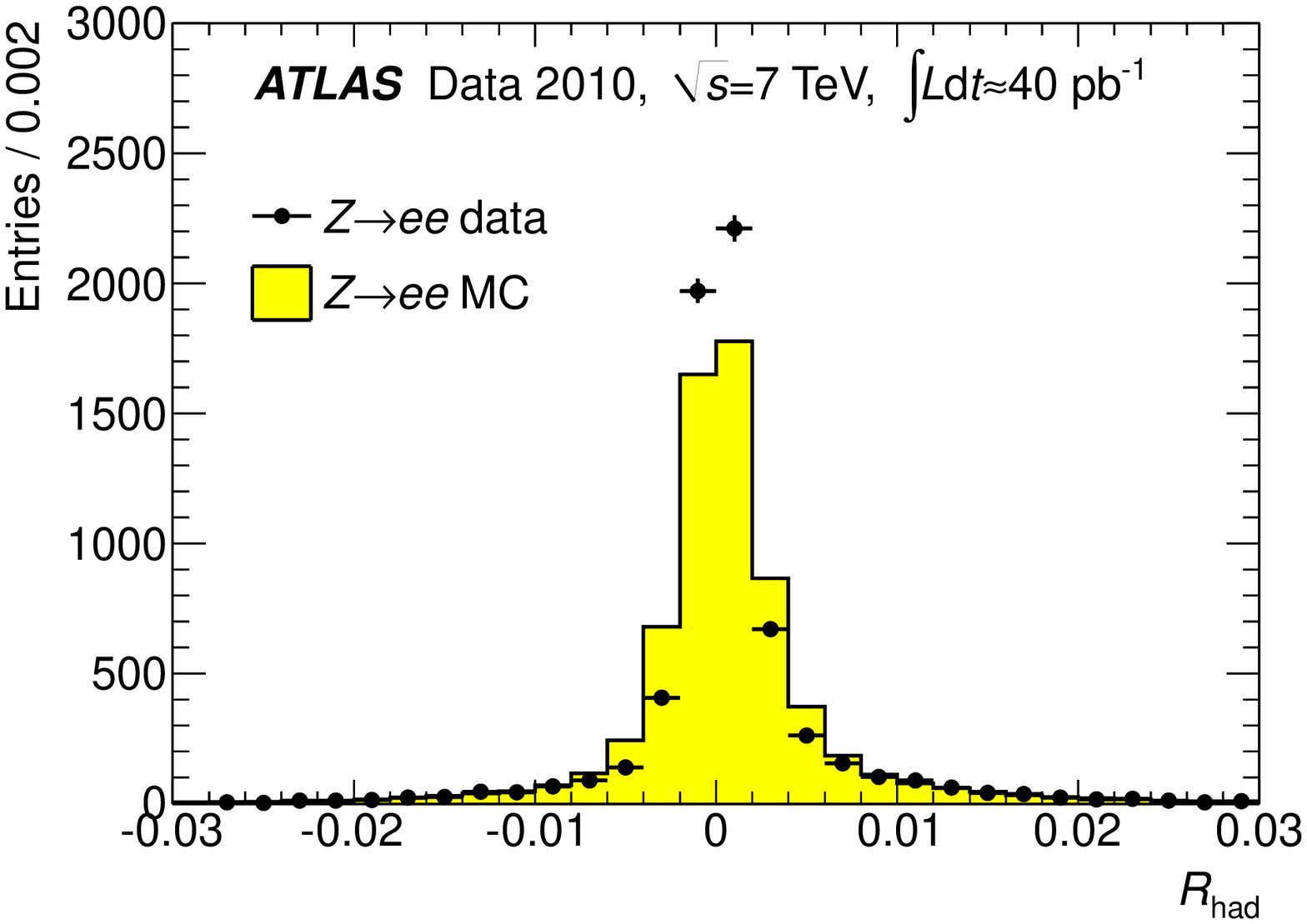}
\includegraphics*[width=0.48\textwidth]{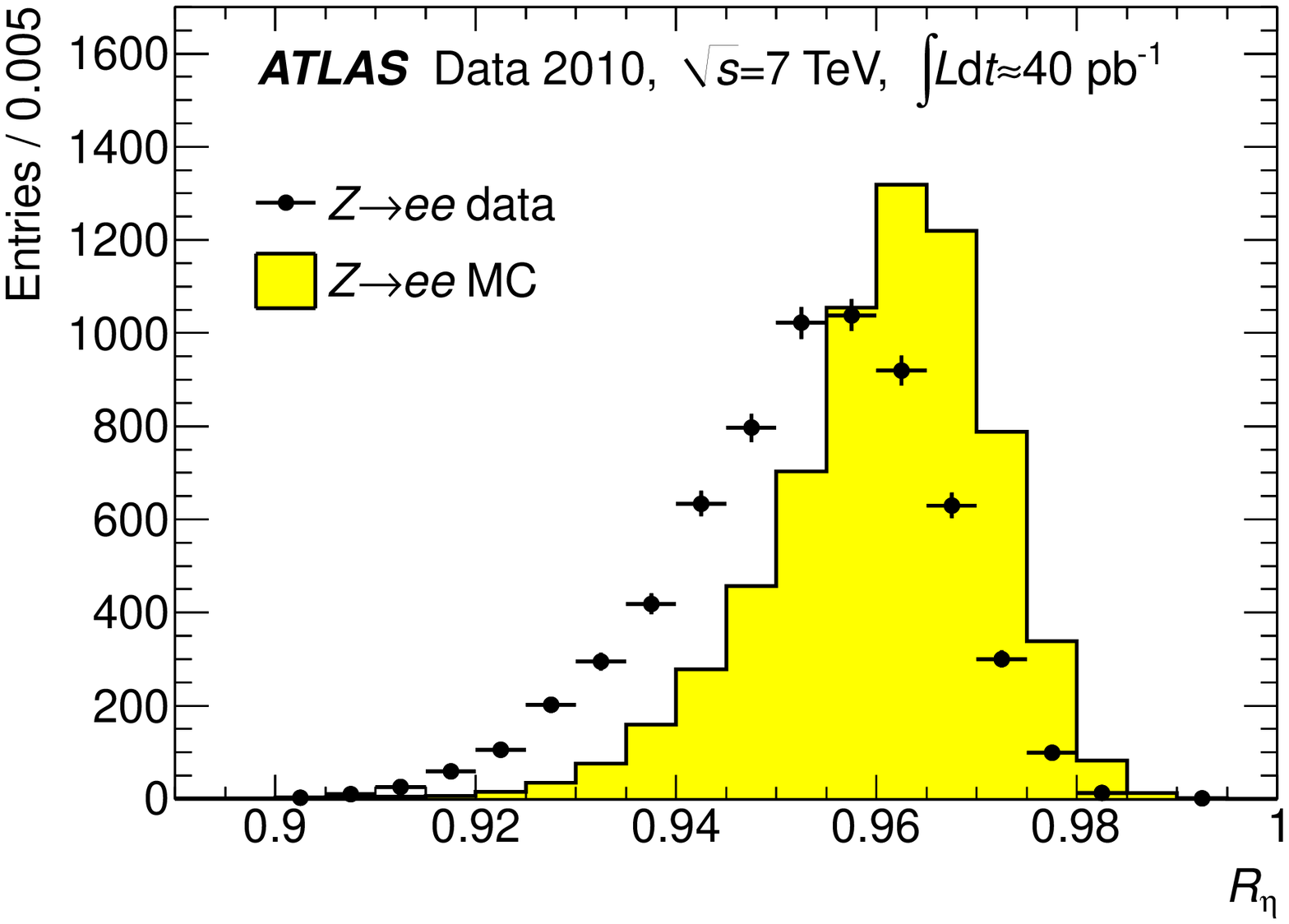}

\includegraphics*[width=0.48\textwidth]{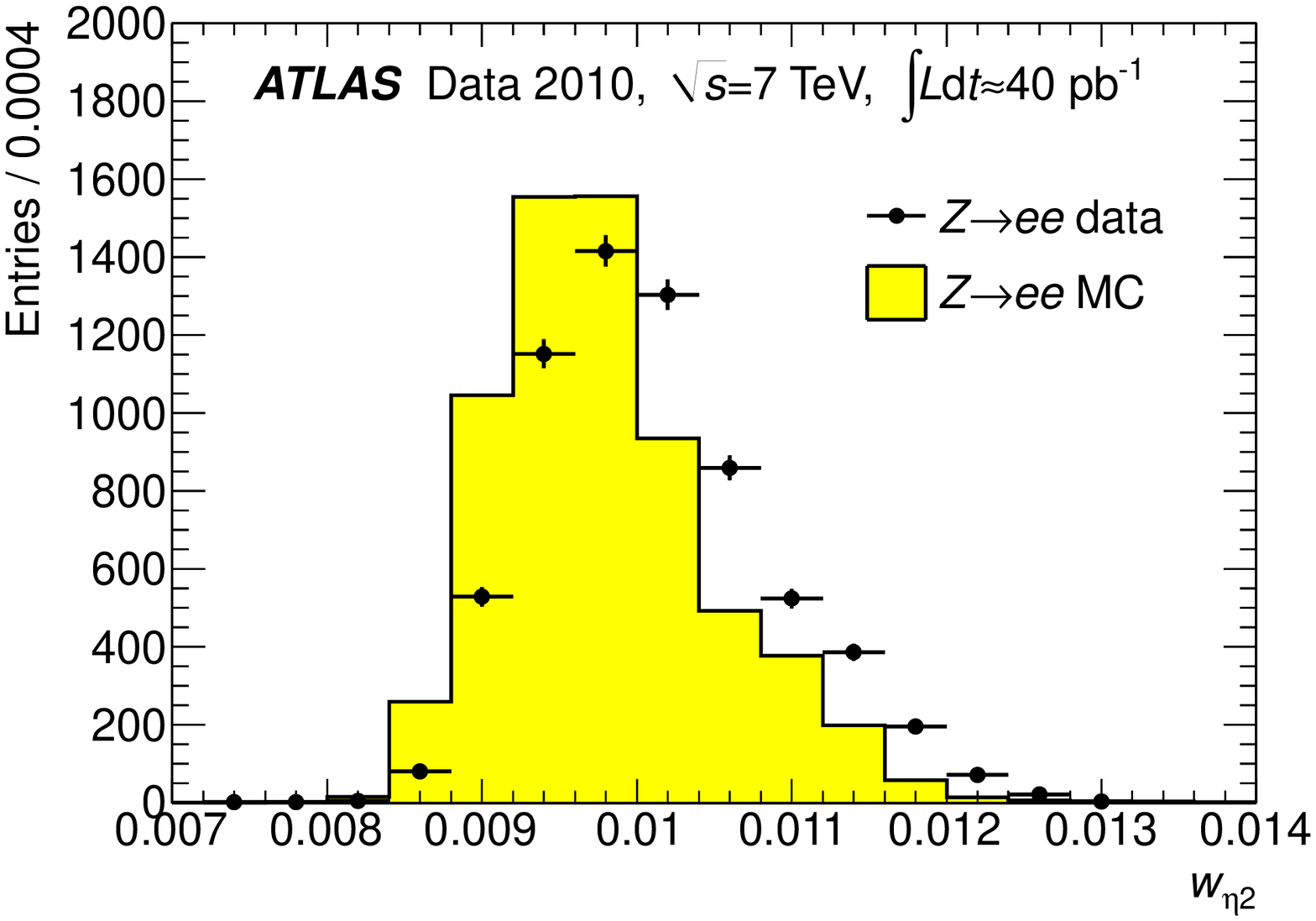}
\includegraphics*[width=0.48\textwidth]{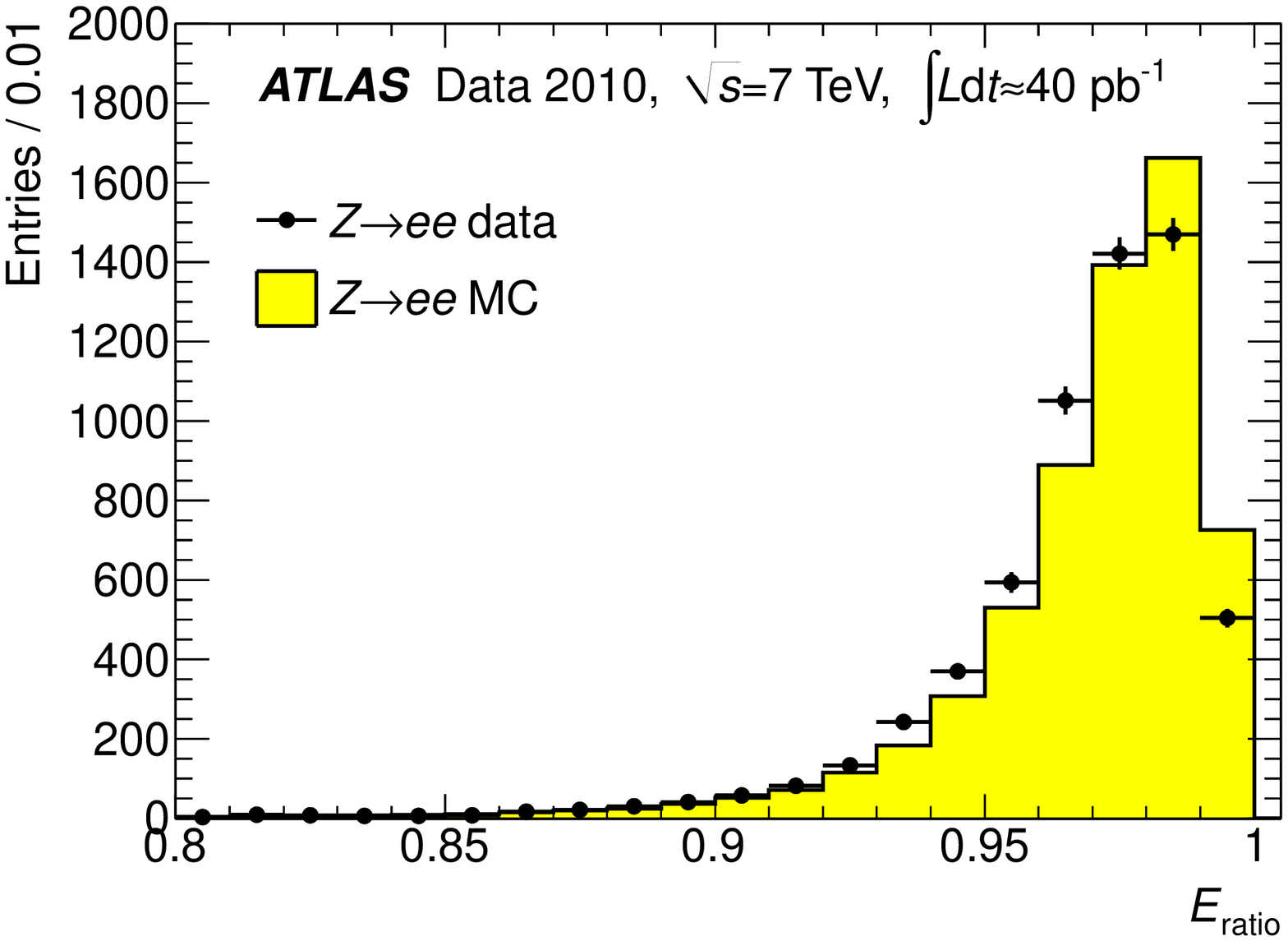}
\end{center} 
\caption {Electron shower shapes from \Zee\ events
for probe electrons in the range $\ET=40-50$~GeV: 
(top left) $R_\mathrm{had}$ hadronic leakage, 
(top right) $R_{\eta}$ and (bottom left) $w_{\eta2}$ middle-layer variables, 
(bottom right) $E_\mathrm{ratio}$ strip-layer variable. 
The data points are plotted as full circles with error bars, 
representing the total statistical and systematic uncertainties. 
The MC predictions, normalised to the number of data entries, are shown
by filled histograms. 
\label{fig:ShowerShapes_Data_MC}}
\end{figure*}

The extracted electron shower shapes from data are compared to the
MC prediction in Figure \ref{fig:ShowerShapes_Data_MC}. 
There are significant differences visible for all
extracted variables. The distributions of the strip and middle layer shapes are wider 
and are also shifted in data towards the background region. 
As a result, somewhat lower \medium\  efficiencies are observed in data compared to~MC. 
Currently, work is ongoing to refine the calorimeter simulation to achieve a better description
of the shower shape distributions.

\paragraph{High threshold TRT hits}

The \tight\ identification cuts listed in Table~\ref{tab:IDcuts} rely 
on more stringent matching cuts between the inner
detector and EM~calorimeter measurements and on additional measurements in the inner
detector. In particular, an advantage of the ATLAS detector is the capability of the TRT
to discriminate against  hadronic fakes over $|\eta|< 2.0$ using information on the ratio
of high threshold transition radiation hits over all hits ($f_\mathrm{HT}$).  

\begin{figure*}
\begin{center}
\includegraphics*[width=0.48\textwidth]{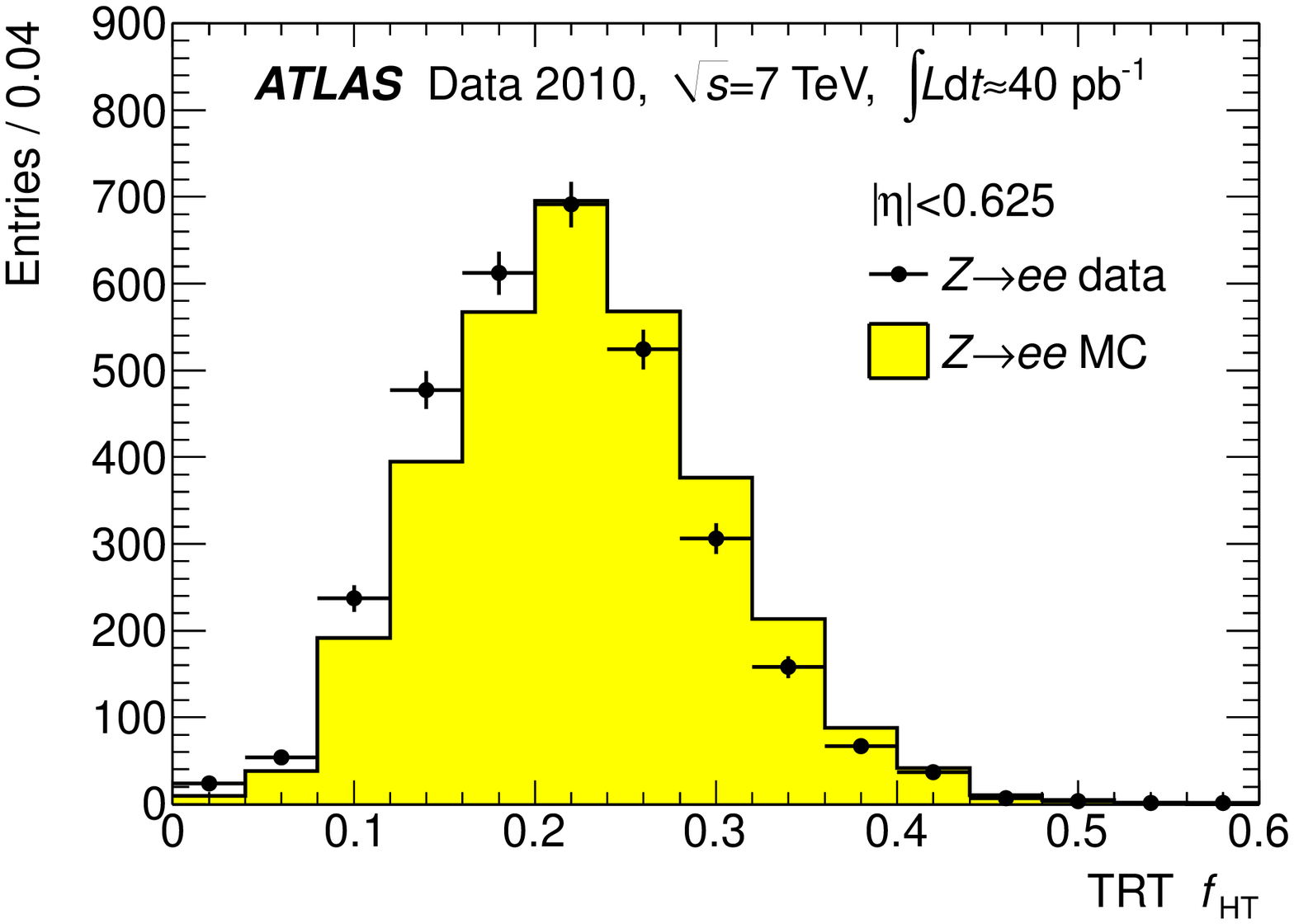}
\includegraphics*[width=0.48\textwidth]{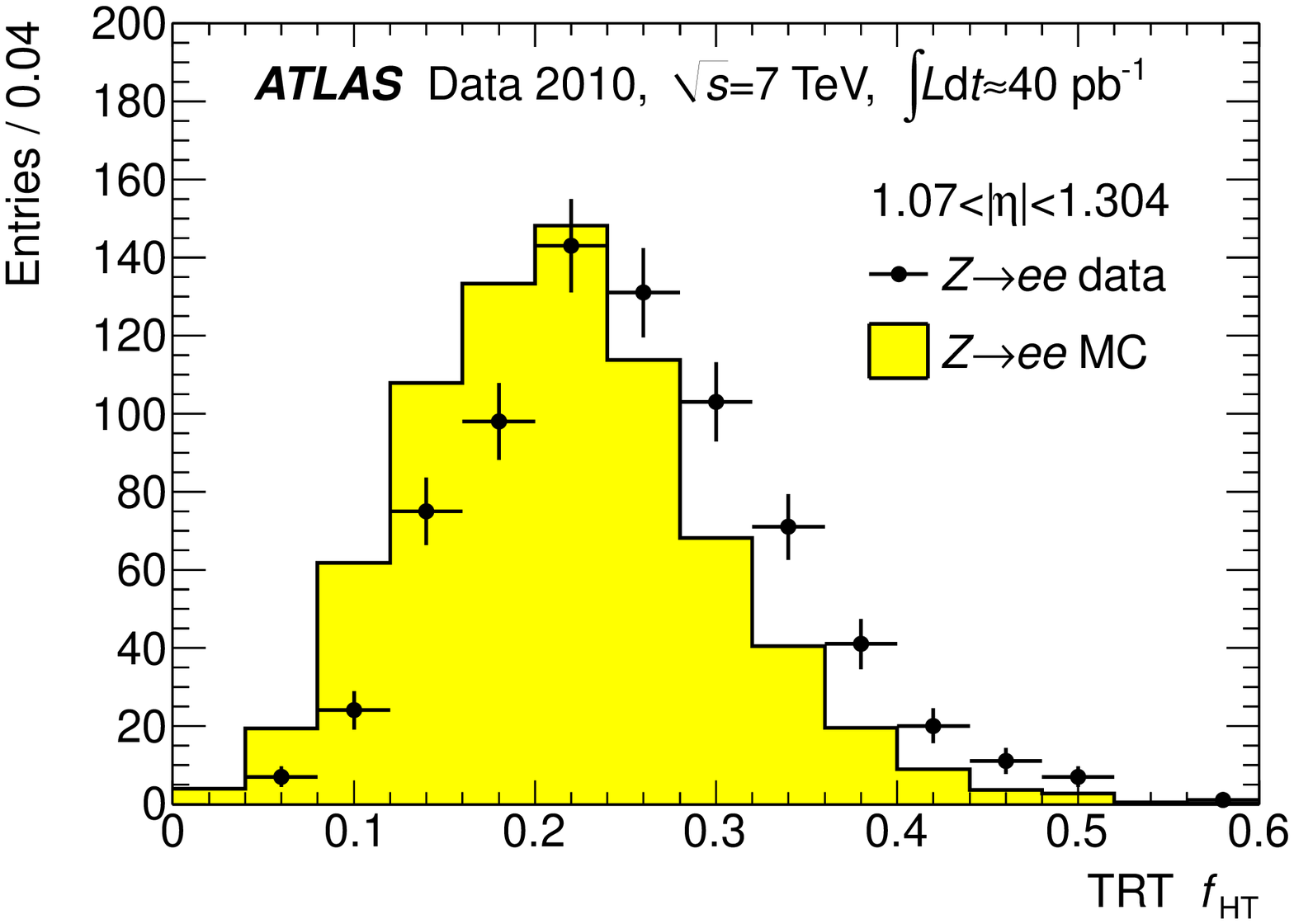}
\end{center} 
\caption {Distributions of the fraction of high-threshold hits in the TRT 
measured from \Zee\ data and compared to MC prediction
for (left) $|\eta|<0.625$ and (right) $1.07<|\eta|<1.304$. 
The data points are plotted as full circles with statistical error bars, 
while the MC predictions, normalised to the number of data entries,  
as filled histograms. }
\label{fig:TRfrac}
\end{figure*}

Figure~\ref{fig:TRfrac} shows the  $f_\mathrm{HT}$ distribution
in two $\eta$-regions for electron candidates from \Zee\ decays, selected
by a \TandP\ analysis and having momenta in the range $10 - 100 \GeV$, 
where the probability for producing high-threshold hits (HT) 
from transition radiation (TR) in the TRT straws is uniform. 
This  probability is in the range of  $0.2 - 0.25$,
to be compared with about $0.05$ for pion candidates 
in the same momentum range~\cite{TR2011conf}.

The HT probability for electrons varies with the radiator type, 
therefore it is expected to be different in the
barrel and endcap regions. It also depends
on the varying incidence angle of the charged particles on the straws.
The observed HT probability as a function of $\eta$ is not modelled perfectly 
in the barrel TRT by the MC simulation,
but the largest effect is the higher than predicted HT probability 
in the TRT endcap wheels.
For $|\eta|>1.0$,
the HT probability in data is measured to be significantly higher 
than in MC,
resulting in a better than expected electron identification performance.

\subsection{Electron identification efficiency in the forward region}
\label{sec:FwdPerf}
The efficiency of electron identification in the forward region outside
the tracking acceptance is studied using \Zee\ events, 
in two bins of pseudorapidity: $2.5 < |\eta| <
3.2$ corresponding to EMEC-IW and $3.2<|\eta|<4.9$ corresponding to the FCal
detectors. 

\subsubsection{Probe selection and background subtraction}

The \tag\ electron is required to be a {\it central} \tight\ electron with
$\ET>25$~GeV  and $|\eta| < 2.47$ excluding the transition region 
$1.37<|\eta|<1.52$, while the \probe\ is a {\it forward} electron 
candidate with $\ET>20$~GeV and $2.5 < |\eta| < 4.9$. 
With this selection, a total of 
5469 pairs in the \mee\ range $59-124$~GeV 
are found in the EMEC-IW, while 
3429 pairs are found in the range $50-160$~GeV 
in the FCal.

The background is subtracted using an unbinned maximum
likelihood fit to the dielectron invariant mass. 
The same methodology is used as in Subsection~\ref{sec:IDPerf}.
The signal is modelled either by  
a Breit-Wigner convolved with a Crystal Ball function or by a MC template.
The background is described either by a template from data requiring that the 
pair fails certain selection cuts or by different analytical functions.

The systematic uncertainties are studied by varying the signal integration
range, the background level via the tag requirements (isolation and \ET\ cut), 
the signal and background shapes and the fit range.
The systematic uncertainties vary between 2.5\% and 4.5\% and
are typically larger for FCal and for \forwardtight\ selection.
The possible bias of the method was also studied by a closure test and 
yielded an additional systematic uncertainty of $3-4$\%.

\subsubsection{Results}
 
Table~\ref{tab:effs_FWD} presents the measured and expected efficiency values. 
The electron identification
efficiency in the forward region is not perfectly 
reproduced by MC. 
This can be explained by the observation that the showers 
are broader and longer in data.
The origin of these discrepancies is under investigation.

\begin{table*}
\caption{Identification efficiencies (in \%) in the
forward region measured from \Zee\ events 
integrated over $\ET > 20$~GeV
and over $2.5 < |\eta| < 3.2$ for EMEC-IW
and over $3.2 < |\eta| < 4.9$ for FCal.
The measured data efficiencies are given together with the expected
efficiencies from MC~simulation and with their ratios. 
For the data measurements and 
for the ratios, the first error corresponds to the statistical and the second to
the systematic uncertainty. 
For the MC~expectations, the statistical uncertainties are about 0.1\%.
}
\label{tab:effs_FWD}
{\small 
\begin{center}
\begin{tabular}{|l|l||c|c|c|} 
\hline
 Detector & Selection &   Data [\%] & MC [\%] &   Ratio \\
\hline\hline
EMEC-IW& {\it Forward loose} & 83.1  $\pm$  1.3 $\pm$  4.6  
                       & 90.7  %$\pm$  0.1  
                       & 0.916  $\pm$  0.014 $\pm$  0.051 \\
       & {\it Forward tight} & 58.2  $\pm$  1.4 $\pm$  3.6 
                       & 72.8  %$\pm$  0.1         
                       & 0.800  $\pm$  0.019 $\pm$  0.050 \\ 
\hline 
FCal   & {\it Forward loose} & 87.5  $\pm$  2.6 $\pm$  7.2  
                       & 89.0  %$\pm$  0.1  
                       & 0.983  $\pm$  0.029 $\pm$  0.081  \\
       & {\it Forward tight} & 53.2  $\pm$  2.3 $\pm$  4.3  
                       & 59.4  %$\pm$  0.2 
                       & 0.896  $\pm$  0.038 $\pm$  0.072 \\ 
\hline
\end{tabular}
\end{center}
}
\end{table*}

\subsection{Reconstruction efficiency of central electrons}
\label{sec:RecoPerf}
In this section, the electron reconstruction efficiencies are studied
with respect to sliding-window clusters in the EM
calorimeter using \Zee\ decays following the 
methodology of Subsection~\ref{sec:IDPerf}.
The reconstruction efficiency defined this way measures the
combined electron track reconstruction and track--cluster matching 
efficiencies. 

\subsubsection{Probe selection and background subtraction}

To measure the electron reconstruction efficiency 
with or without the additional requirements 
on the number of silicon hits on the associated track
introduced in Subsection~\ref{sec:IDProbeSelection}, 
the requirements on the probe electron are released 
to consider all sliding-window EM  clusters.
Using \tight\ tag electrons having $\ET=20-50$~GeV, this leads to almost 20000  
\probes, with $500-4000$ per pseudorapidity bin. 
The \StoB\ ratio in the dielectron mass range $80<\mee<100$~GeV
varies from about 1 (for \medium\ \tags) to $6-10$ (for
\tight\ isolated \tags).

As for the identification efficiency measurement, the average of 
measurements, made with different configurations of the background
level 
and the size of the signal window 
in the dielectron mass, was used to assess the reconstruction efficiencies.
In particular, \medium\ or \tight\ \tags, with or without
track or cluster  isolation requirements, and with or without a cut on the
transverse impact parameter significance, 
and five different integration ranges are considered.
The root mean square of these 80 measurements is assigned as the systematic error on the
reconstruction efficiency due to the stability of the background estimation on
data.

The potential biases of the background subtraction method were also studied in a MC closure
test. The best closure was achieved using an exponential shape to describe the background and a
Breit-Wigner convolved with a Crystal Ball function (to account for detector effects) to
model the signal. The difference between the efficiency estimated using such a fit and the true
efficiency is considered as an additional systematic uncertainty. The
largest bias found in any $\eta$ bin is taken for all bins.
It amounts to 1.5\% (0.5\%) when the requirements on silicon hits on the track are (not) required.

\subsubsection{Results and pseudorapidity dependence}

\begin{figure*}
\centering
\includegraphics*[width=0.48\textwidth]{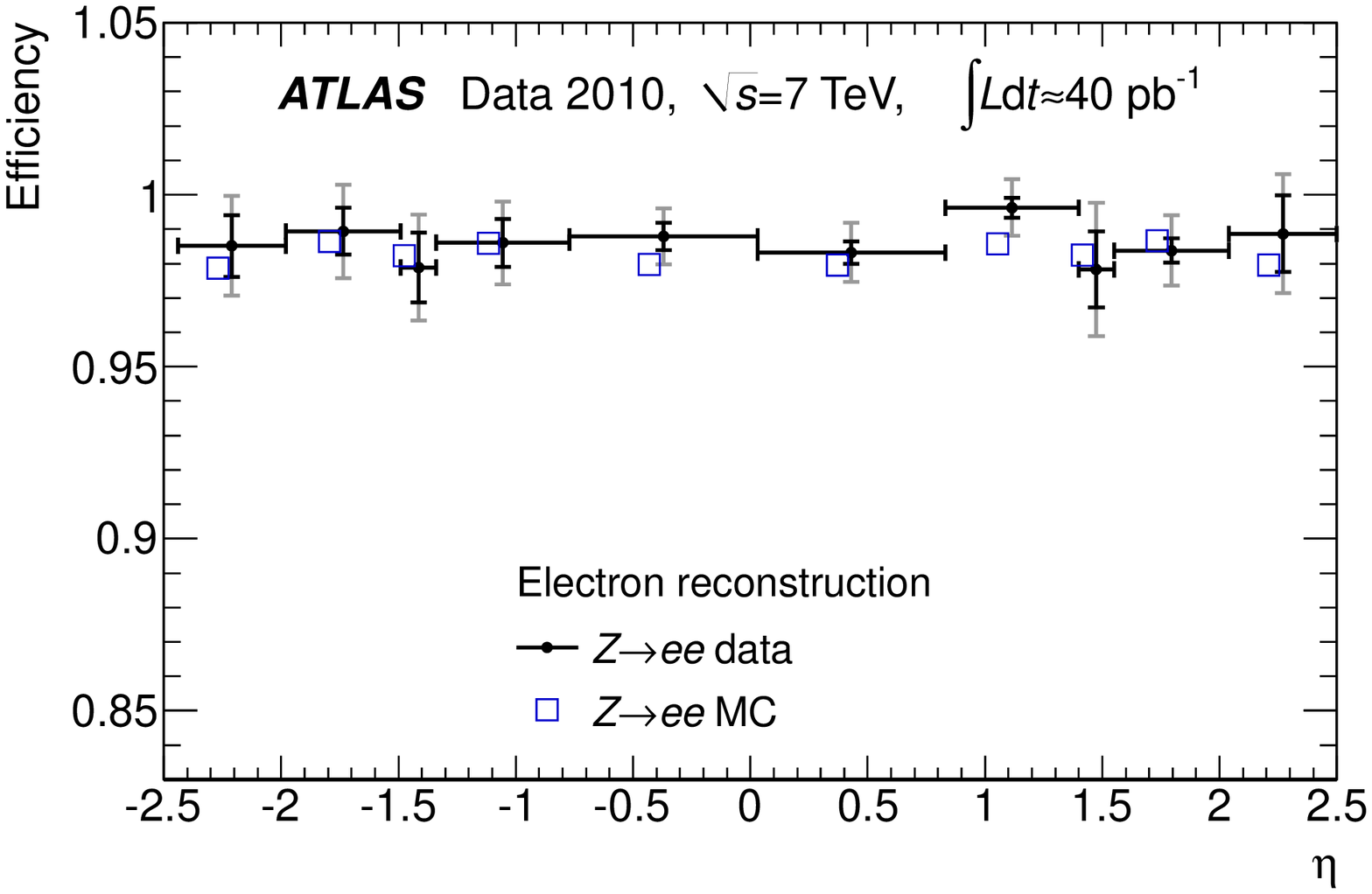} 
\includegraphics*[width=0.48\textwidth]{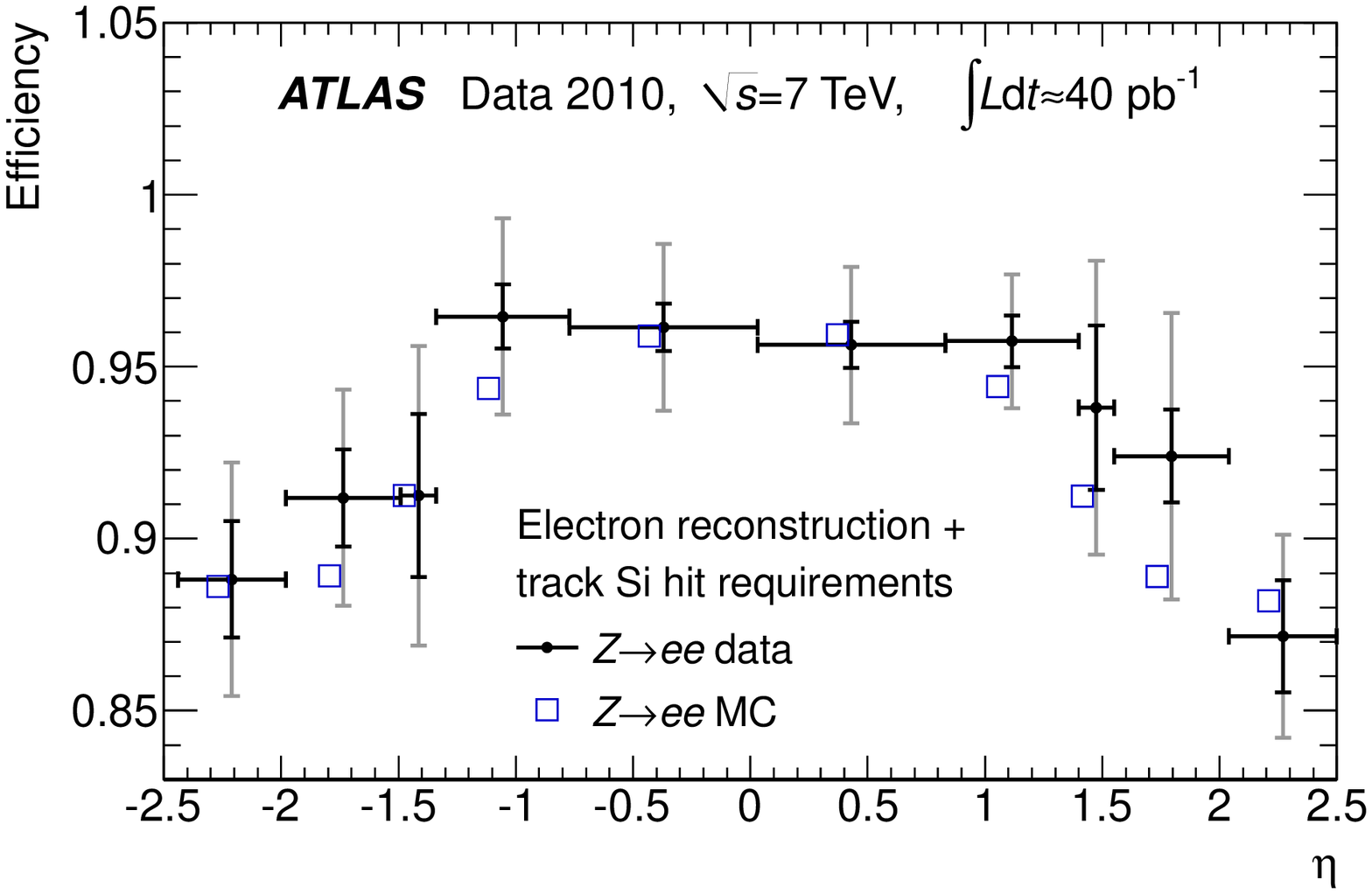}
    \caption {Reconstruction efficiency measured from \Zee\ events and predicted by MC 
    as a function of the cluster pseudorapidity and integrated
    over $20<\ET<50$~GeV
    (left) for electron reconstruction only and 
    (right) after applying requirements on the number of silicon hits on the track.
    The results for the data are shown with their statistical (inner error bars)
    and total (outer error bars) uncertainties. 
    The statistical error on the MC efficiencies plotted as 
    open squares is negligible.
    For clarity, the data and MC points are slightly displaced horizontally in
    opposite directions.
}
\label{fig:RecoZee_EffVsEta}
\end{figure*}

The measured reconstruction efficiency in data, shown in Figure~\ref{fig:RecoZee_EffVsEta}, is
compatible with the MC predictions, though  slightly higher
values are observed in data, 
especially in the region $0.8<|\eta|<2.01$ when
requirements on the numbers of silicon hits on the track are applied. 
The globally averaged efficiencies in the full
pseudorapidity range of $|\eta|<2.47$ are given in Table~\ref{tab:RecoEffi}.
The efficiency loss due to requirements on the numbers of silicon hits 
is smaller than 3\% in the barrel
and reaches almost 10\% in the highest $|\eta|$ bins.

\begin{table*}
\caption{Efficiency (in \%) for electron reconstruction only 
and with requirements on the number of silicon hits on the track, 
measured from \Zee\ events, integrated over~$20<\ET<50$~GeV and 
over $|\eta|<2.47$, excluding the transition region between barrel 
and endcap EM calorimeters at
$1.37<|\eta|<1.52$. 
 The measured data efficiencies are given together with the expected
efficiencies from MC~simulation and with their ratios. For the data measurements and 
for the ratios, the
first error corresponds to the statistical uncertainty and the second one to the systematic
uncertainty. For the MC~expectations, the statistical uncertainties are
negligible.}
\label{tab:RecoEffi}
\begin{center}
\begin{tabular}{|l||l|l|l|}
\hline
Selection & Data [\%] & MC [\%] & Ratio \\
\hline\hline
Electron reconstruction& $ 98.7 \pm 0.1 \pm 0.2 $ & $ 98.3$ & $1.005 \pm 0.001 \pm 0.002$\\
Track silicon hit requirements & $ 94.3 \pm 0.2 \pm 0.8 $ & $ 93.1$ & $1.013 \pm 0.002 \pm 0.008$\\
\hline
\end{tabular}
\end{center}
\end{table*}

 The results for the data are shown with their statistical (inner error bars)
and total (outer error bars) uncertainties. The statistical error on the MC efficiencies 
is negligible.

\subsection{Charge misidentification probability}
\label{sec:ChargeMisID}
Mismeasurement of the charge
happens primarily when the electron interacts early in the detector 
and the EM shower produces several high \pT\ tracks. 
The primary track is then
either not available or 
a different subsequent track
is matched 
to the EM cluster. The
charge  misidentification probability, \chargeMisID,  is defined as
the fraction of electrons with incorrectly measured charge with respect 
to all electrons, and depends on the applied electron
identification cuts. In particular, track quality cuts 
decrease \chargeMisID\ significantly. 

In this study \chargeMisID\ is investigated  comparing same-sign pairs to all
(same-sign and opposite-sign) pairs in \Zee\ events at four levels of electron
identification: reconstruction, silicon hit requirements on the track as
defined in Subsection~\ref{sec:IDProbeSelection}, and the standard \medium\
and \tight\ selections. 

\subsubsection{Probe selection and background subtraction}

To ensure a well measured tag electron charge, the
tag is confined to the barrel region of $|\eta|<1.37$. No correction
is applied for the misidentification of the \tight\ central tag electron.
This increases the measured probability with respect to the ``true'' value
by about 0.2\%.

The selection of same-sign pairs favours background over signal. 
This is especially problematic when studying \chargeMisID\
at early stages of electron identification. 
Additional requirements beyond the standard \Zee\ selection 
described in Subsection~\ref{sec:IDProbeSelection} are necessary. 
To extract the central value for \chargeMisID, a
low missing transverse momentum of $\MET<25$ GeV is required, 
reducing significantly the \Wen\ background.
To assess the systematic uncertainty
due to background contamination,
four other variants of the selection were studied
with different requirements on \MET, 
calorimeter isolation and the tag \ET.
With the standard \Zee\ selection, 
about 1000 \probes\ are found, with a 
\StoB\ ratio of 0.34, in the same-sign sample 
at the reconstruction level in the full pseudorapidity range.
Applying the \MET\ and calorimeter isolation cuts, the
\StoB\ ratio improves to 0.74 but the number of \probes\
drops to 550.
The available statistics is much more limited at \medium\ 
($100-140$ same-sign pairs)
and \tight\ (about 40 same-sign pairs) identification levels, where 
$\StoB=5.5-8$ is achieved. 

The remaining background is subtracted by a template method at early 
identification stages where the available statistics is sufficient,
and by a side-band method at the \medium\ and \tight\ identification levels.
For the fit, 
the background template is derived
from data events where the tag electron candidate fires an EM trigger 
(with no trigger-level electron identification) but fails both the \medium\ 
offline selection and the isolation cut. 
The signal template is obtained from \Zee\ MC. 
The number of signal events is counted within $75<\mee<100$~GeV. 

The systematic uncertainties are estimated by varying the tag requirements, 
the signal and background templates (or the side-bands), 
and the \mee\ signal window, in a way similar to that described
in Subsection~\ref{sec:IDPerf}.

\subsubsection{Results and pseudorapidity dependence}

The results for globally averaged charge misidentification probabilities 
are summarised in Table~\ref{tab:chg_glob_sys}. 
Overall the data-MC agreement
is good. The measurement in the data tends to be slightly lower
than the MC prediction.

\begin{table*}
\caption{Charge misidentification probabilities (in \%) at different levels of electron
identification from \Zee\ events, integrated
over $|\eta|<2.47$ excluding the transition region between barrel and \endcap\ EM calorimeters at
$1.37<|\eta|<1.52$ and over~$\ET\ >20$~GeV.
 The measured data efficiencies are given together with the expected
efficiencies from MC~simulation. For the data measurements, the
first error corresponds to the statistical uncertainty and the second one to the systematic
uncertainty. For the MC~expectations, the statistical uncertainties are negligible.
}
\label{tab:chg_glob_sys}
{\footnotesize
\begin{center}
\begin{tabular}{|l||l|l|l|}
\hline
Selection & Data [\%] & MC [\%] \\
\hline\hline
Electron reconstruction & $ 2.17\pm 0.25\pm 0.28 $  & $ 2.73 $\\
Track silicon hit requirements  & $ 1.13\pm 0.21\pm 0.16 $  & $ 1.28 $\\
{\it Medium} identification        & $ 1.04\pm 0.11\pm 0.14 $  & $ 1.20 $\\
{\it Tight} identification        & $ 0.37\pm 0.07\pm 0.11 $  & $ 0.50 $\\
\hline
\end{tabular}
\end{center}
}
\end{table*}

\begin{figure*}
\begin{center}   
\includegraphics[width=0.49\textwidth]{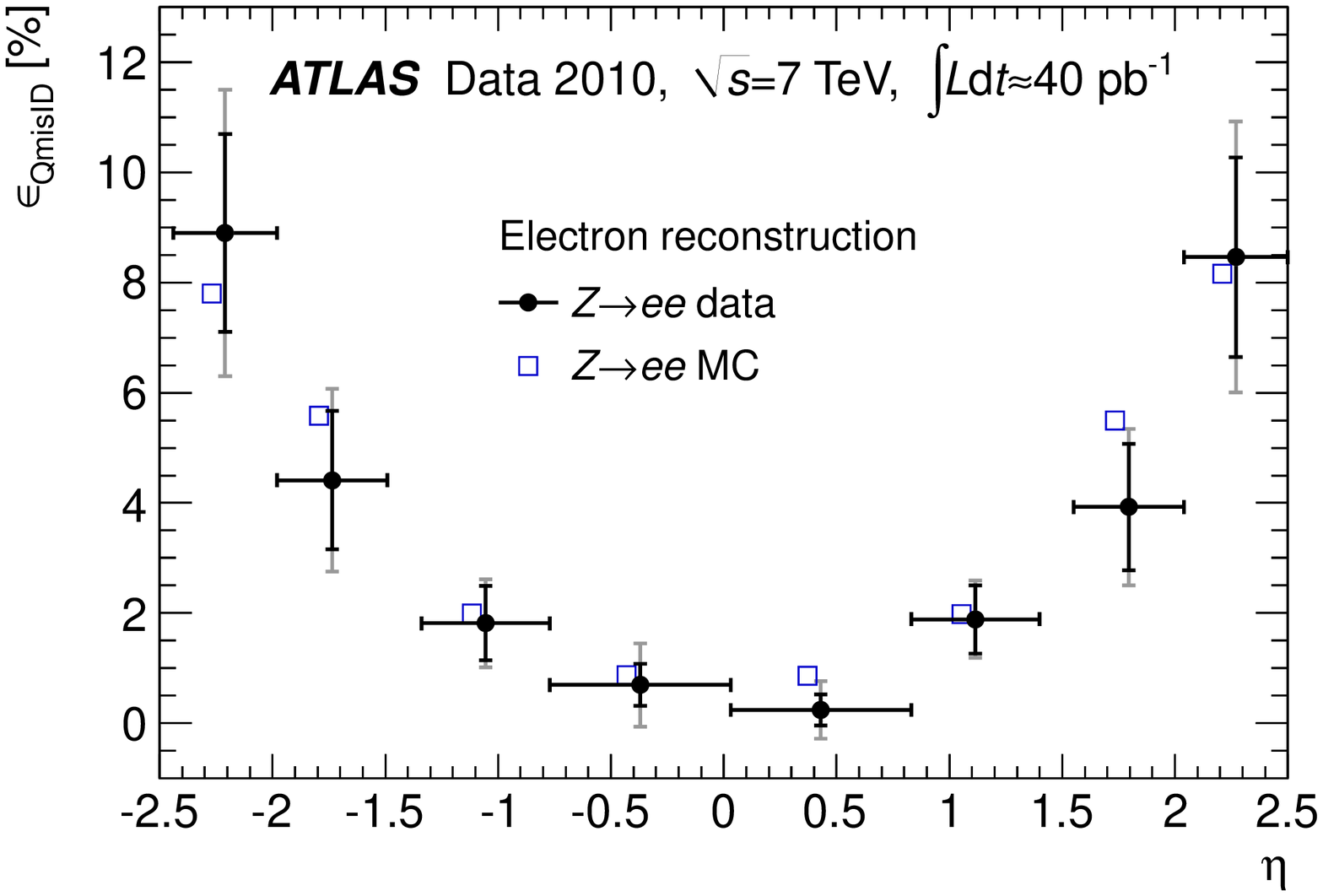}
\includegraphics[width=0.49\textwidth]{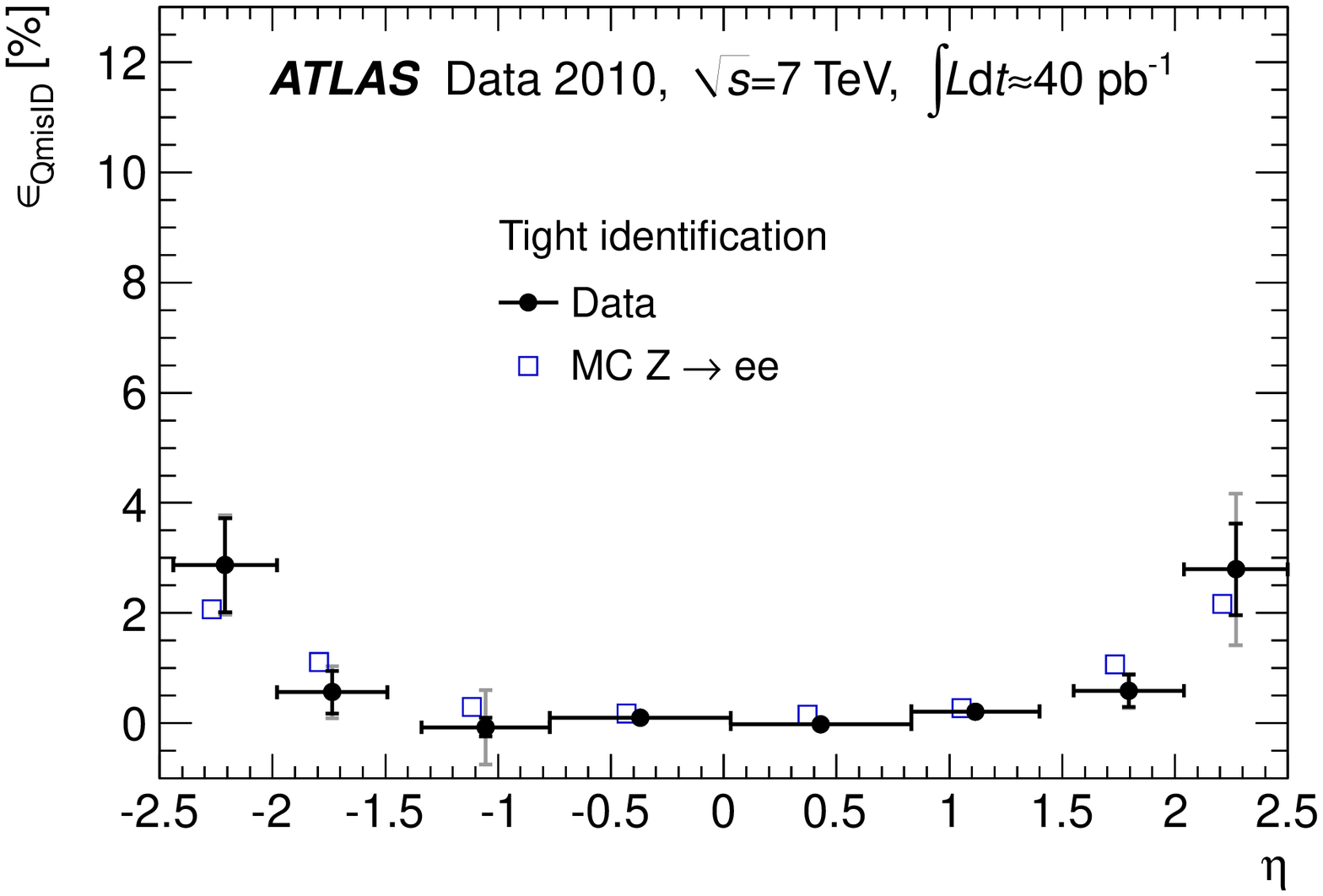}
\caption{Electron charge misidentification probability
measured from \Zee\ events as a function of pseudorapidity and 
integrated over $\ET>20$~GeV 
(left) after electron reconstruction
and (right) after \tight\ selection. 
Data points are shown with
statistical (inner error bars) and
total uncertainties (outer error bars).
The MC expectation is indicated by open squares.
For clarity, the data and MC points are slightly displaced horizontally in opposite directions.
}
\label{fig:chg_eta}
\end{center}
\end{figure*}

The same techniques are applied in bins of electron probe pseudorapidity. 
The results are displayed in Figure~\ref{fig:chg_eta} at the two extreme 
levels of selection: after electron reconstruction only and after \tight\
identification.

The measurements are repeated separately for the cases of positive (negative) tag
electrons, measuring \chargeMisID\
predominantly for true negative (positive) probes.
The results for the different charges agree within uncertainties. 

These measurements, even if limited in precision, do not show any
significant difference between the charge misidentification probability in data and MC.
An \chargeMisID\ of about 0.5\% is observed in the barrel 
and up to 8\% at high $\eta$ for candidates at the
reconstruction level. The measured probability decreases to about 0.2\% in the barrel
and around 2\% in the endcaps after \tight\ identification cuts.

\subsection{Electron trigger efficiency}
\label{sec:TriggerPerf}
The trigger efficiency is defined as the fraction of identified offline
electrons that fire a given trigger. Here, the
\medium\ and \tight\ selections are considered as offline benchmarks, 
for which the most commonly used triggers were
designed to have close to 100\% efficiency in the plateau \ET-region,
starting typically about 5 GeV above the trigger threshold. The main
sources of inefficiency are readout problems of the L1 system,  lower
reconstruction efficiency (especially for tracking)  at trigger level due to
timing constraints, and small differences of the electron identification
variables between trigger and offline~\cite{TriggerPaper2010,TrigEGConf2010}.

In 2010, events with high-\pT\ electrons were primarily  selected by the {\it
e15\_medium} and {\it e20\_loose} triggers, which require an electron
candidate reconstructed at the event filter (EF) 
level with $\ET>15$ and 20~GeV passing the \medium\
and \loose\ identification cuts, respectively.  In this section, 
their efficiency measurements using \Wen\ and \Zee\ decays are reported.

\subsubsection{Probe selection}

To measure the trigger efficiency, electron probes in the range
$\ET>15$~GeV are checked for a match to an EF electron fulfilling the
trigger selection. The angular distance $\Delta R$ between the trigger and offline electron 
candidates is computed using the
tracking variables. It is
required to be smaller than 0.15. This loose cut results
in a 100\% matching efficiency. Note that, while all three levels of the 
trigger have to be implicitly satisfied, no particular matching is required
between the offline electron and L1 or L2 trigger objects.
% which seeded the EF reconstruction. 

\Wen\ and \Zee\ candidates are selected following Subsection~\ref{sec:IDPerf}.
The \medium\ or \tight\ requirement on the probe electron candidate increases
significantly the purity of the sample. For example, in the \Zee\ channel
the background fraction of \tight--\medium\ pairs is below 1\%.
Therefore, no background subtraction is applied when obtaining the central values
of the trigger efficiency measurements.

Systematic uncertainties due to the tag requirements, 
the \mee\ requirement in the probe definition for the \Zee\ channel,
the background contamination, the energy-scale uncertainty
and the trigger--offline matching requirement
have been studied and found to be less than 0.1\% in total.

%\subsubsection{\texorpdfstring{Results and \ET\ dependence}{Results and ET dependence}}
\subsubsection{Results and \ET\ dependence}

\begin{figure*} 
\begin{center}
\includegraphics[width=0.49\textwidth]{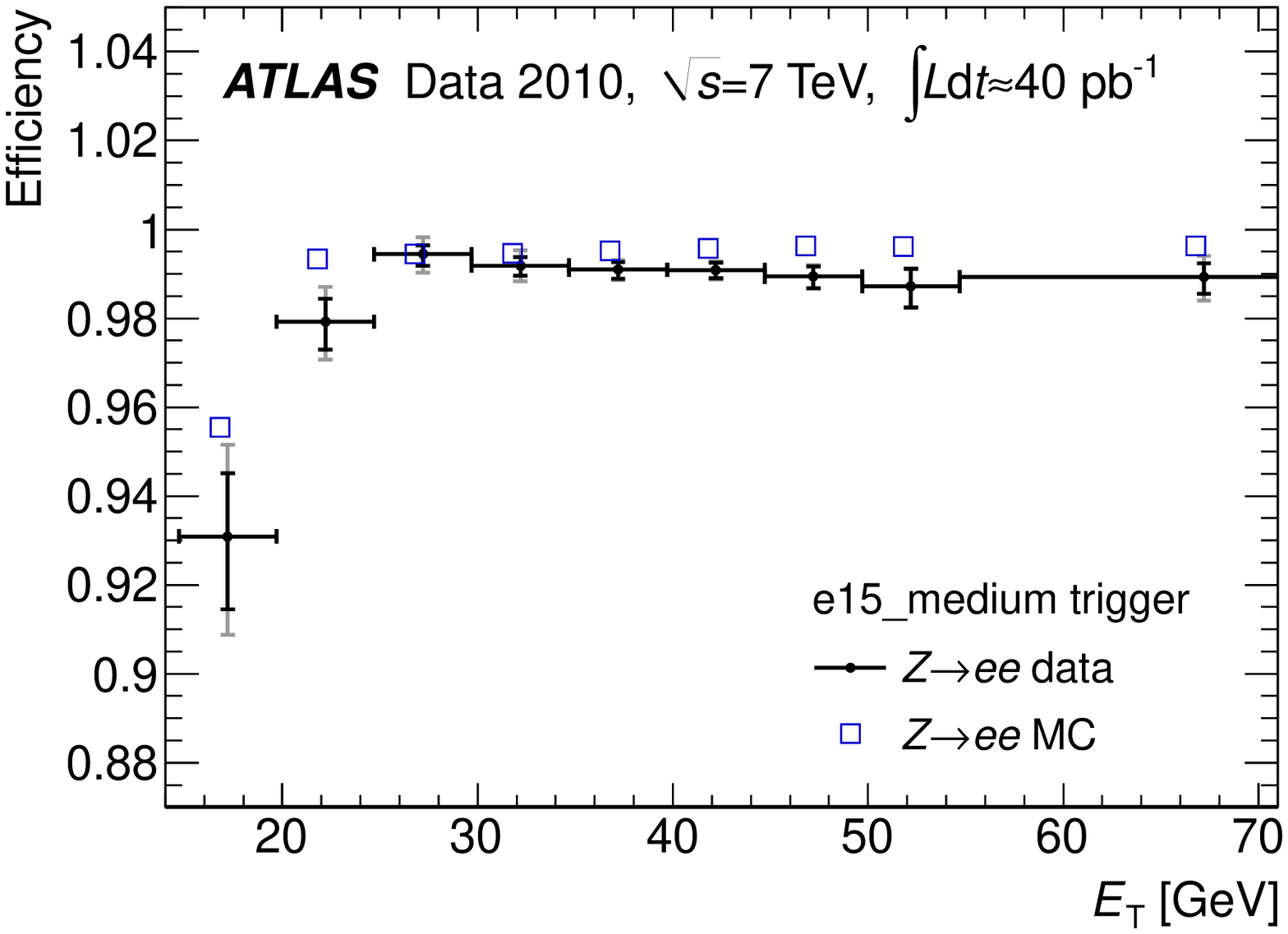}
\includegraphics[width=0.49\textwidth]{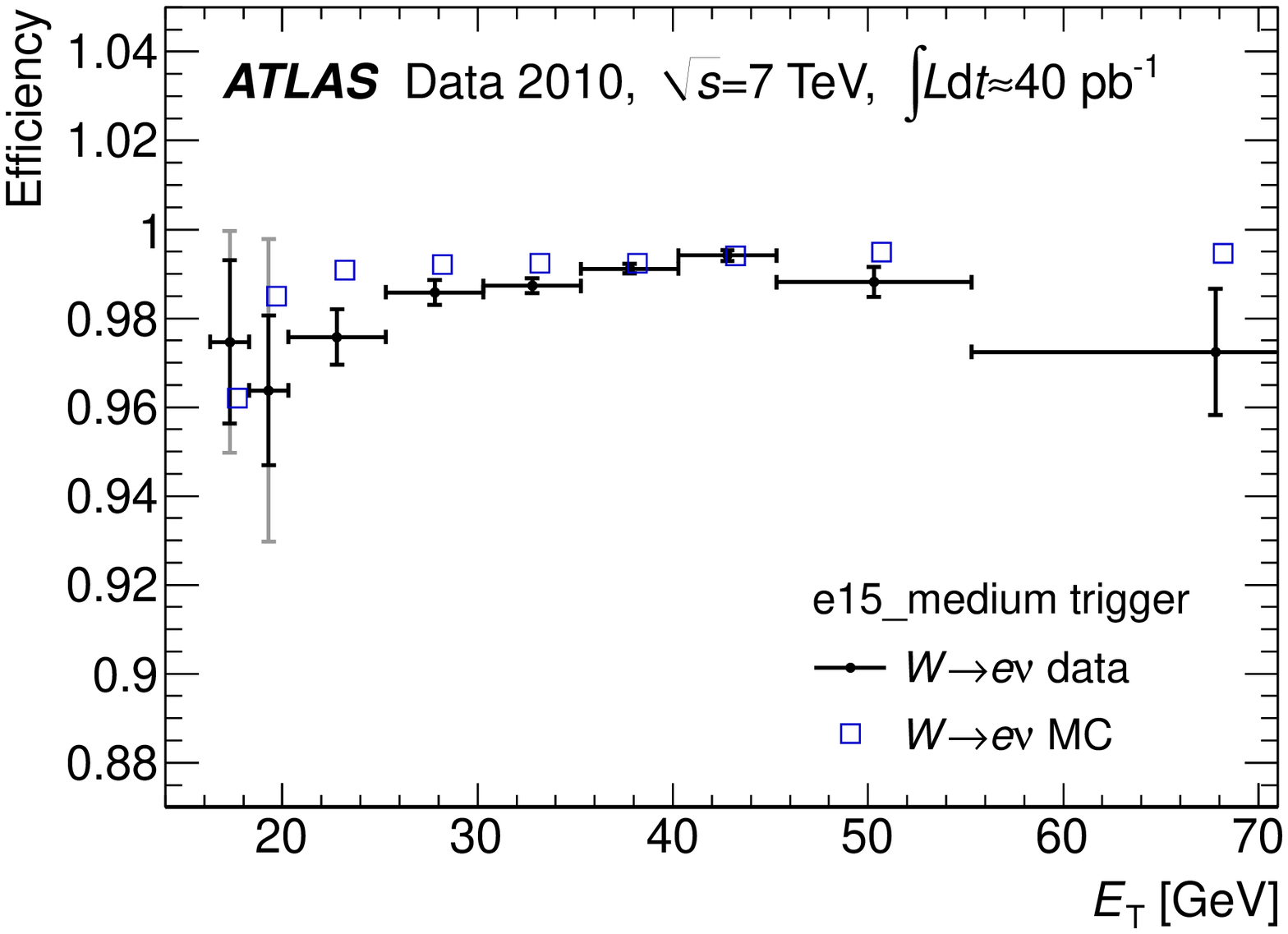}

\includegraphics[width=0.49\textwidth]{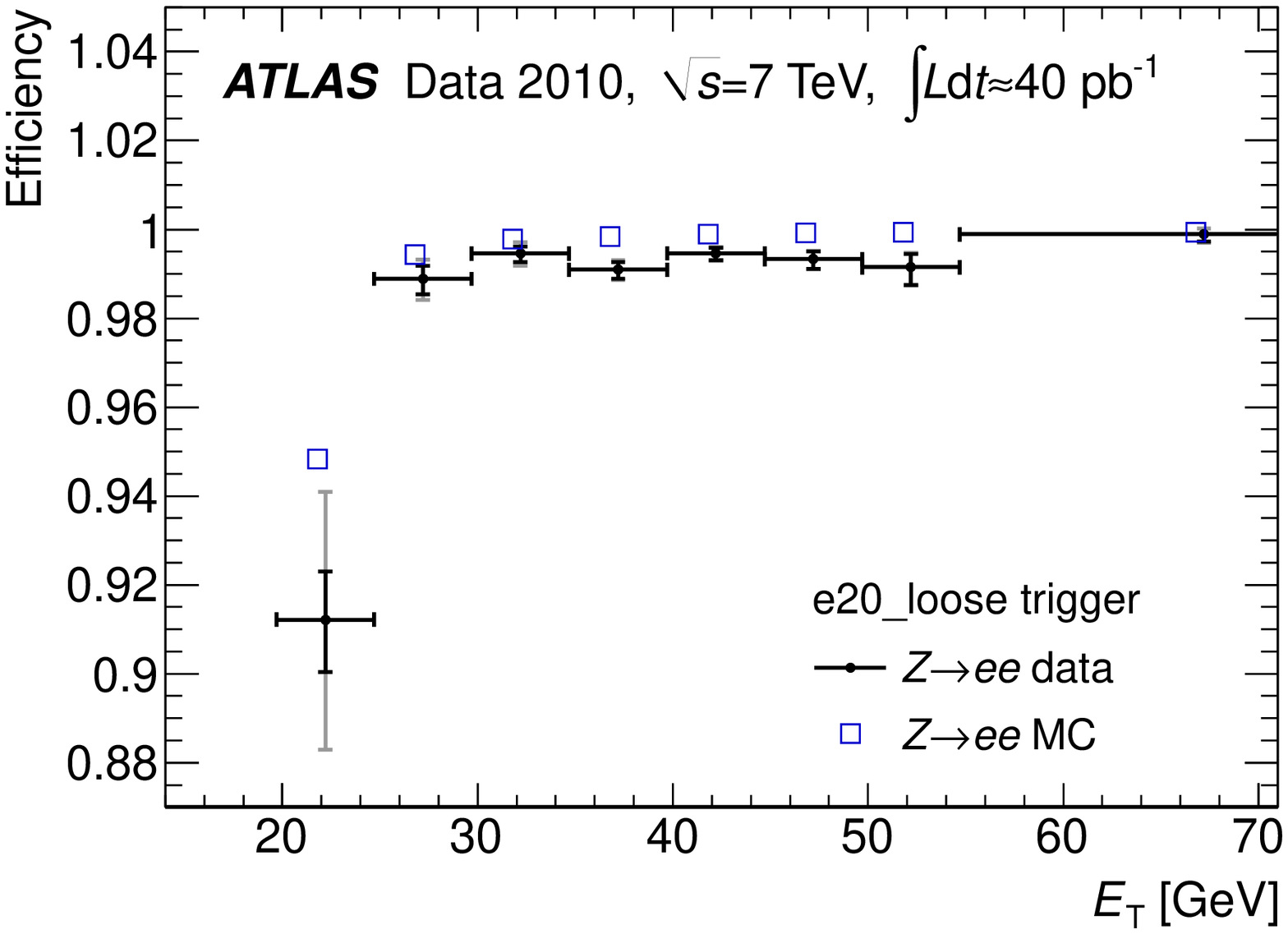}
\includegraphics[width=0.49\textwidth]{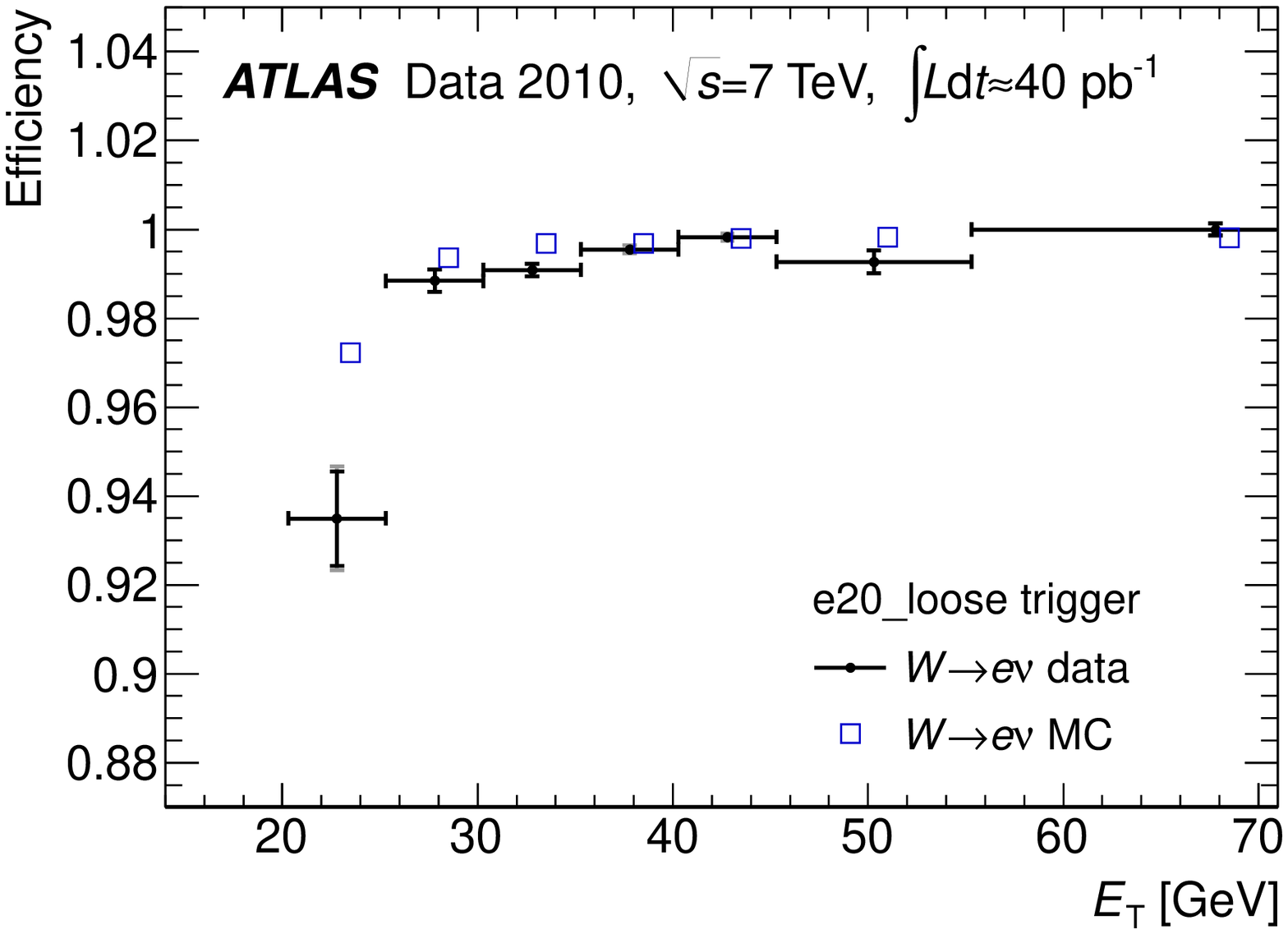}
\end{center}
\caption{Efficiency with respect to offline \tight\ electrons
for (top) e15\_medium and (bottom) e20\_loose triggers 
measured from (left) \Zee\ and (right) \Wen\ events
as a function of the offline electron \ET\ and
integrated over $|\eta|<2.47$ excluding the transition region 
between the barrel and \endcap\ EM calorimeters.
The results for the data are shown with their statistical (inner error bars)
and total (outer error bars) uncertainties. 
The statistical error on the MC efficiencies plotted as 
open squares is negligible.
For clarity, the data and MC points are slightly displaced horizontally in opposite directions.
} 
\label{fig:Trigger_efficiency}
\end{figure*}

Figure~\ref{fig:Trigger_efficiency} shows the trigger efficiency as a function
of the offline \ET\ of \tight\ probe electrons for the e15\_medium
and e20\_loose triggers. As expected, both triggers are very efficient 
in the plateau region starting 5 GeV above
the trigger threshold. 

The integrated efficiencies in the plateau region are summarized in
Table~\ref{tab:Trigger_IntEff} together with the data/MC efficiency ratios.
As correctly predicted by the MC, the trigger
efficiency is slightly higher with respect to the offline \tight\ selection than to
the \medium\ one. This is mainly due to the $E/p$ cut present in the \tight\ 
selection; it rejects electrons with a large amount of bremsstrahlung radiation
which are less efficiently reconstructed by the fast L2 tracking algorithm. 
The \Wboson\ and \Zboson\ results are compatible 
for all four trigger--offline selection combinations.

\begin{table*}
\caption{Efficiency (in \%) for the e15\_medium (e20\_loose) 
trigger measured from \Wen\ and \Zee\ events, integrated
over $|\eta|<2.47$ excluding the transition region between barrel and \endcap\ EM calorimeters at
$1.37<|\eta|<1.52$ and over \ET{}$>20$ (25) GeV. 
The measured data efficiencies are given together with the expected
efficiencies from MC~simulation and with their ratios. 
For the data measurements and 
for the ratios, the error corresponds to the statistical uncertainty. 
The systematic errors are below 0.1\%. 
For the MC~expectations, the statistical uncertainties are negligible.}
\label{tab:Trigger_IntEff}
\begin{center}
{\small
\begin{tabular}{|l|l|l||c|c|c|}
\hline 
Trigger & Probe & Channel & Data [\%] & MC [\%] & Ratio \\
\hline\hline
e15\_medium & Offline \medium & \Wen & $98.48\pm 0.08$ & $98.76$ & $0.997 \pm 0.001$ \\
                           & & \Zee & $98.67\pm 0.10$ & $99.24$ & $0.994 \pm 0.001$ \\
\cline{2-6} 
            & Offline \tight  & \Wen & $98.96\pm 0.07$ & $99.30$ & $0.997 \pm 0.001$ \\
                           & & \Zee & $99.02\pm 0.09$ & $99.54$ & $0.995 \pm 0.001$ \\
\hline 
e20\_loose  & Offline \medium & \Wen & $99.28\pm 0.05$ & $99.52$ & $0.998 \pm 0.001$ \\
                           & & \Zee & $99.11\pm 0.08$ & $99.73$ & $0.994 \pm 0.001$ \\
\cline{2-6} 
            & Offline \tight  & \Wen & $99.42\pm 0.05$ & $99.69$ & $0.997 \pm 0.001$ \\
                           & & \Zee & $99.33\pm 0.08$ & $99.83$ & $0.995 \pm 0.001$ \\
\hline 
\end{tabular}
}
\end{center}
\end{table*}

The small difference in the trigger efficiency behaviour between data and 
MC could be explained by the presence of dead L1 trigger towers\footnote{
These dead L1 trigger towers were repaired in the $2010-2011$ LHC winter shutdown.} 
not simulated in MC (typically well below the per mille level),
differences at the few \% level in the electron
energy-scale calibration introduced by the offline data reprocessing, and
differences in the distribution of identification variables between 
data and MC as discussed in Subsection~\ref{sec:Shapes}.

\section{Conclusions}
\label{sec:Conclusion}
The performance of the ATLAS detector for electrons in 2010 was presented,
using \Wen, \Zee\ and \Jee\ decays in $pp$ collision data.

An inter-alignment of the inner detector and the EM calorimeter 
has been performed and resulted in a track--cluster matching accuracy
close to the MC expectation. Further improvements are in progress, in particular for
$\phi$ in the \endcap\ regions covering $1.52<|\eta|<2.47$. 

The electron energy scale has been determined in bins of pseudorapidity with a
precision of $0.3-1.6$\% in the central region over $|\eta|<2.47$ 
and $2-3$\% in the forward regions over $2.5<|\eta|<4.9$, with
a residual non-uniformity in $\phi$ below 1\% in the central region. After
applying the 2010 in-situ calibration, the constant term of the energy
resolution is measured to be 
($1.2 \pm 0.1 \mathrm{(stat)} \pm 0.3 \mathrm{(syst)}$)\%
in the barrel EM calorimeter covering $|\eta|<1.37$, 
increasing to 1.8\% in the endcaps and to about 3\% in the forward regions.   
With the additional statistics being collected in 2011, the energy-scale
will be determined in $(\eta, \phi)$ bins
and the knowledge of the material in front of the calorimeter will be improved.
The EM calorimeter constant term should therefore be determined more accurately
and should decrease towards its design value of 0.7\%.   

Precise measurements as a function of~$\eta$ and~\ET\ have been 
performed for a variety of components of the electron selection efficiency
in the central region over $|\eta|<2.47$.
The electron identification efficiency has been measured with a total 
accuracy better than 1\% for the highest-statistics bin of $\ET=35-40$~GeV using \Wen\
events, 
and to about 10\% for the lowest-statistics bin of $\ET=15-20$~GeV using \Wen\ and \Jee\
events. 

The differences between calorimeter shower shapes measured in data and predicted by MC
have been an ongoing topic of study since the first runs collecting cosmic-ray 
events~\cite{ATLASCosmicPerf,ATLAS900Perf}. 
These are now precisely measured  for $|\eta|<2.47$
using the \Zee\ channel which
allows to extract unbiased distributions for the electron probes. 

Other important components of the electron selection efficiency
have been determined with good accuracy in the \Zee\ channel, 
even though they are more difficult to extract: 
the electron reconstruction efficiency, 
the efficiency of the track silicon hit requirements, 
and the probability of electron charge misidentification.
The trigger efficiency measurements
have established very
high plateau efficiencies of the electron triggers used in 2010.

In the forward region over $2.5 < |\eta| < 4.9$, despite the difficulty of the
measurements without any tracking information and with non-optimal EM~calorimeter
measurements, the clear signal observed from \Zee\ decays has been used to also measure
the electron identification efficiencies with reasonable accuracy. 
The
disagreements between data and MC are found to be larger in this region.

In parallel, work is ongoing to measure precisely the material in the detector
and to refine the description of the detector material,
the simulation of the EM shower development in the calorimeter,
and the transition radiation production in the TRT.
This will ultimately improve the description of the data 
by the MC.  

The accuracy of all efficiency measurements will benefit from the much larger statistics
available in~2011. 
Two-dimensional measurements in (\ET,$\eta$) space with
finer $\eta$ granularity will be obtained with accuracies better than 1\%,
allowing a more precise identification of
the sources of the different \ET-dependence of the efficiencies in data and
MC.

In the low-\ET\ range, the \Jee\ measurements require 
a substantial increase in statistics 
to measure the reconstruction
and identification efficiencies in the low-\ET\ region, important
for Higgs-boson searches. In this region, the material effects are large, 
the energies are closer to the reconstruction threshold, 
and the identification cuts are stringent. 

In the high-\ET\ range, above that explored in this paper, much higher
statistics of \Wen\ and \Zee\ decays are required to extend the measurements to
a region important for exotic searches where the efficiencies are expected to
become asymptotically flat with~\ET. 

Overall, the performance of the ATLAS inner detector and  EM~calorimeters has
been firmly established using the limited electron statistics  
from \Wboson, \Zboson\ and \Jpsi\ decays 
obtained in 2010 at $\sqrt{s}=7$~TeV corresponding to
about 40~\invpb. The agreement between the
measurements in data and the predictions of the MC is generally good, 
leading to only
small corrections of the MC electron performance estimates 
in physics analyses.

\section{Acknowledgements}

We thank CERN for the very successful operation of the LHC, as well as the
support staff from our institutions without whom ATLAS could not be
operated efficiently.

We acknowledge the support of ANPCyT, Argentina; YerPhI, Armenia; ARC,
Australia; BMWF, Austria; ANAS, Azerbaijan; SSTC, Belarus; CNPq and FAPESP,
Brazil; NSERC, NRC and CFI, Canada; CERN; CONICYT, Chile; CAS, MOST and
NSFC, China; COLCIENCIAS, Colombia; MSMT CR, MPO CR and VSC CR, Czech
Republic; DNRF, DNSRC and Lundbeck Foundation, Denmark; ARTEMIS, European
Union; IN2P3-CNRS, CEA-DSM/IRFU, France; GNAS, Georgia; BMBF, DFG, HGF, MPG
and AvH Foundation, Germany; GSRT, Greece; ISF, MINERVA, GIF, DIP and
Benoziyo Center, Israel; INFN, Italy; MEXT and JSPS, Japan; CNRST, Morocco;
FOM and NWO, Netherlands; RCN, Norway; MNiSW, Poland; GRICES and FCT,
Portugal; MERYS (MECTS), Romania; MES of Russia and ROSATOM, Russian
Federation; JINR; MSTD, Serbia; MSSR, Slovakia; ARRS and MVZT, Slovenia;
DST/NRF, South Africa; MICINN, Spain; SRC and Wallenberg Foundation,
Sweden; SER, SNSF and Cantons of Bern and Geneva, Switzerland; NSC, Taiwan;
TAEK, Turkey; STFC, the Royal Society and Leverhulme Trust, United Kingdom;
DOE and NSF, United States of America.

The crucial computing support from all WLCG partners is acknowledged
gratefully, in particular from CERN and the ATLAS Tier-1 facilities at
TRIUMF (Canada), NDGF (Denmark, Norway, Sweden), CC-IN2P3 (France),
KIT/GridKA (Germany), INFN-CNAF (Italy), NL-T1 (Netherlands), PIC (Spain),
ASGC (Taiwan), RAL (UK) and BNL (USA) and in the Tier-2 facilities
worldwide.

%%%%%%%%%%%%%%%%%%%%%%%%%%%%%%%%%%%%%%%%%%%%%%%%%%%%%%%%%%%%%%%%%%%%%%%%%%%%%%%
% Bibliography
%%%%%%%%%%%%%%%%%%%%%%%%%%%%%%%%%%%%%%%%%%%%%%%%%%%%%%%%%%%%%%%%%%%%%%%%%%%%%%

\bibliography{egPaper}

\providecommand{\href}[2]{#2}\begingroup\raggedright\begin{thebibliography}{10}

\bibitem{detpaper}
{The ATLAS Collaboration}, {\em {The ATLAS Experiment at the CERN Large Hadron
  Collider}\/},  JINST {\bf 3} (2008)  S08003.

\bibitem{cscbook}
{The ATLAS Collaboration}, {\em {Expected performance of the ATLAS experiment :
  detector, trigger and physics}\/},  CERN-OPEN-2008-20  ,
\href{http://arxiv.org/abs/0901.0512}{{\tt arXiv:0901.0512 [hep-ex]}}.
%%CITATION = 0901.0512;%%.

\bibitem{Pythia}
T.~Sjostrand, S.~Mrenna, and P.~Z. Skands, {\em {PYTHIA 6.4 Physics and
  Manual}\/},  \href{http://dx.doi.org/10.1088/1126-6708/2006/05/026}{JHEP {\bf
  0605} (2006)  026},
\href{http://arxiv.org/abs/hep-ph/0603175}{{\tt arXiv:hep-ph/0603175}}.
%%CITATION = HEP-PH/0603175;%%.

\bibitem{ATLASGeant4}
{The ATLAS Collaboration}, {\em {The ATLAS Simulation Infrastructure}\/},
  \href{http://dx.doi.org/10.1140/epjc/s10052-010-1429-9}{Eur. Phys. J. {\bf
  C70} (2010)  823--874},
\href{http://arxiv.org/abs/1005.4568}{{\tt arXiv:1005.4568 [physics.ins-det]}}.
%%CITATION = 1005.4568;%%.

\bibitem{Geant4}
S.~Agostinelli et al., {\em {GEANT4: A simulation toolkit}\/},
\href{http://dx.doi.org/10.1016/S0168-9002(03)01368-8}{Nucl. Instrum. Meth.
  {\bf A506} (2003)  250--303}.
%%CITATION = NUIMA,A506,250;%%.

\bibitem{MinbiasPaper1}
{The ATLAS Collaboration}, {\em {Charged-particle multiplicities in pp
  interactions at $\sqrt{s} = 900$~GeV measured with the ATLAS detector at the
  LHC}\/},  \href{http://dx.doi.org/10.1016/j.physletb.2010.03.064}{Phys. Lett.
  {\bf B688} (2010)  21--42}, \href{http://arxiv.org/abs/1003.3124}{{\tt
  arXiv:1003.3124 [hep-ex]}}.

\bibitem{MinbiasPaper2}
{The ATLAS Collaboration}, {\em {Charged-particle multiplicities in pp
  interactions measured with the ATLAS detector at the LHC}\/},
  \href{http://dx.doi.org/10.1088/1367-2630/13/5/053033}{New J. Phys. {\bf 13}
  (2011)  053033}, \href{http://arxiv.org/abs/1012.5104}{{\tt arXiv:1012.5104
  [hep-ex]}}.

\bibitem{MaterialK0s}
{The ATLAS Collaboration}, {\em {Study of the Material Budget in the ATLAS
  Inner Detector with $K^0_S$ decays in collision data at $\sqrt{s}=900$
  GeV}\/},  ATLAS-CONF-2010-019.

\bibitem{MaterialHadronicInt}
{The ATLAS Collaboration}, {\em {A measurement of the material in the ATLAS
  inner detector using secondary hadronic interactions}\/},  JINST {\bf 7}
  (2012)  P01013, \href{http://arxiv.org/abs/1110.6191}{{\tt arXiv:1110.6191
  [hep-ex]}}.

\bibitem{MaterialConversions}
{The ATLAS Collaboration}, {\em {Photon conversion at $\sqrt{s}=900$~GeV
  measured with the ATLAS detector}\/},  ATLAS-CONF-2010-007.

\bibitem{MaterialEnergyFlow}
{The ATLAS Collaboration}, {\em {Probing the material in front of the ATLAS
  electromagnetic calorimeter with energy flow from $\sqrt{s}=7$ TeV minimum
  bias events}\/},  ATLAS-CONF-2010-037.

\bibitem{TriggerPaper2010}
{The ATLAS Collaboration}, {\em {Performance of the ATLAS Trigger System in
  2010}\/},  Eur. Phys. J. {\bf C72} (2012)  1849,
  \href{http://arxiv.org/abs/1110.1530}{{\tt arXiv:1110.1530 [hep-ex]}}.

\bibitem{TrigEGConf2010}
{The ATLAS Collaboration}, {\em {Performance of the Electron and Photon Trigger
  in p-p Collisions at $\sqrt{s} = 7$ TeV with the ATLAS Detector at the
  LHC}\/},  ATLAS-CONF-2011-114.

\bibitem{ElectronNote}
{ATLAS Collaboration}, {\em {Expected electron performance in the ATLAS
  experiment}\/},  ATL-PHYS-PUB-2011-006.

\bibitem{TopoClusters}
{W. Lampl et al.}, {\em {Calorimeter Clustering Algorithms : Description and
  Performance}\/},  ATL-LARG-PUB-2008-002.

\bibitem{LArPaper}
{The ATLAS Collaboration}, {\em {Readiness of the ATLAS Liquid Argon
  Calorimeter for LHC Collisions}\/},  Eur. Phys. J. {\bf C70} (2010)
  723--753, \href{http://arxiv.org/abs/0912.2642}{{\tt arXiv:0912.2642
  [physics.ins-det]}}.

\bibitem{PhotonNote}
{The ATLAS Collaboration}, {\em {Expected photon performance in the ATLAS
  experiment}\/},  ATL-PHYS-PUB-2011-007.

\bibitem{atlastdr}
{The ATLAS Collaboration}, {\em {ATLAS detector and physics performance:
  Technical Design Report, Vol. 1}}.
\newblock Technical Design Report ATLAS. CERN, Geneva, 1999.
\newblock CERN/LHCC 99-14, ATLAS TDR 14.

\bibitem{Aharrouche:2006nf}
M.~Aharrouche et al., {\em {Energy linearity and resolution of the ATLAS
  electromagnetic barrel calorimeter in an electron test-beam}\/},
  \href{http://dx.doi.org/10.1016/j.nima.2006.07.053}{Nucl.Instrum.Meth. {\bf
  A568} (2006)  601--623}, \href{http://arxiv.org/abs/physics/0608012}{{\tt
  arXiv:physics/0608012}}.

\bibitem{Aharrouche:2007nk}
J.~Colas et al., {\em {Response Uniformity of the ATLAS Liquid Argon
  Electromagnetic Calorimeter}\/},
  \href{http://dx.doi.org/10.1016/j.nima.2007.08.157}{Nucl.Instrum.Meth. {\bf
  A582} (2007)  429--455}, \href{http://arxiv.org/abs/0709.1094}{{\tt
  arXiv:0709.1094 [physics.ins-det]}}.

\bibitem{Aharrouche:2010zz}
M.~Aharrouche et al., {\em {Measurement of the response of the ATLAS liquid
  argon barrel calorimeter to electrons at the 2004 combined test-beam}\/},
  \href{http://dx.doi.org/10.1016/j.nima.2009.12.055}{Nucl.Instrum.Meth. {\bf
  A614} (2010)  400--432}.

\bibitem{Aubert:2002mw}
B.~Aubert et al., {\em {Performance of the ATLAS electromagnetic calorimeter
  end-cap module 0}\/},
  \href{http://dx.doi.org/10.1016/S0168-9002(03)00344-9}{Nucl.Instrum.Meth.
  {\bf A500} (2003)  178--201}.

\bibitem{Aubert:2002dm}
B.~Aubert et al., {\em {Performance of the ATLAS electromagnetic calorimeter
  barrel module 0}\/},
  \href{http://dx.doi.org/10.1016/S0168-9002(03)00345-0}{Nucl.Instrum.Meth.
  {\bf A500} (2003)  202--231}.

\bibitem{JetCleaning2010conf}
{The ATLAS Collaboration}, {\em {Data-quality requirements and event cleaning
  for jets and missing transverse energy reconstruction with the ATLAS detector
  in Proton-Proton collisions at a center-of-mass energy of
  $\sqrt{s}=7$~TeV}\/},  ATLAS-CONF-2010-038.

\bibitem{Banfi:731840}
D.~Banfi et al., {\em {Cell response equalization of the ATLAS electromagnetic
  calorimeter without the direct knowledge of the ionization signals}\/},  J.
  Instrum. {\bf 1} (2006)  P08001.

\bibitem{Collard:1058294}
C.~Collard et al., {\em Prediction of signal amplitude and shape for the ATLAS
  electromagnetic calorimeter\/},  ATL-LARG-PUB-2007-010.

\bibitem{Abreu:2010zzc}
H.~Abreu et al., {\em {Performance of the electronic readout of the ATLAS
  liquid argon calorimeters}\/},
\href{http://dx.doi.org/10.1088/1748-0221/5/09/P09003}{JINST {\bf 5} (2010)
  P09003}.
%%CITATION = JINST,5,P09003;%%.

\bibitem{CrystalBall1}
{M.J. Oreglia}, {\em {A study of the reactions $\psi^\prime \rightarrow \gamma
  \gamma \psi$}\/},  {Ph.D. thesis, SLAC-R-236 (1980), Appendix D}.

\bibitem{CrystalBall2}
{J.E. Gaiser}, {\em {Charmonium spectroscopy from radiative decays of the
  J/$\psi$ and $\psi^\prime$}\/},  {Ph.D. thesis, SLAC-R-255 (1982), Appendix
  F}.

\bibitem{Zedometry}
{The ALEPH, DELPHI, L3, OPAL and SLD Collaborations}, {\em Precision
  electroweak measurements on the Z resonance\/},  Physics Reports {\bf 427}
  (2006)  257--454, \href{http://arxiv.org/abs/hep-ex/0509008}{{\tt
  arXiv:hep-ex/0509008}}.

\bibitem{TPMethod-CDF}
{The CDF Collaboration}, {\em {First measurement of inclusive W and Z cross
  sections from Run II of the Fermilab Tevatron Collider}\/},  Phys. Rev. Lett.
  {\bf 94} (2005)  091803, \href{http://arxiv.org/abs/hep-ex/0406078}{{\tt
  arXiv:hep-ex/0406078}}.

\bibitem{TPMethod-D0}
{The D0 Collaboration}, {\em {Measurement of the shape of the boson rapidity
  distribution for $p\bar{p} \rightarrow Z/\gamma^* \rightarrow e^+e^- + X$
  events produced at $\sqrt{s}$ of 1.96 TeV}\/},  Phys. Rev. {\bf D76} (2007)
  012003, \href{http://arxiv.org/abs/hep-ex/0702025}{{\tt
  arXiv:hep-ex/0702025}}.

\bibitem{CDFErrors}
{C. Blocker}, {\em {Treatment of Errors in Efficiency Calculations}\/},
  CDF/MEMO/STATISTICS/PUBLIC/7168, 2004.

\bibitem{JpsimmPaper}
{The ATLAS Collaboration}, {\em {Measurement of the differential cross-sections
  of inclusive, prompt and non-prompt J/$\psi$ production in proton-proton
  collisions at $\sqrt{s} = 7$ TeV}\/},  Nucl. Phys. {\bf B850} (2011)
  387--444, \href{http://arxiv.org/abs/1104.3038}{{\tt arXiv:1104.3038
  [hep-ex]}}.

\bibitem{sPlot:2005}
{M. Pivk and F.R. Le Diberder}, {\em {sPlot : a statistical tool to unfold data
  distributions}\/},  Nucl. Istrum. Meth. {\bf A555} (2005)  356--369,
  \href{http://arxiv.org/abs/physics/0402083}{{\tt arXiv:physics/0402083}}.

\bibitem{TR2011conf}
{The ATLAS Collaboration}, {\em {Particle identification performance of the
  ATLAS Transition Radiation Tracker}\/},  ATLAS-CONF-2011-128.

\bibitem{ATLASCosmicPerf}
{The ATLAS Collaboration}, {\em Studies of the performance of the ATLAS
  detector using cosmic-ray muons\/},  Eur. Phys. J. {\bf C71} (2011)  1593,
  \href{http://arxiv.org/abs/1011.6665}{{\tt arXiv:1011.6665
  [physics.ins-det]}}.

\bibitem{ATLAS900Perf}
{The ATLAS Collaboration}, {\em Performance of the ATLAS Detector using First
  Collision Data\/},  JHEP {\bf 1009} (2010)  056,
  \href{http://arxiv.org/abs/1005.5254}{{\tt arXiv:1005.5254 [hep-ex]}}.

\end{thebibliography}\endgroup


\providecommand{\href}[2]{#2}\begingroup\raggedright\endgroup

%%%%%%%%%%%%%%%%%%%%%%%%%%%%%%%%%%%%%%%%%%%%%%%%%%%%%%%%%%%%%%%%%%%%%%%%%%%%%%%
% Author list
%%%%%%%%%%%%%%%%%%%%%%%%%%%%%%%%%%%%%%%%%%%%%%%%%%%%%%%%%%%%%%%%%%%%%%%%%%%%%%

\clearpage
\onecolumn
% ATLAS Collaboration author list for 24-JUN-2011
% Data extracted on 14-Jul-2011 for paperid 101
%\documentclass[11pt]{article}
%\usepackage{a4wide}\begin{document}
\begin{flushleft}
{\Large The ATLAS Collaboration}

\bigskip

G.~Aad$^{\rm 48}$,
B.~Abbott$^{\rm 111}$,
J.~Abdallah$^{\rm 11}$,
A.A.~Abdelalim$^{\rm 49}$,
A.~Abdesselam$^{\rm 118}$,
O.~Abdinov$^{\rm 10}$,
B.~Abi$^{\rm 112}$,
M.~Abolins$^{\rm 88}$,
H.~Abramowicz$^{\rm 153}$,
H.~Abreu$^{\rm 115}$
% ...
\bigskip

% ...
$^{10}$ Institute of Physics, Azerbaijan Academy of Sciences, Baku, Azerbaijan\\
$^{11}$ Institut de F\'isica d'Altes Energies and Departament de F\'isica de la Universitat Aut\`onoma  de Barcelona and ICREA, Barcelona, Spain\\
% ...
$^{48}$ Fakult\"{a}t f\"{u}r Mathematik und Physik, Albert-Ludwigs-Universit\"{a}t, Freiburg i.Br., Germany\\
$^{49}$ Section de Physique, Universit\'e de Gen\`eve, Geneva, Switzerland\\
% ...
$^{88}$ Department of Physics and Astronomy, Michigan State University, East Lansing MI, United States of America\\
% ...
$^{111}$ Homer L. Dodge Department of Physics and Astronomy, University of Oklahoma, Norman OK, United States of America\\
$^{112}$ Department of Physics, Oklahoma State University, Stillwater OK, United States of America\\
% ...
$^{115}$ LAL, Univ. Paris-Sud and CNRS/IN2P3, Orsay, France\\
% ...
$^{118}$ Department of Physics, Oxford University, Oxford, United Kingdom\\
% ...
$^{153}$ Raymond and Beverly Sackler School of Physics and Astronomy, Tel Aviv University, Tel Aviv, Israel\\
% ...
\end{flushleft}

%\end{document}

\end{document}